\documentclass{aa}
\usepackage{txfonts}
\usepackage{graphicx}
\begin{document}
\title{MHD mode coupling in the neighbourhood of a 2D null point}

\author{J.~A. McLaughlin   \and A.~W. Hood }


\offprints{J.~A. McLaughlin,  \email{james@mcs.st-and.ac.uk}}

\institute{School of Mathematics and Statistics, University of St
Andrews, KY16 9SS, UK}




\date{Received 5 May 2006 / Accepted 4 August 2006}
\authorrunning{McLaughlin \& Hood}
\abstract {At this time there does not exist a robust set of rules
connecting low and high $\beta$ waves across the $\beta \approx 1$
layer. The work here contributes specifically to what happens when
a low $\beta$ fast wave crosses the $\beta \approx 1$  layer  and
transforms into high $\beta$ fast and slow waves.} {The nature of
fast and slow magnetoacoustic waves is investigated in a finite
$\beta$ plasma in the neighbourhood of a two-dimensional null
point.} {The linearised equations are solved in both polar and
cartesian forms with a two-step Lax-Wendroff numerical scheme.
Analytical work (e.g. small $\beta$ expansion and WKB
approximation) also complement the work.} {It is found that when a
finite gas pressure is included in magnetic equilibrium containing
an X-type null point, a fast wave is attracted towards the null by
a refraction effect and that a slow wave is generated as the wave
crosses the $\beta \approx 1$ layer. Current accumulation occurs
close to the null and along nearby separatrices. The fast wave can
now \emph{pass through the
origin} due to the non-zero sound speed, an effect not previously
seen in related papers but clear seen for larger values of $\beta$.
Some of the energy can now leave the region of the null point
and there is again generation of a slow wave component  (we
find that the fraction of the incident wave converted to a slow
wave is proportional to $\beta$). We conclude that there are two
competing phenomena; the refraction effect (due to the variable
Alfv\'en speed) and the contribution from the non-zero sound
speed.} {These experiments illustrate the importance of the
magnetic topology and of the location of the $\beta \approx 1$
layer in the system.}

\maketitle

\section{Introduction}\label{section1}
MHD wave motions have recently been observed in the Sun's atmosphere 
with the SOHO and TRACE satellites, see for example the detection of
slow  magnetoacoustic waves by Berghmans \& Clette, (1999) and De 
Moortel \emph{et al.} (2000) and of fast magnetoacoustic waves by
Nakariakov \emph{et al.} (1999).  It is clear that the local coronal 
magnetic field plays a key role in
determining their propagation properties.

To begin to understand this 
inhomogeneous magnetised  environment, it is useful to look at the
structure (topology) of the magnetic field itself.  Potential field 
extrapolations of the coronal magnetic field can be made from
photospheric magnetograms. Such extrapolations show the existence of 
an important feature of the topology; \emph{null points}. Null
points are points in the field where the Alfv\'en speed is zero. 
Detailed investigations of the coronal magnetic field, using such
potential field calculations, can be found in \cite{Beveridge2002} 
and \cite{Brown2001}.

Building on the earlier work of Bulanov and Syrovatskii (1980) and 
Craig and Watson (1992),
McLaughlin \& Hood (2004) found that for a single 2D null point, the 
fast magnetoacoustic wave was attracted to the null and the wave
energy accumulated there. In addition, they found that the
Alfv\'en wave energy accumulated along the separatrices, the 
topological feature that separates regions with different magnetic 
flux
connectivity.
Their paper looked at
MHD wave propagation in a $\beta=0$ plasma. The aim of this paper is 
to extend their model to include plasma pressure (finite $\beta$
plasma). The most obvious effect is the introduction of
slow magnetoacoustic waves. The fast wave can now also pass through 
the null point (as we have a non-zero fast wave speed there due to
the finite sound speed) and, thus, perhaps carry wave
energy away from that area. There could also be
coupling and wave conversion near the location where the sound speed 
and Alfv\'en speed become comparable in magnitude.
However, the exact nature of
such coupling in MHD
is unknown and will be looked at here. The behaviour of
the Alfv\'en waves is unaffected by a finite $\beta$ (the plasma 
pressure plays no role in its propagation) and so the description by
\cite{McLaughlin2004}  remains valid in this linear, 2D regime.

Waves in the neighbourhood of a single 2D null point have been investigated
by various authors. \cite{Bulanov1980} provided a
detailed discussion of the propagation of fast and Alfv\'en waves
using cylindrical symmetry. \cite{CraigWatson1992} mainly consider 
the radial propagation of the $m=0$ mode (where $m$ is the
azimuthal wavenumber) using a mixture of analytical and numerical
solutions. They show that the propagation of the $m=0$ wave
towards the null point generates an exponentially large increase in
the current density and that magnetic resistivity dissipates this current in
a time related to $\log { \eta }$. Craig and McClymont (1991, 1993) 
investigate the normal mode solutions for both $m=0$ and $m\ne 0$
modes with resistivity included. Again, they emphasise that the 
current builds up as the inverse square of the radial distance from 
the
null point. All these investigations were carried out using 
cylindrical models in which the generated waves encircled the null 
point. In
a sense, there is nowhere else for the wave to propagate except into 
the null point.

A very detailed and comprehensive set of 2D numerical simulations
of wave propagation in a stratified magneto-atmosphere was
conducted by \cite{Rosenthal2002} and \cite{Bogdan2003}. In these
simulations, an oscillating piston generated both fast and slow
MHD waves on a lower boundary and sent these waves up into the
stratified, magnetised plasma. Their calculations showed that
there was coupling between the fast and slow waves, and that this
coupling was confined to a thin layer where the sound speed and
the Alfv\'en velocity are comparable in magnitude, i.e. where the
plasma-beta approaches unity. Away from this conversion zone, the
waves were decoupled as either the magnetic pressure or plasma
pressure dominated. {{In this, their papers and ours have a similar goal; to
see how the topology affects the propagation of MHD waves, in a 2D 
system where the
ratio of the sound speed to the Alfv\'en speed varys along every 
magnetic line of force.}}

Other authors have also looked at MHD mode coupling. \cite{Cally1997} 
describes 2D simulations in which both $f$-modes and $p$-modes
are (partially) converted to slow magnetoacoustic gravity waves, due 
to strong gravitational stratification. \cite{Ineke2004}
investigated driving slow waves on the boundary of a 2D geometry with 
a horizontal density variation. They found coupling between
slow and fast waves and phase mixing of the slow waves. The coupling 
of different wave modes has also been investigated
by \cite{Ferraro1958}, \cite{Zhugzhd} (with Meijer G-functions) and 
\cite{Cally2001} (with hypergeometric $_2F_3$ functions).
All these works considered mode coupling through a gravitational 
stratification (vertical density inhomogenity) {and, in 
particular,
investigated the propagation of waves from a high $\beta$ to a low 
$\beta$ plasma. As waves propagate towards a null point, it is
the nature of low $\beta$ to high $\beta$ propagation that is important}.
Finally, the
coupling of fast waves and Alfv\'en waves has been investigated by 
\cite{Parker1991} (linear with a density gradient) and by
\cite{Valery1997} (nonlinear excitation).

The paper has the following outline. In Section \ref{sec:1.1}, the
basic equations are described and the importance of the plasma
$\beta$ discussed. The results for the fast and slow
magnetoacoustic waves are presented for a reference value of the
plasma $\beta$, namely $\beta_0=0.25$ in Section \ref{sec:2.1}.
Section \ref{Interpretion} provides an interpretation of our
results, with a discussion of mode conversion in Section
\ref{modeconversion}. The robustness of our results and a clearer
demonstration of mode conversion is shown in Section \ref{sec:3.1}
for $\beta_0=2.25$.  {{The conclusions are given in Section 7 and the appendices provide analytical approximations.}}

\begin{figure}[t]
\begin{center}
\includegraphics[width=2.0in]{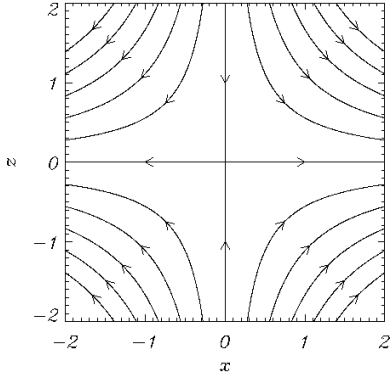}
\caption{The equilibrium equilibrium magnetic field and coordinate system.}
\label{figureone}
\end{center}
\end{figure}

\section{Basic Equations}\label{sec:1.1}
The usual MHD equations appropriate to the solar corona are used, 
with pressure and resistivity included. Hence,
\begin{eqnarray*}
   \rho \left[ {\partial {\bf{v}}\over \partial t} + \left( 
{\bf{v}}\cdot\nabla \right) {\bf{v}} \right] &=& \frac{1}{\mu}
\left(\nabla \times {\bf{B}}\right)\times {\bf{B}}- \nabla p \; 
,\label{eq:2.1a} \\
{\partial {\bf{B}}\over \partial t} &=& \nabla \times 
\left({\bf{v}}\times {\bf{B}}\right ) + \eta \nabla^2{\bf{B}} \; 
,\label{eq:2.1b} \\
{\partial \rho\over \partial t} + \nabla \cdot \left (\rho 
{\bf{v}}\right) &=& 0 \; , \label{eq:2.1c} \\
{\partial p\over \partial t} + \left( \bf{v} \cdot \nabla \right) p 
&=& - \gamma p \nabla \cdot \bf{v}\; , \label{eq:2.1d}
\end{eqnarray*}
where $\rho$ is the mass density, ${\bf{v}}$ is the plasma
velocity, ${\bf{B}}$ the magnetic induction (usually called the
magnetic field), $p$ is the plasma pressure, $ \mu = 4 \pi \times 
10^{-7} \/\mathrm{Hm^{-1}}$  the magnetic permeability,
$\eta = 1/\mu\sigma$ is the magnetic diffusivity $\left( 
\mathrm{m}^2\mathrm{s}^{-1}\right)$
and $\sigma$ the electrical conductivity.

\subsection{Basic equilibrium}\label{sec:1.2}
The basic potential magnetic field structure is taken as a simple
two dimensional, X-type neutral point. There are a lot of complicated
effects including mode conversion and coupling; and a 2D geometry
allows one to understand these effects better, before the extension 
to 3D. Therefore, the equilibrium
magnetic field (see Figure \ref{figureone}) is taken as
\begin{equation}\label{eq:2.2}
{\bf{B}}_0 = B \left({x\over a}, 0, -{z\over a}\right),
\end{equation}
where $B$ is a characteristic field strength and $a$ is the
length scale for magnetic field variations. Obviously, this
configuration is no longer valid far from the null point, as the 
field strength tends to infinity. However, \cite{mcl2006}
looked at a magnetic field with a field strength that decays far from the null
(again for a $\beta=0$ plasma) and they found that the key results from
\cite{McLaughlin2004} remain valid close to the null.

As in \cite{McLaughlin2004}, the linearised MHD equations are used
to study the nature of the wave propagation near the null point.
Using subscript $0$ for equilibrium quantities and $1$ for
perturbed quantities, the linearised equation of motion becomes
\begin{equation}\label{eq:2.3}
  \rho_0 \frac{\partial \mathbf{v}_1}{\partial t} = \left(\frac{ 
\nabla \times \mathbf{B}_1}{\mu} \right) \times \mathbf{B}_0 - \nabla 
p_1\; ,
\end{equation}
the linearised induction equation
\begin{equation}\label{eq:2.4}
{\partial {\bf{B}}_1\over \partial t} = \nabla \times
     ({\bf{v}}_1 \times {\bf{B}}_0)  + \eta \nabla^2 {{\bf{B}}_1}\; ,
\end{equation}
the linearised equation of mass continuity
\begin{equation}\label{eq:2.5}
\frac{\partial \rho_1} {\partial t} + \nabla\cdot\left( \rho_0 
\mathbf{v} _1 \right) =0 \; ,
\end{equation}
and the adiabatic energy equation
\begin{equation}\label{eq:2.6}
\frac{\partial p_1} {\partial t} = -\gamma p_0 \left( \nabla \cdot 
\mathbf{v} _1  \right) \; .
\end{equation}

We will not discuss equation (\ref{eq:2.5}) further as it can be
solved once we know $\mathbf{v} _1$. We assume the background
density and gas pressure are uniform and labelled as $\rho_0$ and
$p_0$ respectively. However, we note that a spatial variation in
$\rho _0$ can cause phase mixing (\cite{Heyvaerts1983},
\cite{DeMoortel1999}, \cite{Hood2002}).

\subsection{Coordinate system and non-dimensionalisation}\label{sec:1.2b}

{The linearised velocity, $\mathbf{v}_1$, is split into 
components parallel and perpendicular to the
equilibrium magnetic field. Thus,}
\begin{eqnarray*}
\mathbf{v}_1 = {V}_\parallel \left( \frac {\mathbf{B}_0 } {\sqrt 
{\mathbf{B}_0 \cdot \mathbf{B}_0} } \right) - {V}_\perp
\left( \frac { \nabla A_0 } {\sqrt {\mathbf{B}_0 \cdot \mathbf{B}_0} 
} \right) + v_y \: {\hat{\bf{y}}}
\end{eqnarray*}
where $A_0=-Bxz/a$ is the $y$-component of the vector magnetic
potential. The terms in brackets are unit vectors. {Splitting 
the velocity
into components  parallel and
perpendicular to the equilibrium magnetic field makes MHD
mode detection and interpretation easier. For example, in a low 
$\beta$ plasma,} the
slow wave is guided by the magnetic field and has a velocity
component that is mainly field-aligned. This makes perfect sense
when $\beta \ll 1$ but its usefulness is less clear when $\beta
\gg 1$ near the null point.

To aid the numerical calculation, our
primary variables are considered to be ${\rm{v}}_ \perp = \sqrt
{\mathbf{B}_0 \cdot \mathbf{B}_0} {V}_\perp $ and
${\rm{v}}_\parallel = \sqrt {\mathbf{B}_0 \cdot \mathbf{B}_0}
V_\parallel$.

{Taking  $v_y =0$ means} we do not
consider the Alfv\'en wave here as the  description by
\cite{McLaughlin2004} is still valid (the plasma pressure plays no
role in its propagation).

{We non-dimensionalise all variables by setting} 
${\rm{\bf{v}}}_1 = {\rm{v}}_0 {\mathbf{v}}_1^*$, ${\rm{v}}_\perp =
{\rm{v}}_0 B {\rm{v}}_\perp^*$,${\rm{v}}_\parallel =  {\rm{v}}_0 B 
{\rm{v}}_\parallel^*$,  ${\mathbf{B}}_0 = B {\mathbf{B}}_0^*$,
${\mathbf{B}}_1 = B {\mathbf{B}}_1^*$, $x = a x^*$, $z=az^*$, $p_1 = 
p_0 p_1^*$, $\nabla = \frac{1}{a}\nabla^*$,
$t={t}_0 t^*$, $A_0=a B A_0^*$ and $\eta = \eta_0$, where we let * 
denote a dimensionless quantity and
${\rm{v}}_0$, $B$, $a$,$p_0$, ${t}_0$ and $\eta_0$
are constants with the dimensions of the variable they are scaling. 
We then set $\frac {B}{\sqrt{\mu \rho _0 } } ={\rm{v}}_0$ and
${\rm{v}}_0 =  \frac{a}{{t}_0}$ (i.e. we measure our speed in units 
of ${\rm{v}}_0$, which can be thought of as a
constant background Alfv\'{e}n speed). We also set
$\frac {\eta_0 {t}_0 } {a^2} =R_m^{-1}$, where $R_m$ is the magnetic 
Reynolds number, and set $ {\beta_0} = \frac {2 \mu p_0}{B^2}$,
where $\beta_0$ is the plasma $\beta$ at a distance unity from the 
origin (see Section \ref{sec:1.4}).
Thus, we generate {the non-dimensionalised versions of 
equations} (\ref{eq:2.3}),
(\ref{eq:2.4}) and (\ref{eq:2.6}) and under these scalings,
$t^*=1$ (for example) refers to $t={t}_0=  \frac{a}{{\rm{v}}_0}$;
i.e. the time taken to travel a distance $a$ at the 
{reference} background
Alfv\'en speed. For the rest of this paper, we drop the star
indices; the fact that they are now non-dimensionalised is
understood.

\subsection{Linearised equations}\label{sec:1.3}
The linearised equations are:
\begin{eqnarray}
  \frac{\partial {{\rm{v}}_\perp}}{\partial t} &=& v_A^2 \left( x,z 
\right) \left( \frac{\partial b_z}
{\partial x} - \frac{\partial b_x}{\partial z}  \right) - \frac 
{\beta_0}{2} \left( z \frac{ \partial p_1}
{\partial x}  + x   \frac{ \partial p_1}{\partial z} \right) 
\nonumber          \label{a} \\
\frac{\partial {{\rm{v}}_\parallel}}{\partial t} &=& - \frac 
{\beta_0}{2}  \left( x \frac{ \partial p_1}
{\partial x} -z \frac{ \partial p_1}{\partial z} \right) \label{b} 
\nonumber    \\
\frac{\partial b_x}{\partial t} &=& -\frac{\partial 
{{\rm{v}}_\perp}}{\partial z} + \frac {1}{R_m}
  \left( \frac{\partial^2 b_x}{\partial x^2} + \frac{\partial^2 
b_x}{\partial z^2} \right) \label{c}         \nonumber     \\
\frac{\partial b_z}{\partial t} &=& \; \;   \frac{\partial 
{{\rm{v}}_\perp}}{\partial x} + \frac {1}{R_m}
\left( \frac{\partial^2 b_z}{\partial x^2}+ \frac{\partial^2 
b_z}{\partial z^2} \right) \label{d}         \nonumber     \\
\frac{\partial p_1 }{\partial t} &=& \frac {-\gamma}{x^2+z^2} \left[ 
\left(  x \frac{\partial {{\rm{v}}_
\parallel} }{\partial x} - z \frac{\partial {{\rm{v}}_\parallel} 
}{\partial z} \right) - 2 \frac {x^2-z^2}{x^2+z^2} \;
{{\rm{v}}_\parallel}  \right.\nonumber\\
&+&  \left. \left( z \frac{\partial {{\rm{v}}_\perp} 
}{\partial x}  + x \frac {\partial {{\rm{v}}_\perp} }
{\partial z} \right) - \frac {4 x z} {x^2+z^2} \; {{\rm{v}}_\perp} 
\right] \label{e}
\end{eqnarray}
where the dimensionless Alfv\'{e}n speed, $v_A \left( x,z \right) 
=\sqrt{x^2+z^2}$.
Note that by taking $\beta_0=0$ and ignoring resistivity, $\left(R_m 
\to \infty\right)$, we recover the equations discussed in
\cite{McLaughlin2004}.
\subsection{Plasma $\beta$}\label{sec:1.4}
The parameter of key importance in equations (\ref{e}) is $\beta_0$. 
The plasma $\beta$ is defined as the ratio of the thermal
plasma pressure to the magnetic pressure. In most parts of the 
corona, the plasma $\beta$ is much less than unity but,
near null points, the magnetic field strength is small (and is zero 
at the null) and
the plasma $\beta$ becomes large. There is also coupling between the 
perpendicular and parallel velocity components when
$\beta_0\ne =0$ and this coupling is most effective where the sound 
speed and the Alfv\'en velocity are comparable in magnitude.
\cite{Bogdan2003} call this zone the magnetic canopy or the  $\beta 
\approx 1$ layer.

The plasma $\beta$ varies throughout the whole region and
\begin{eqnarray}
  \beta=\frac{2 \mu p_0}{B^2}\frac{1}{x^2+z^2}\;\Rightarrow\;
\beta=\frac{\beta_0}{x^2+z^2} =\frac{\beta_0}{r^2}\;,  \label{night1}
\end{eqnarray}
where $r^2=x^2+z^2$. Therefore, the $\beta=1$ layer occurs at a 
radius $r = \sqrt{\beta_0}$,
where the gas pressure is equal to the magnetic pressure.
However, in a uniform plasma, it is not the $\beta=1$ layer that is 
most important, but instead it is
where the sound speed is equal to the Alfv\'en speed, i.e. $c_s=v_A$. 
Recalling that $c_s=
\sqrt{\frac{\gamma p_0}{\rho_0}}=\sqrt{\frac{\gamma \beta_0}{2} 
\frac{B^2}{\mu \rho_0}  }$
(which implies $c_s^* = \sqrt{\frac{\gamma }{2}\beta_0 }$ in 
non-dimensional units), we see that for the present
non-uniform equilibrium $c_s=v_A\:$ at a radius  $r= 
\sqrt{\frac{\gamma}{2} \beta_0}$ and it is when the incoming wave 
passes
through this layer that the mode coupling occurs.
Of course, the difference between the $\beta=1$ layer at  $r = 
\sqrt{\beta_0}$  and the $c_s=v_A\:$ layer at
$r= \sqrt{\frac{\gamma}{2} \beta_0}$  is very small, and, hence, it 
is easier to refer to this as the $\beta \approx 1$ layer.

{Finally, we note that the basic fast and slow wave speeds for 
this equilibrium are given in dimensionless form as}
\begin{equation}
c_{fast}^2 = {\gamma\over 2}\beta_0 + r^2\;, \qquad c_{slow}^2 = {r^2 
\gamma \beta_0/2\over \gamma \beta_0/2 + r^2}.\label{wavespeeds}
\end{equation}

\section{MHD wave propagation with $\beta_0=0.25$}\label{sec:2.1}
In this section, the linearised MHD equations, namely equations 
(\ref{e}), are solved numerically using a two-step Lax-Wendroff 
scheme.
In dimensionless units, the numerical domain is $-4\leq x \leq 4$ and 
$-4\leq z \leq 4$ and initially we consider a single wave
pulse coming in from the top boundary due to a disturbance in the 
perpendicular component of the
velocity. For the single wave pulse, the boundary conditions are 
chosen such that:
\begin{eqnarray}
{\rm{v}_\perp}(x, 4) &=& \left\{\begin{array}{cl}
{\sin { \omega t } } & {\mathrm{for} \; \;0 \leq t \leq \frac 
{\pi}{\omega} } \\
{0} & { \mathrm{otherwise} }
\end{array} \right. \; , \label{BC} \quad  {\rm{v}_\parallel} (x,4) = 
0  \;,  \\
\frac {\partial {\rm{v}_\perp} } {\partial x } | _{x=-4} &=& 0 \; , 
\quad \frac {\partial {\rm{v}_\perp} } {\partial x }  | _{x=4} = 0
\; , \quad \frac {\partial {\rm{v}_\perp} } {\partial z }  | _{z=-4} 
= 0 \; ,\nonumber \\
\frac {\partial {\rm{v}_\parallel} } {\partial x } | _{x=-4} &=& 0 \; , \quad
\frac {\partial {\rm{v}_\parallel} } {\partial x }  | _{x=4} = 0 \; , 
\quad \frac {\partial {\rm{v}_\parallel} }
{\partial z }  | _{z=-4}  = 0 \; \nonumber .
\end{eqnarray}
Tests show that the central behaviour is largely unaffected by the
choices of side and bottom boundary conditions. The other boundary
conditions on the perturbed magnetic field follow from the
remaining equations and the solenoidal condition.

{The location of the upper boundary is not important and it is 
the value of $\beta_0$ that
determines the distance the fast wave travels before encountering the 
mode conversion region}. {Initially}, we set
$\beta_0=0.25$ and $R_m=10^3$. We also take $\omega=2\pi$, as in 
\cite{McLaughlin2004}.
The results for $ {\rm{v}_\perp}$ can be seen in
Figures \ref{fig:P3.1} and \ref{electricsix} and for 
${\rm{v}_\parallel}$ in Figure \ref{fig:P3.2}.

\subsection{Initial disturbance in the perpendicular component}
  In the zero $\beta$ limit (see McLaughlin \& Hood 2004), the wave 
pulse given above is a fast MHD disturbance,
but for the present problem this is no longer true. In the  system 
presented here, the
fast wave is no longer purely described by a perpendicular
component: it is {\textit{predominately}} given by the
perpendicular component but also has a smaller parallel component.
Thus, the single wave pulse described above is a
{\textit{disturbance in the perpendicular component}}, as opposed
to simply a fast MHD disturbance. This difference in definition
will be important when we come to interpret the different wave
types in Section \ref{Interpretion}.
In addition, it can be seen that the governing equations (\ref{e})
that $\rm{v}_\perp$ acts as a driver for
$\rm{v}_\parallel$. Since the value of the plasma $\beta$ is small on 
the upper boundary, $\rm{v}_\perp$ initially behaves
in the same manner as the zero $\beta$ case. However, the parallel 
component of the velocity is now driven
by the perpendicular component.

Thus,the general solution for $\rm{v}_\parallel$ will consist of two 
parts; a {\emph{complementary function}}, corresponding to a slow
mode disturbance
and a {\emph{particular integral}}, due to the fast mode driver term.
These two parts to the  $\rm{v}_\parallel$ wave are clearly seen in 
the simulations. However, it
is clearer if we first discuss the behaviour of the $\rm{v}_\perp$ 
component before moving on to the parallel component.

\subsection{Behaviour of  $\rm{v}_\perp$}\label{sec:2.2}
\begin{center}
\begin{figure*}[t]
\begin{center}
\hspace{0.01in}
\includegraphics[width=1.5in]{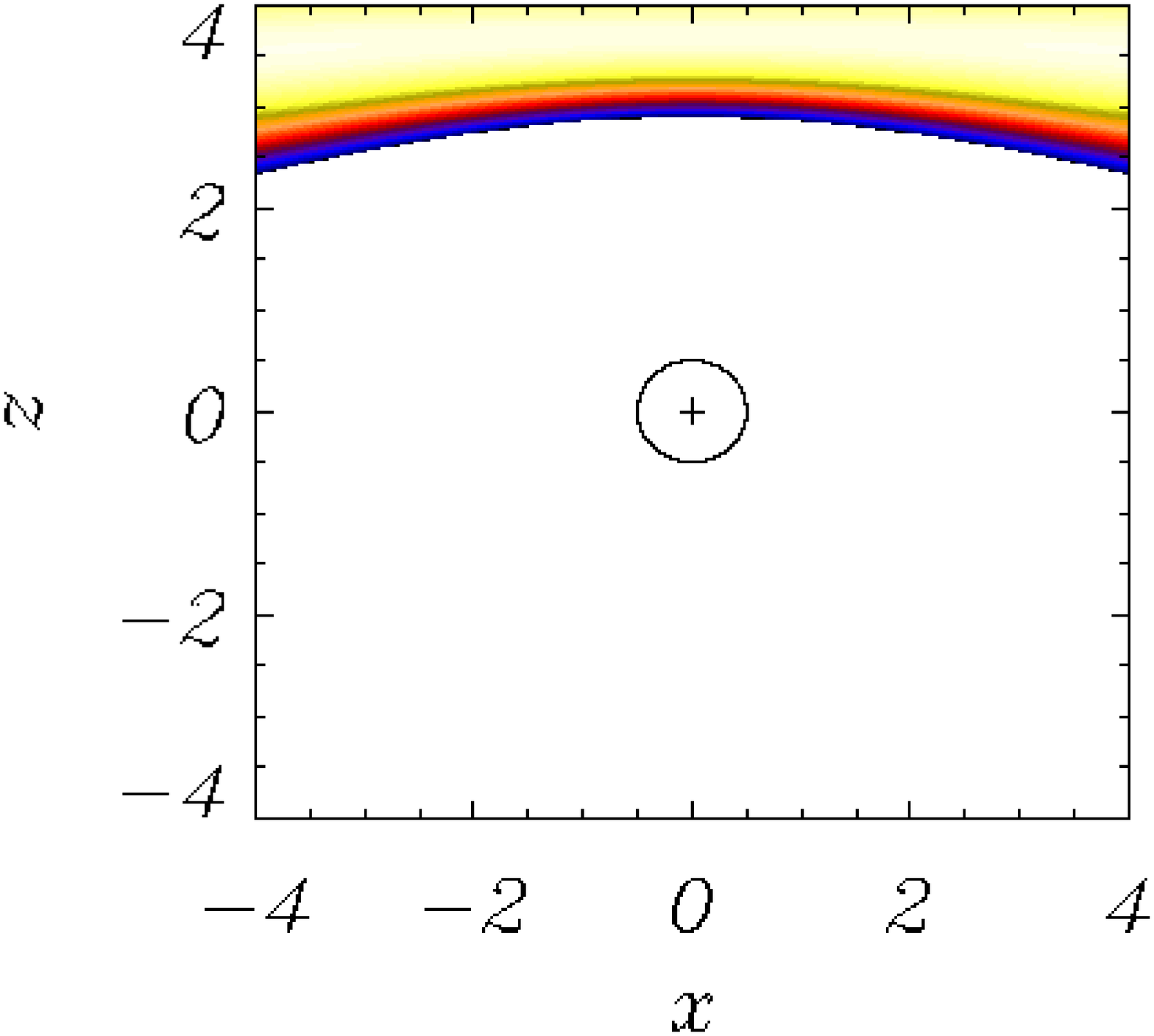}
\hspace{0.08in}
\includegraphics[width=1.5in]{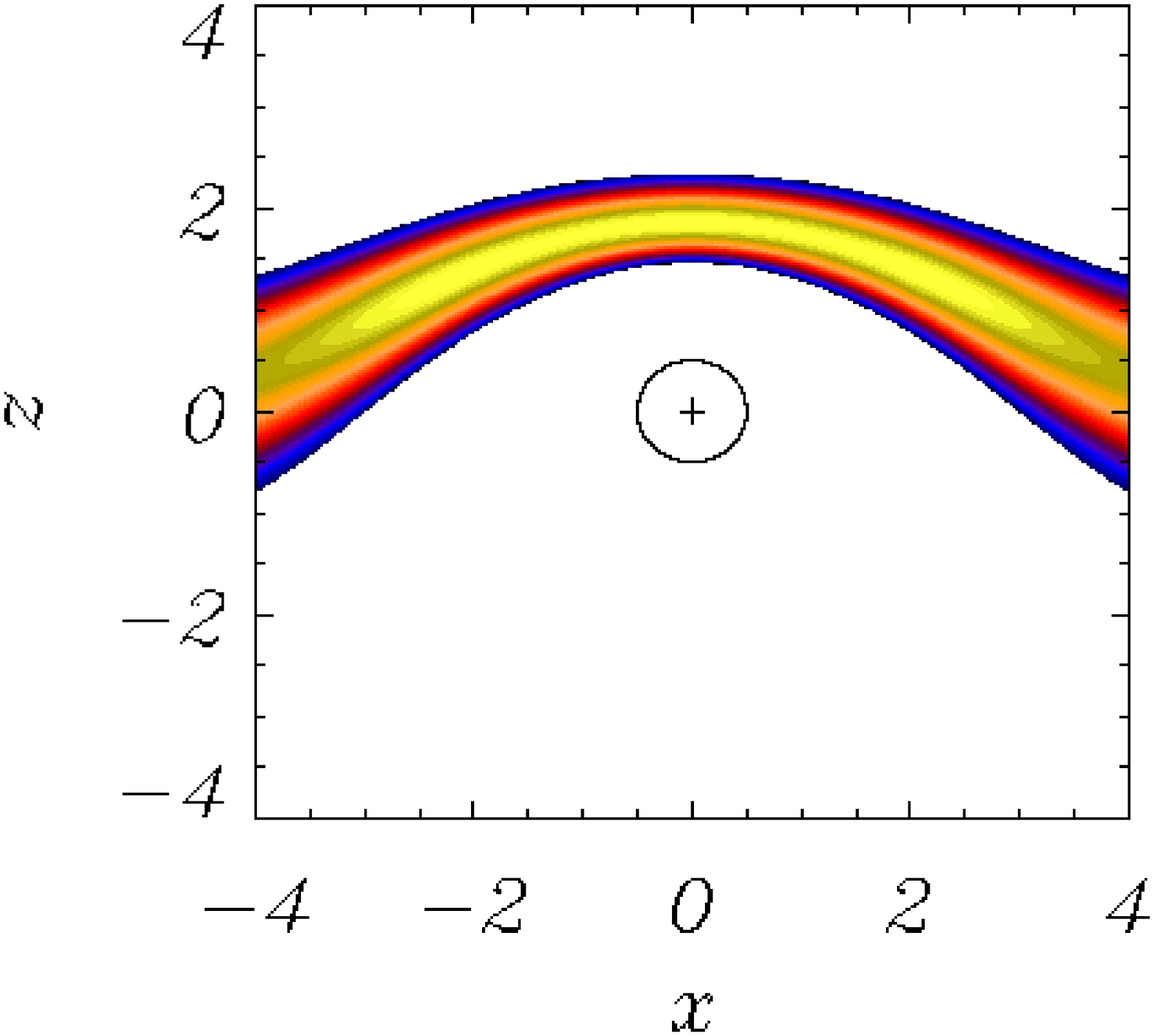}
\hspace{0.08in}
\includegraphics[width=1.5in]{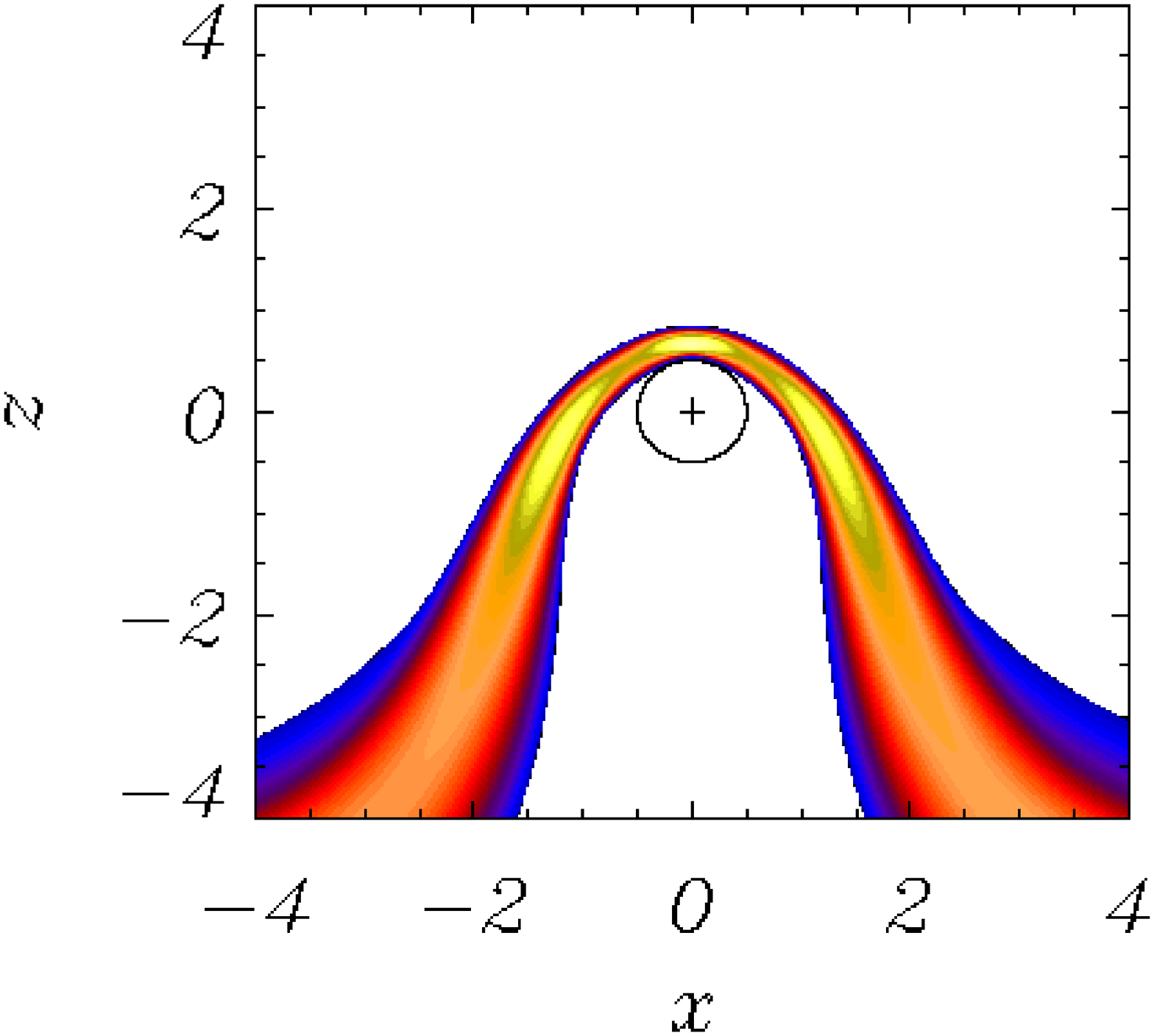}\\
\vspace{0.08in}
\hspace{0.01in}
\includegraphics[width=1.5in]{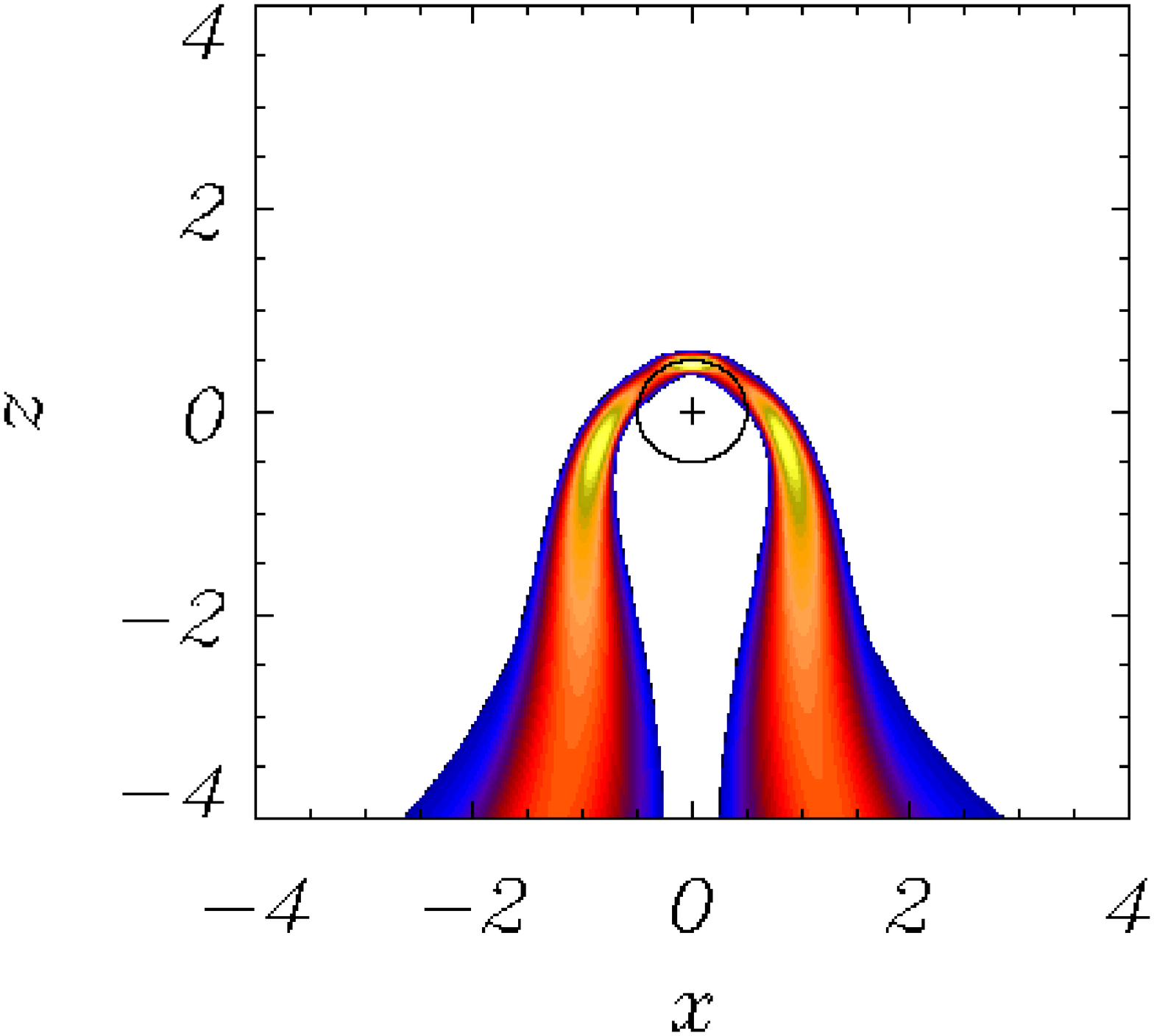}
\hspace{0.08in}
\includegraphics[width=1.5in]{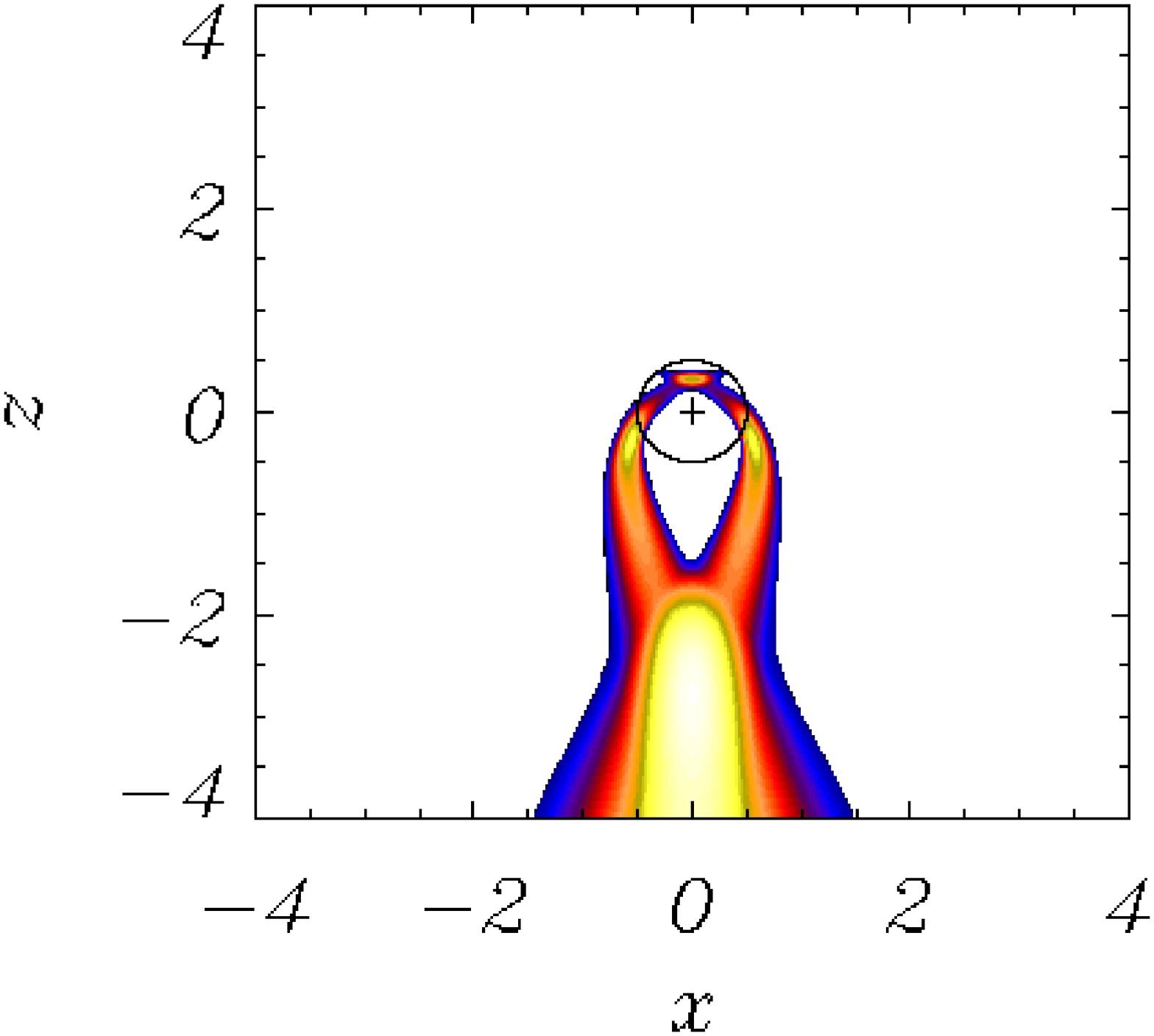}
\hspace{0.08in}
\includegraphics[width=1.5in]{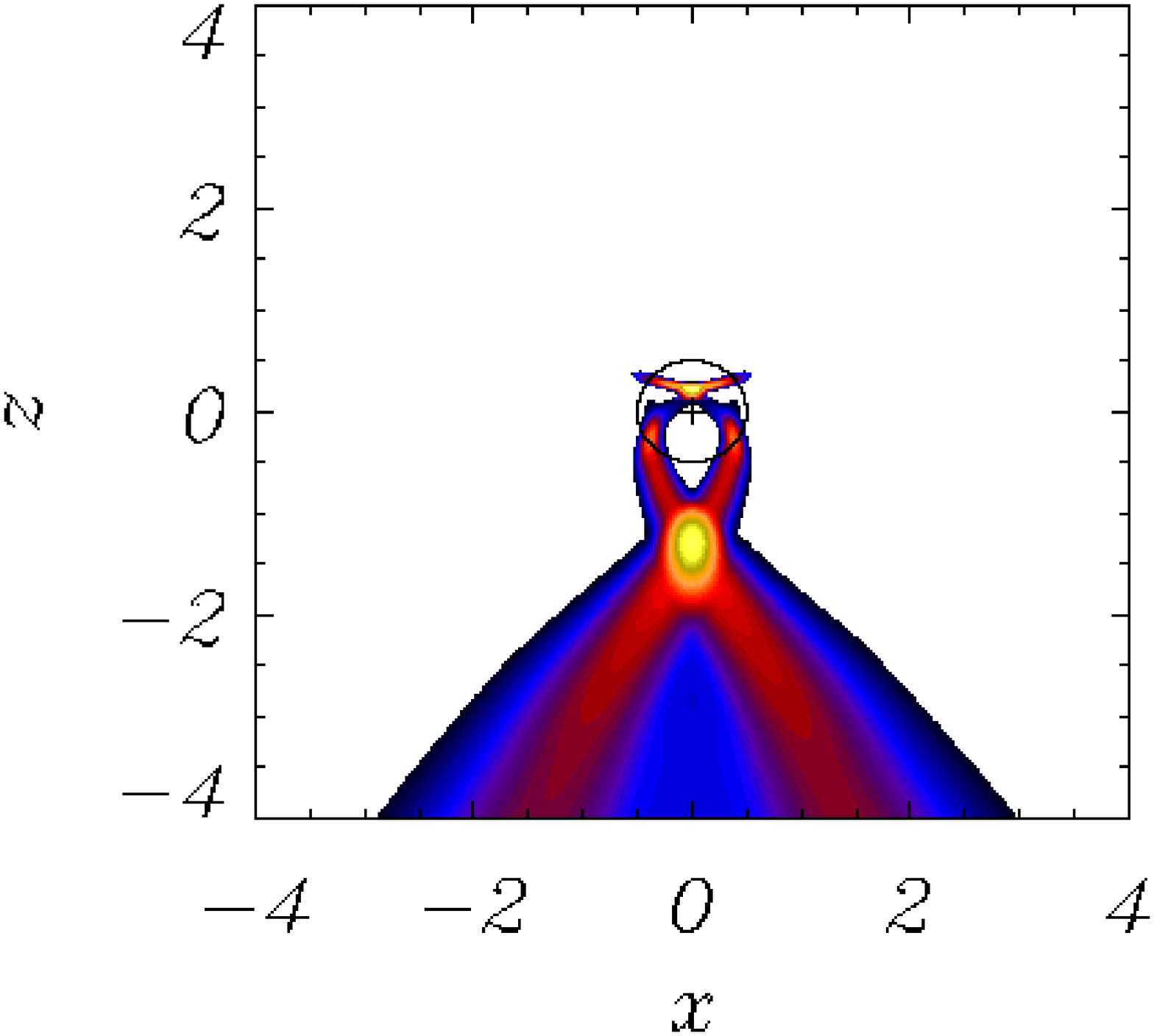}\\
\vspace{0.08in}
\hspace{0.5in}
\includegraphics[width=1.5in]{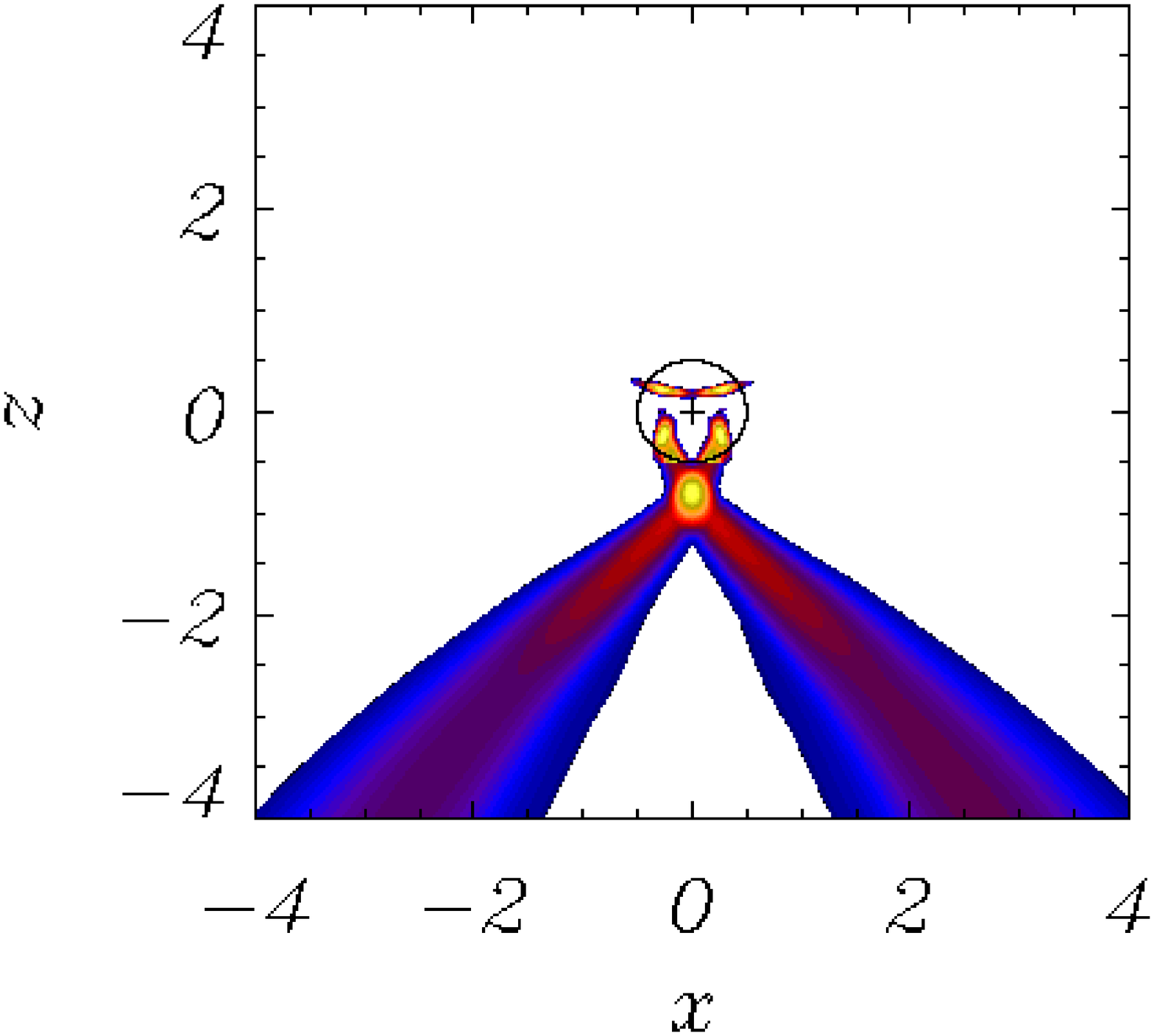}
\hspace{0.08in}
\includegraphics[width=1.5in]{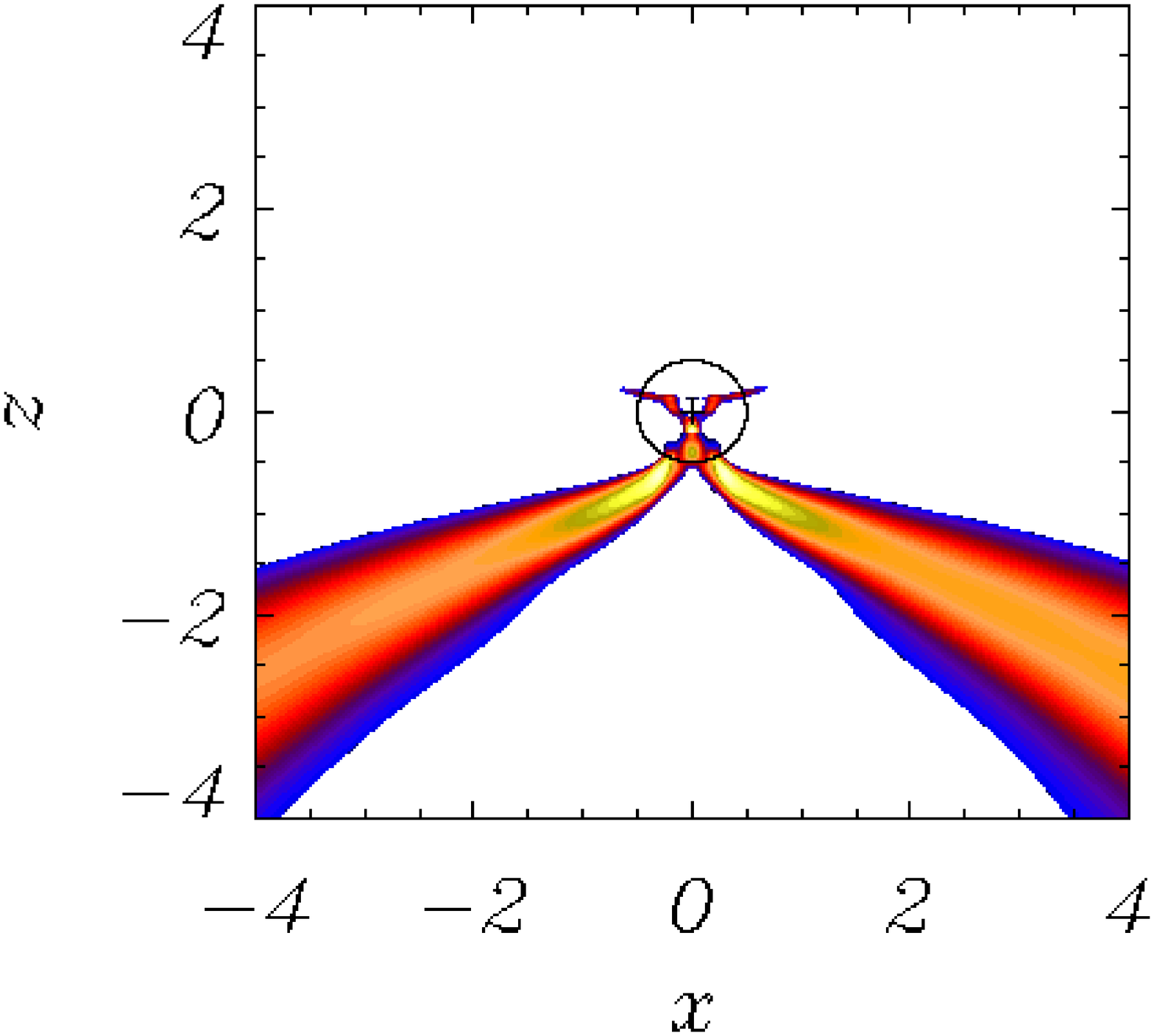}
\hspace{0.08in}
\includegraphics[width=1.9811in]{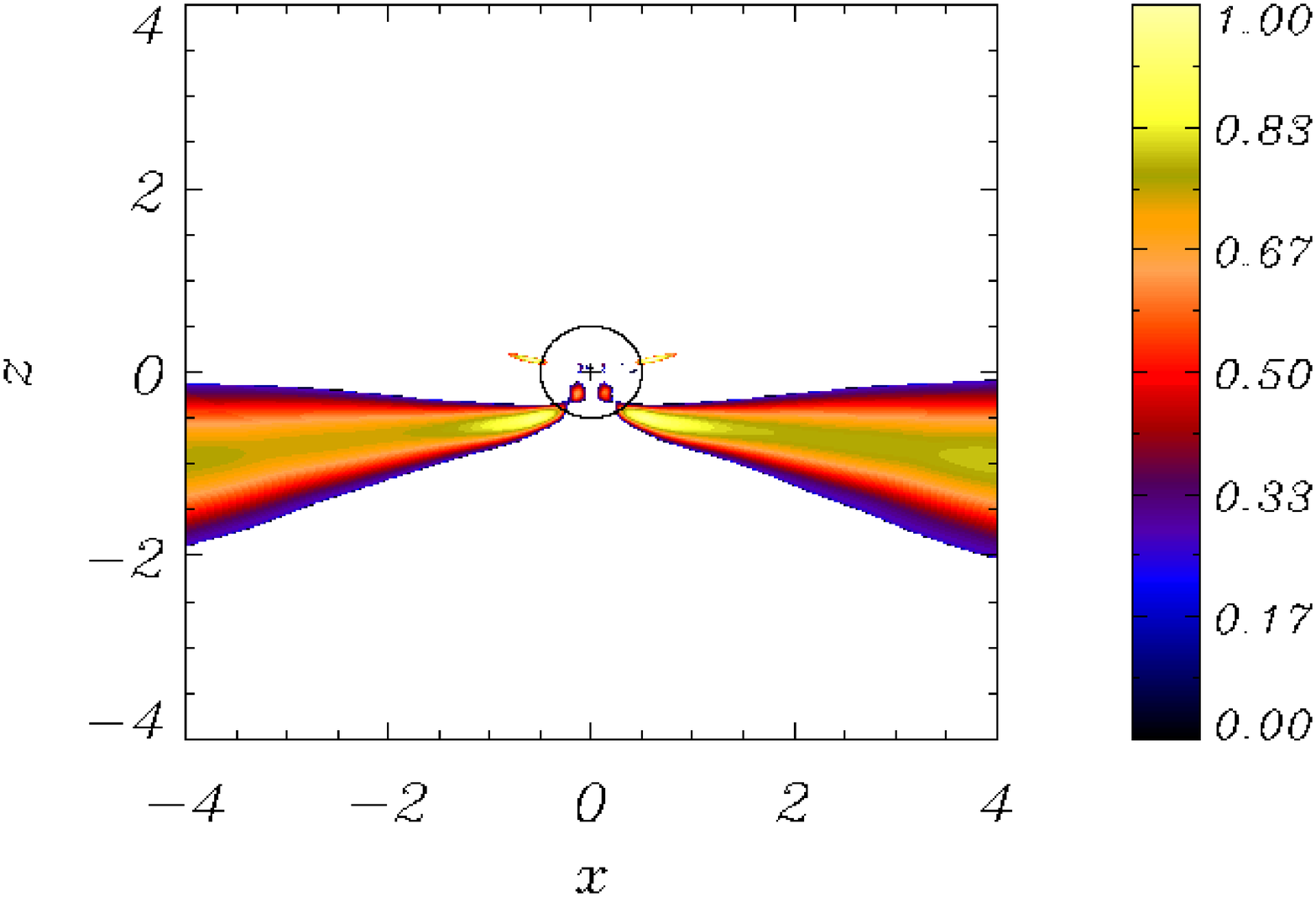}
\caption{Contours of $\rm{v}_\perp$ for  numerical simulation for a 
wave sent in from the upper boundary for $-4 \leq x \leq 4 $
and $\beta_0=0.25$ and its resultant propagation at times $(a)$ 
$t$=0.33, $(b)$ $t$=1.0,
$(c)$ $t$=2.0, $(d)$ $t$=2.33, $(e)$ $t$=2.67, $(f)$ $t$=3.0, $(g)$ 
$t=$3.33, $(h)$ $t$=3.67 and $(i)$ $t$=4.0, labelling from
top left to bottom right. The black circle indicates the position of 
the $c_s=v_A\:$ layer. The cross denotes the null
point in the magnetic configuration.}
\label{fig:P3.1}
\end{center}
\end{figure*}
\end{center}

\begin{center}
\begin{figure*}
\begin{center}
\vspace{0.4in}
\hspace{0.05in}
\includegraphics[width=1.6in]{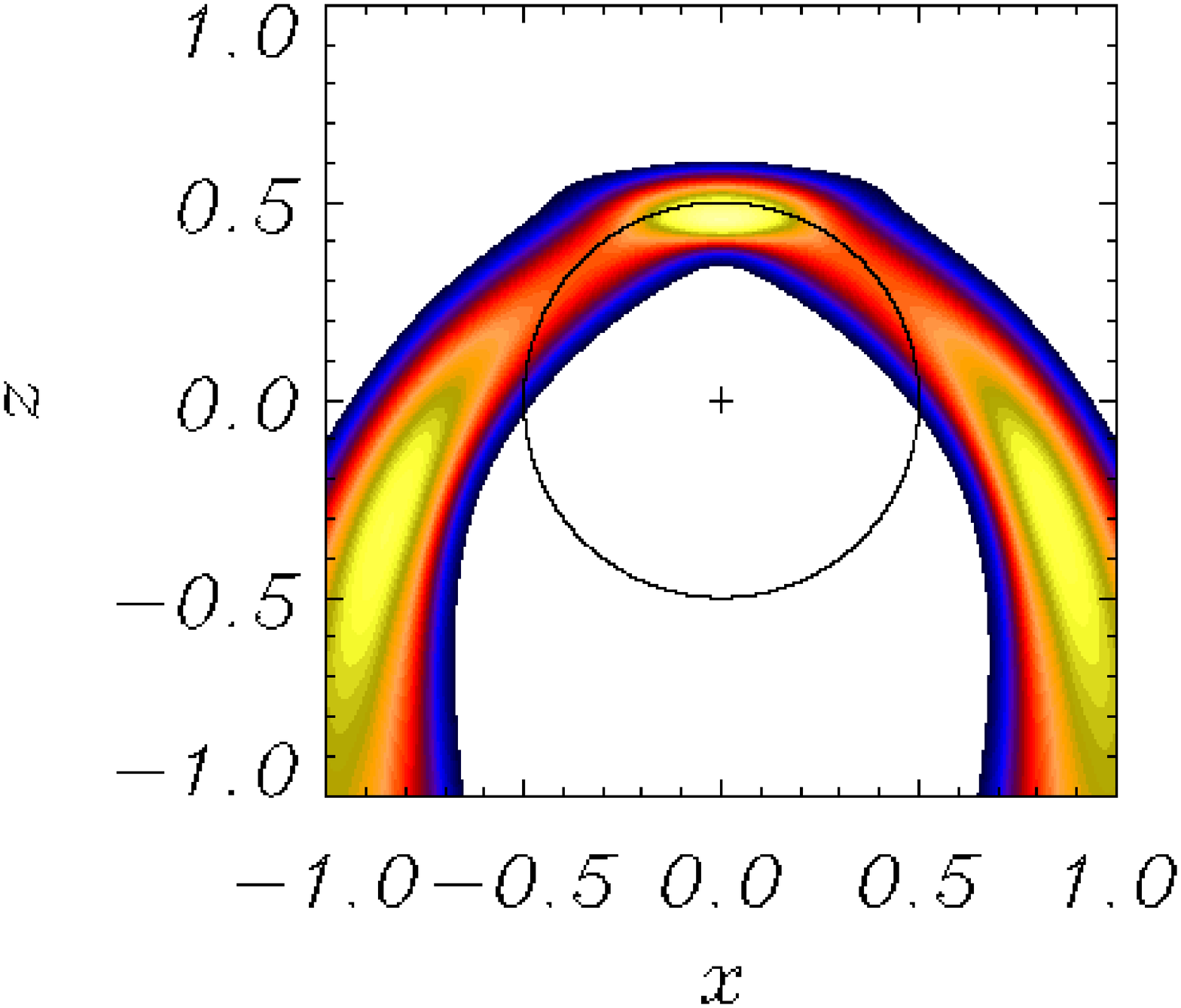}
\hspace{0.04in}
\includegraphics[width=1.6in]{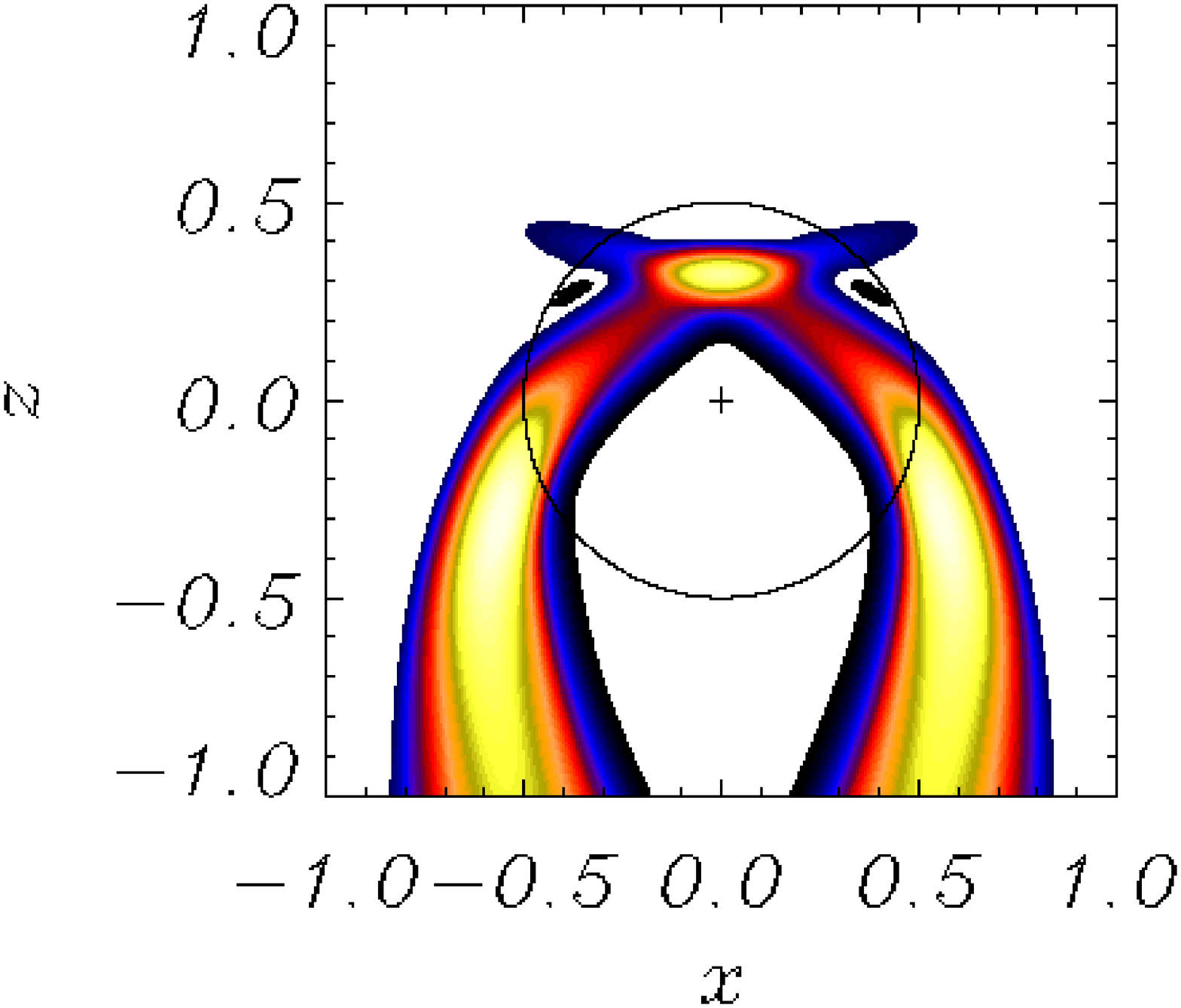}
\hspace{0.04in}
\includegraphics[width=1.6in]{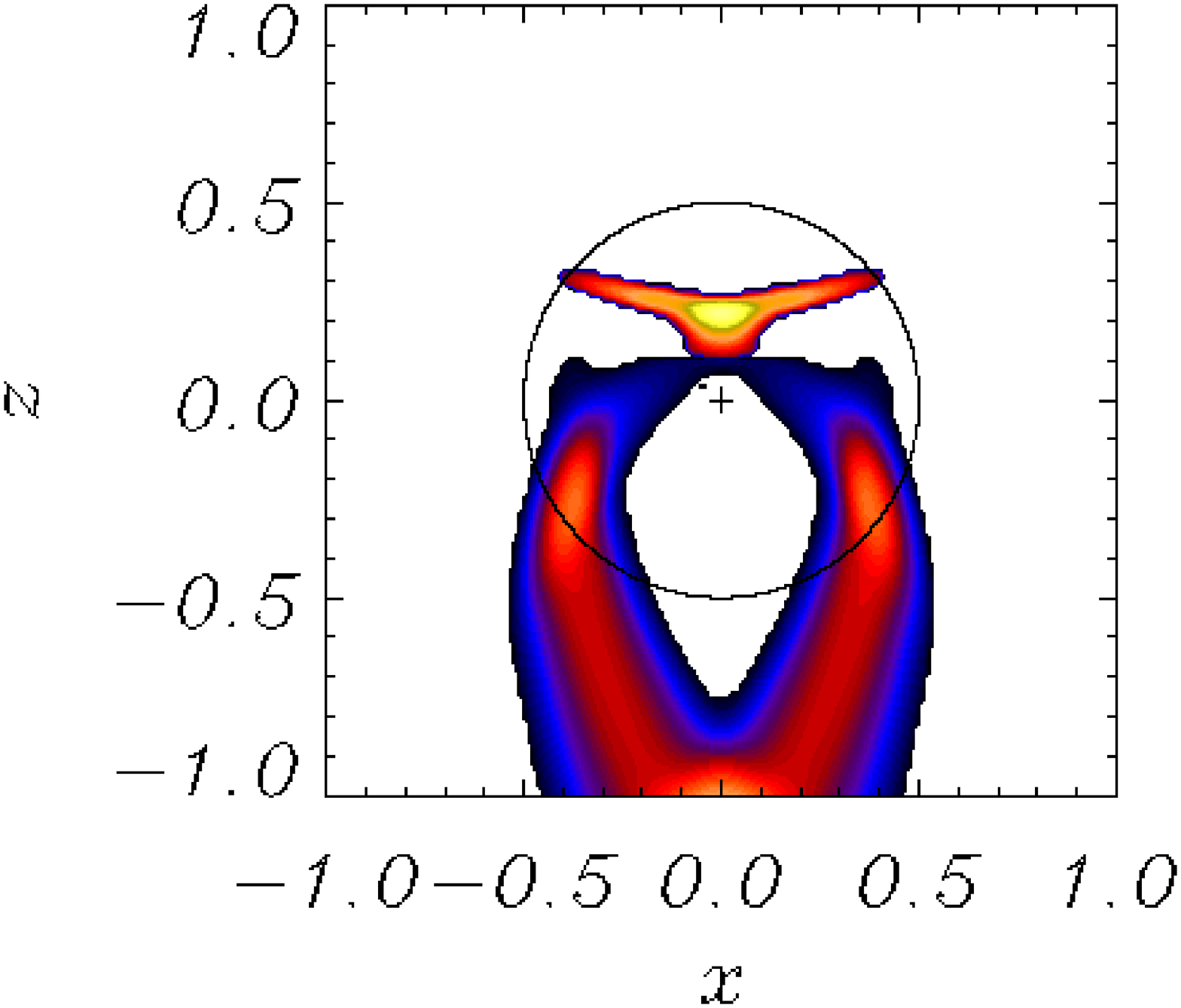}
\caption{Blow-up subfigures of  $\rm{v}_\perp$ from Figure 
\ref{fig:P3.1} at times $(a)$ $t$=2.33,  $(b)$ $t$=2.67  and  $(c)$
$t$=3.0, labelling left to right.}
\label{electricsix}
\end{center}
\end{figure*}
\end{center}

We find that the $\rm{v}_\perp$ disturbance travels towards the
neighbourhood of the null point and begins to wrap around it. This
is due to refraction caused by the spatially varying Alfv\'en speed, 
$v_A(x,z)$, as noted by
\cite{Nakariakov1995} and seen in \cite{McLaughlin2004} (further
references are cited in the latter). Once the wave reaches the 
$c_s=v_A\:$ layer
(denoted by the black circle around the null in Figure 
\ref{fig:P3.1}) the nature of the wave changes;
part of the wave now appears to spread out along the field lines 
(this part also moves
slower than the rest of the wave). Meanwhile, the majority of the 
wave outside the circle continues to
refract about the null. This wrapping effect repeats, and each time 
part of the wave
crosses the  $c_s=v_A\:$ layer, part of it is transformed to the 
field-guided wave.
Figure \ref{electricsix} shows a blow-up of subfigures $(d)$ to $(f)$ 
during the time of the first crossing.
In this figure, we can clearly see the splitting of the wave  as it 
nears and then crosses the
$c_s=v_A\:$ layer. Thus, we can see that there are three main types 
of wave behaviour in the
perpendicular component; outside the  $c_s=v_A\:$ layer we 
predominatrely see the refraction effect, whereas inside  we see
both fast and slow wave behaviour; the latter which appears as a 
field-guided wave. We shall give a fuller description of this 
phenomenon in
Section \ref{sec:3.1} and  give an interpretation of these different 
wave types in Section \ref{Interpretion}.

\subsection{Behaviour of  $\rm{v}_\parallel$}\label{sec:2.3}
\begin{center}
\begin{figure*}[t]
\begin{center}
\hspace{0.01in}
\includegraphics[width=1.5in]{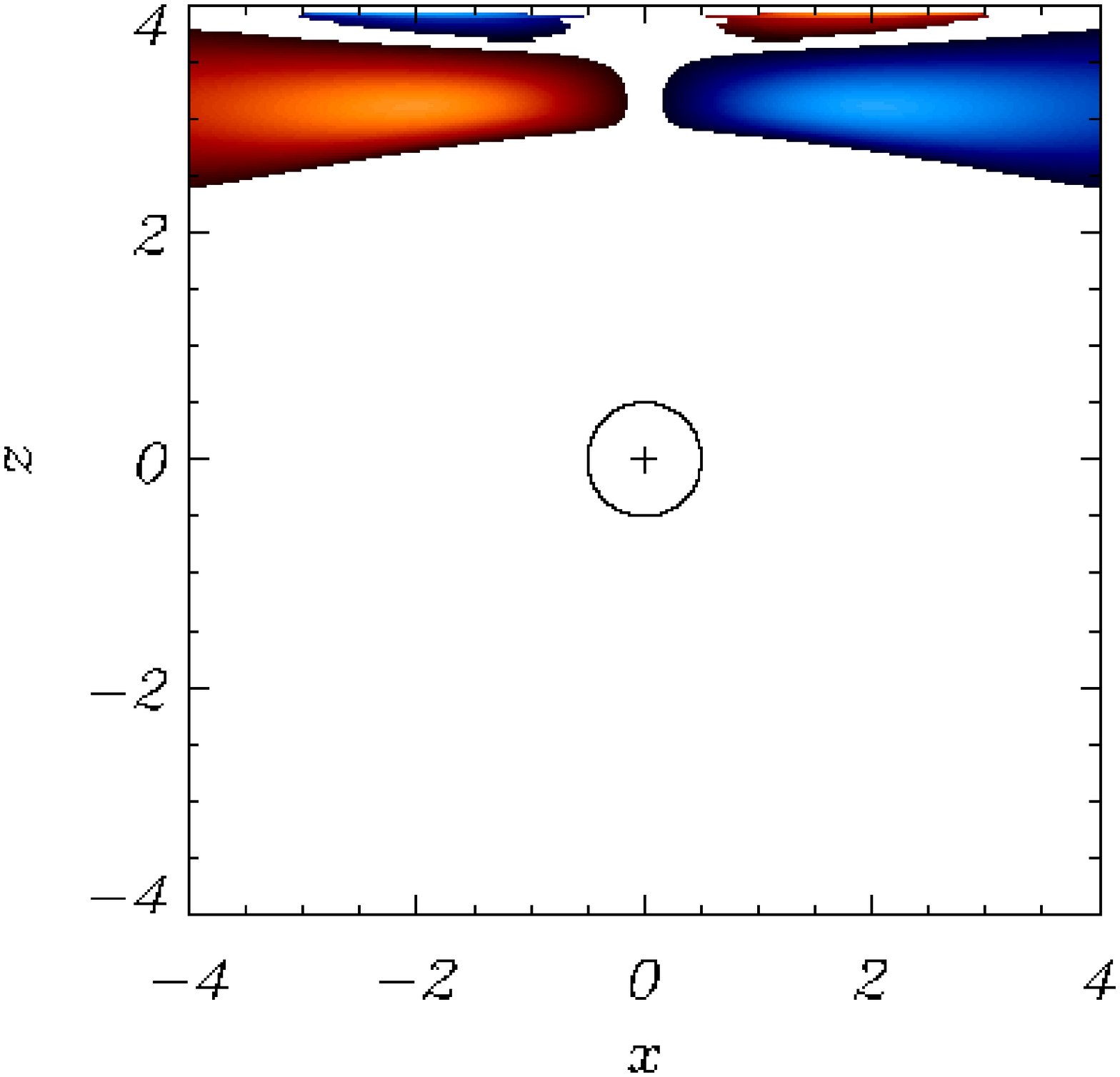}
\hspace{0.05in}
\includegraphics[width=1.5in]{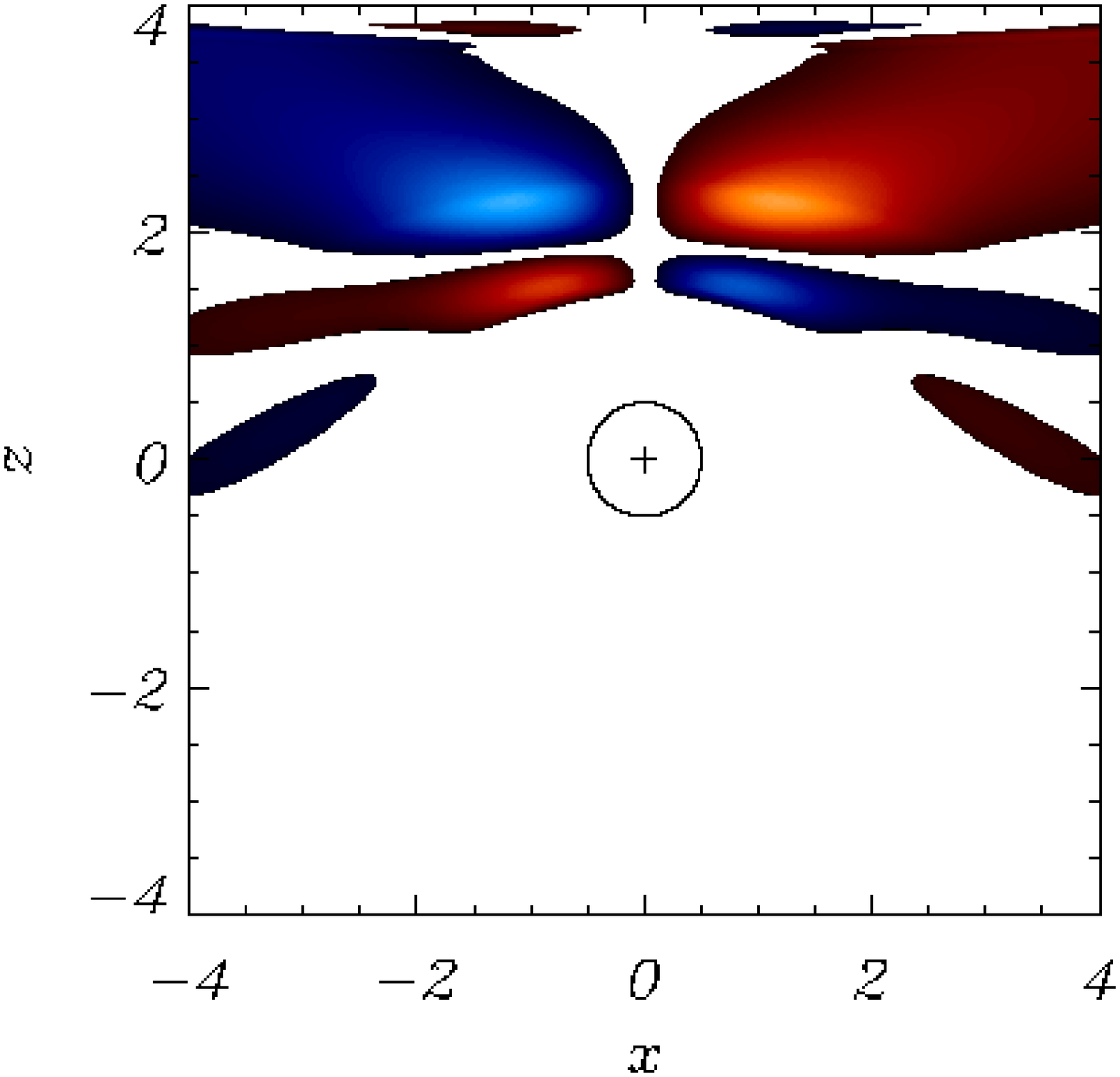}
\hspace{0.05in}
\includegraphics[width=1.5in]{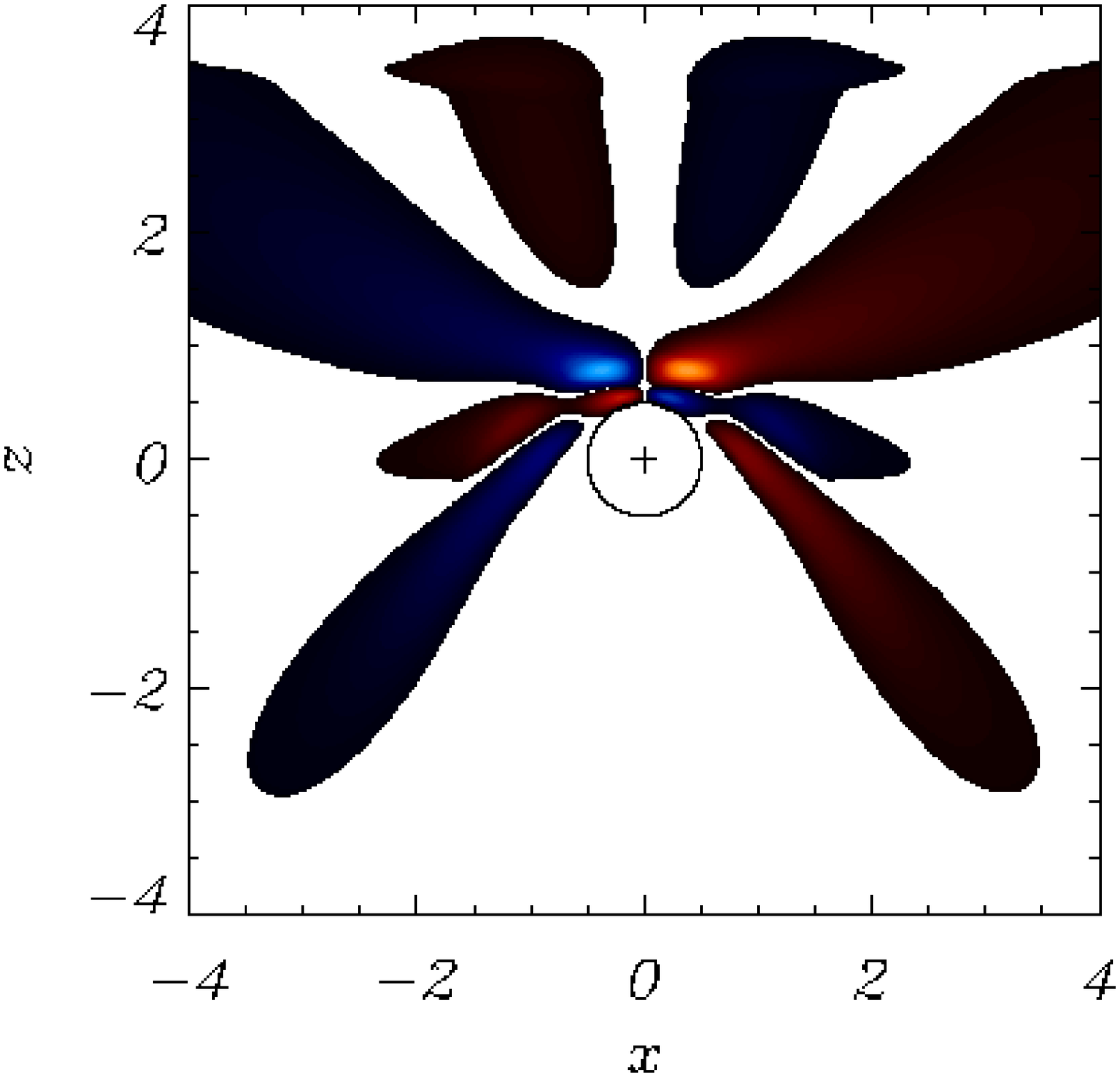}\vspace{0.05in}\\
\vspace{0.05in}
\hspace{0.01in}
\includegraphics[width=1.5in]{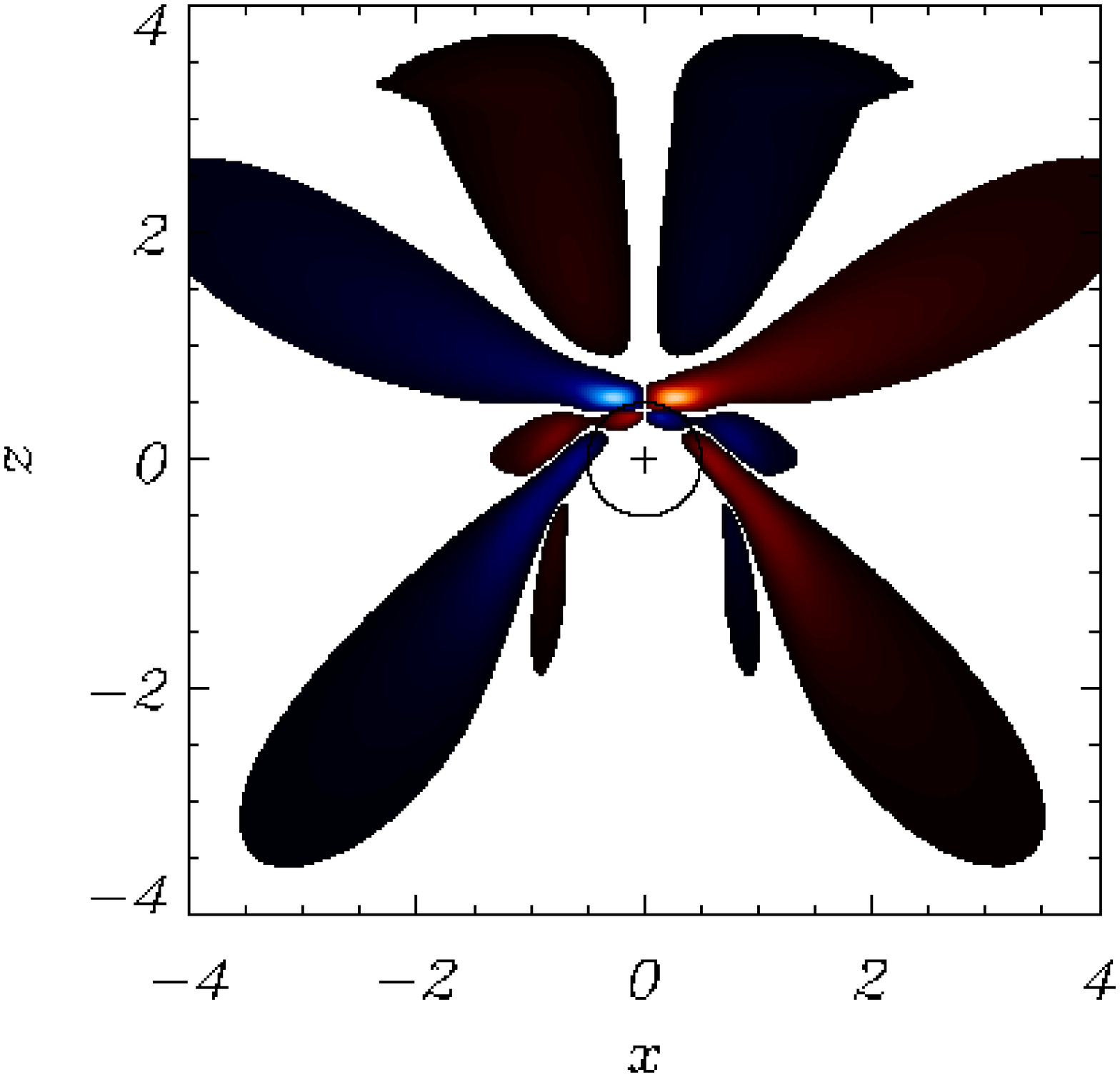}
\hspace{0.05in}
\includegraphics[width=1.5in]{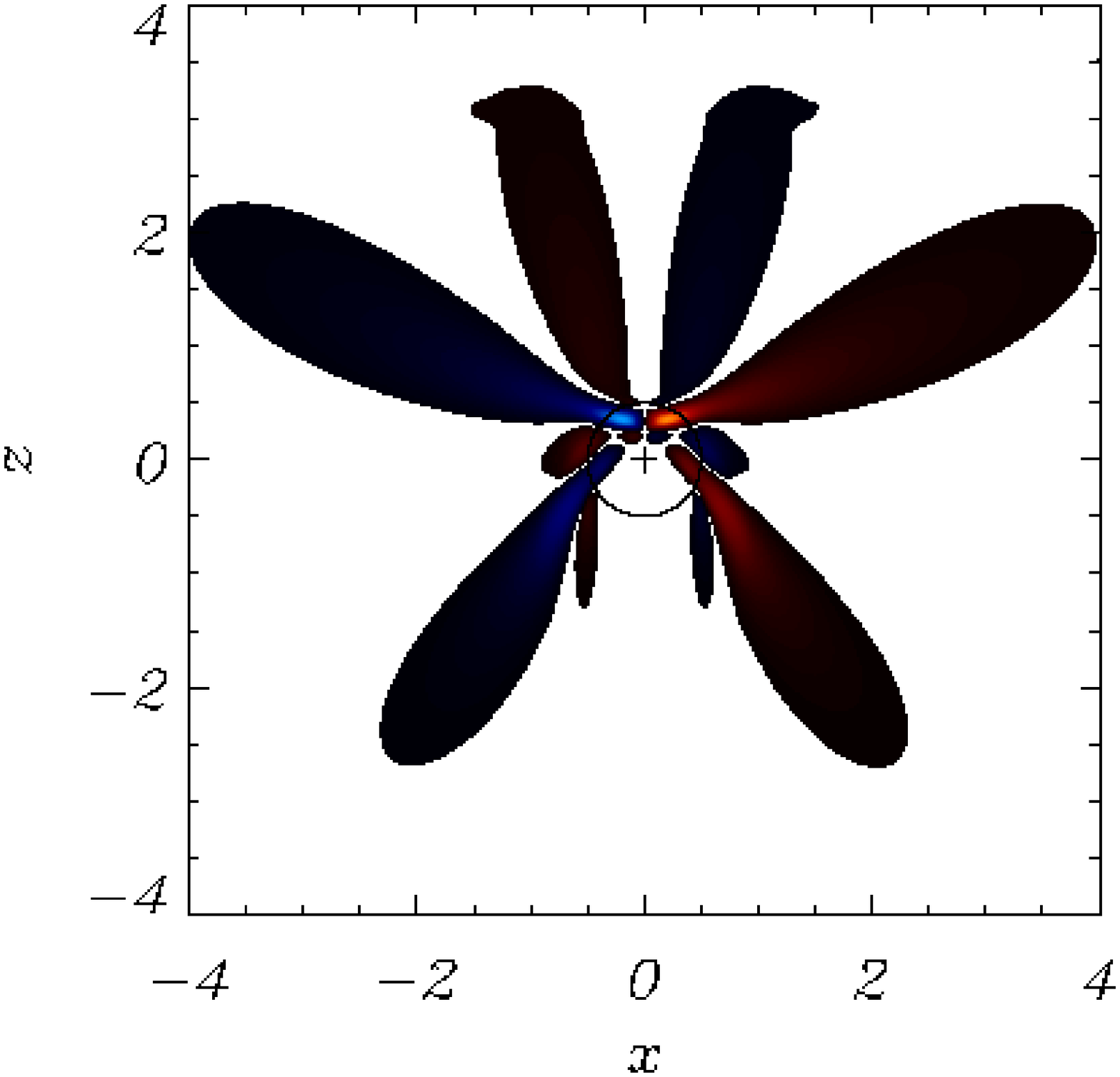}
\hspace{0.05in}
\includegraphics[width=1.5in]{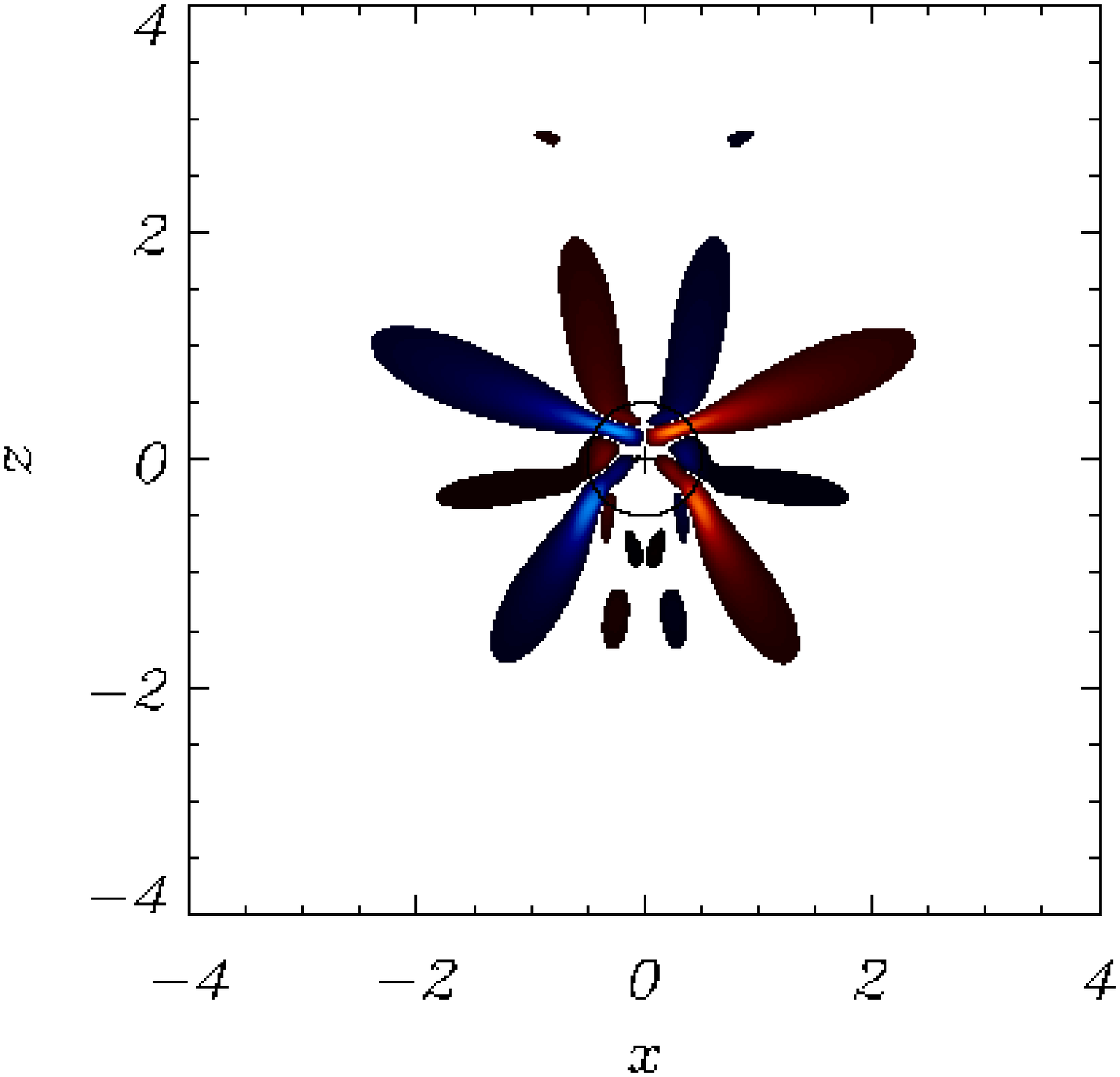}\\
\vspace{0.05in}
\hspace{0.61in}
\includegraphics[width=1.5in]{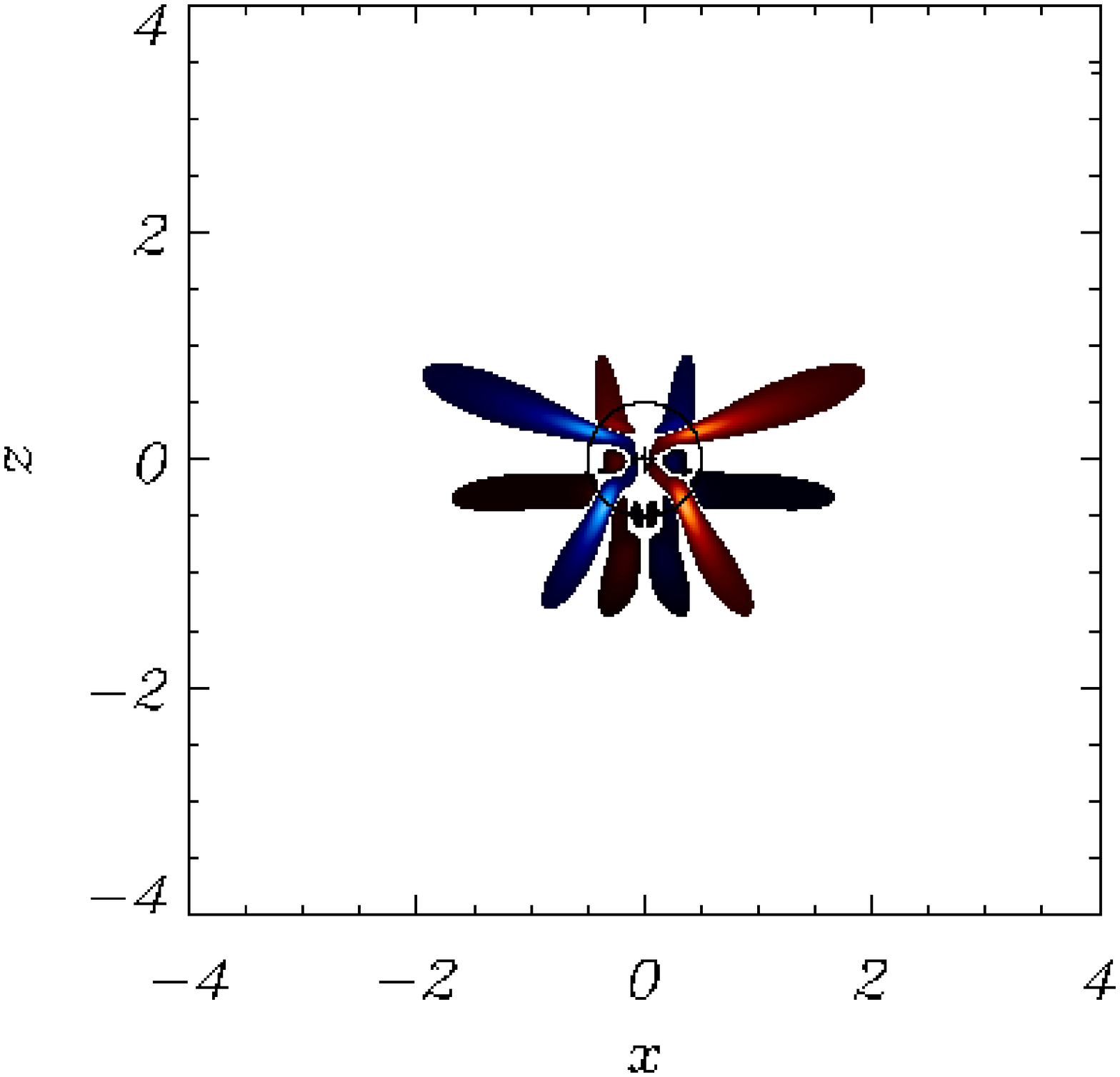}
\hspace{0.05in}
\includegraphics[width=1.5in]{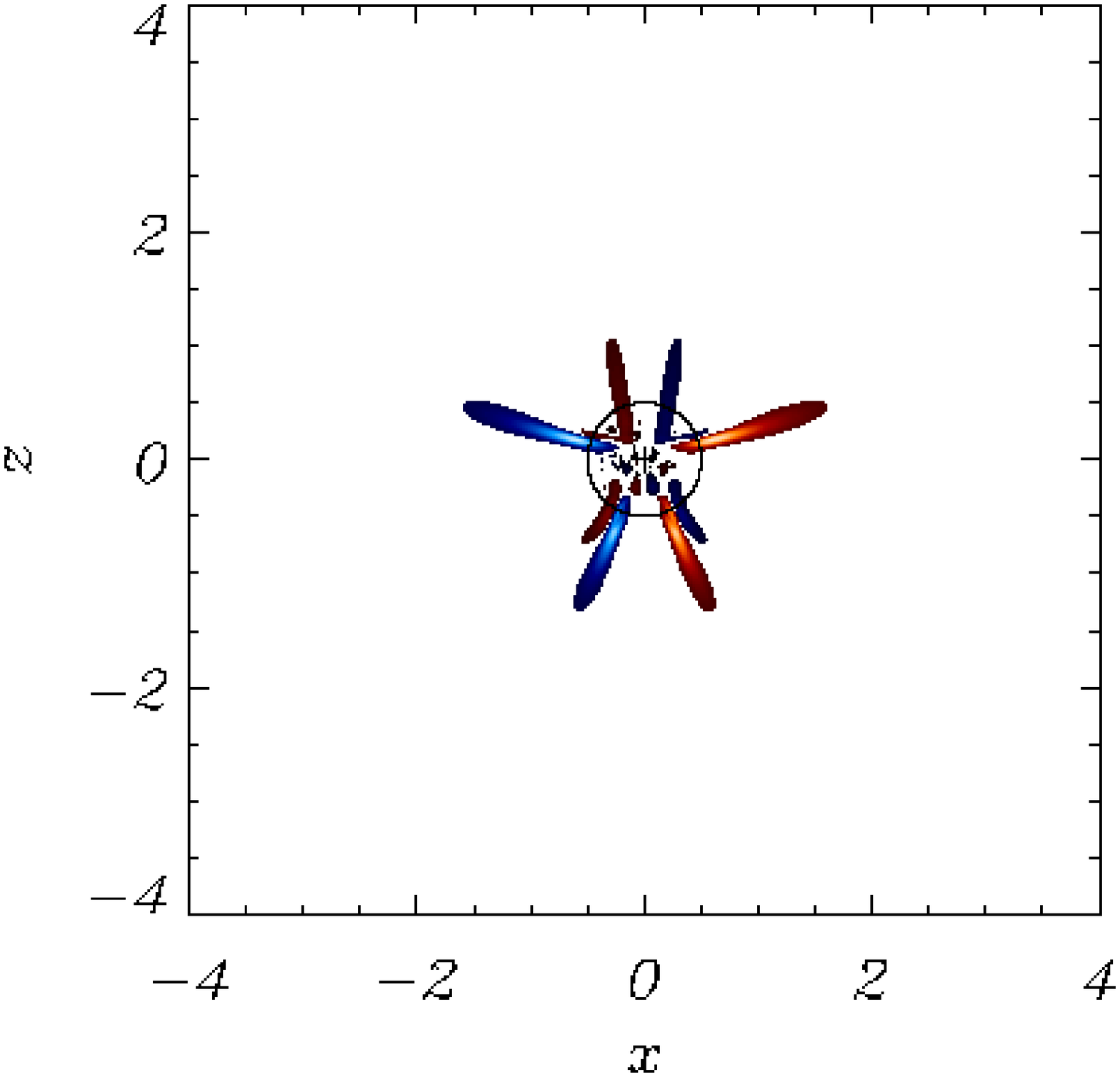}
\hspace{0.05in}
\includegraphics[width=2.15in]{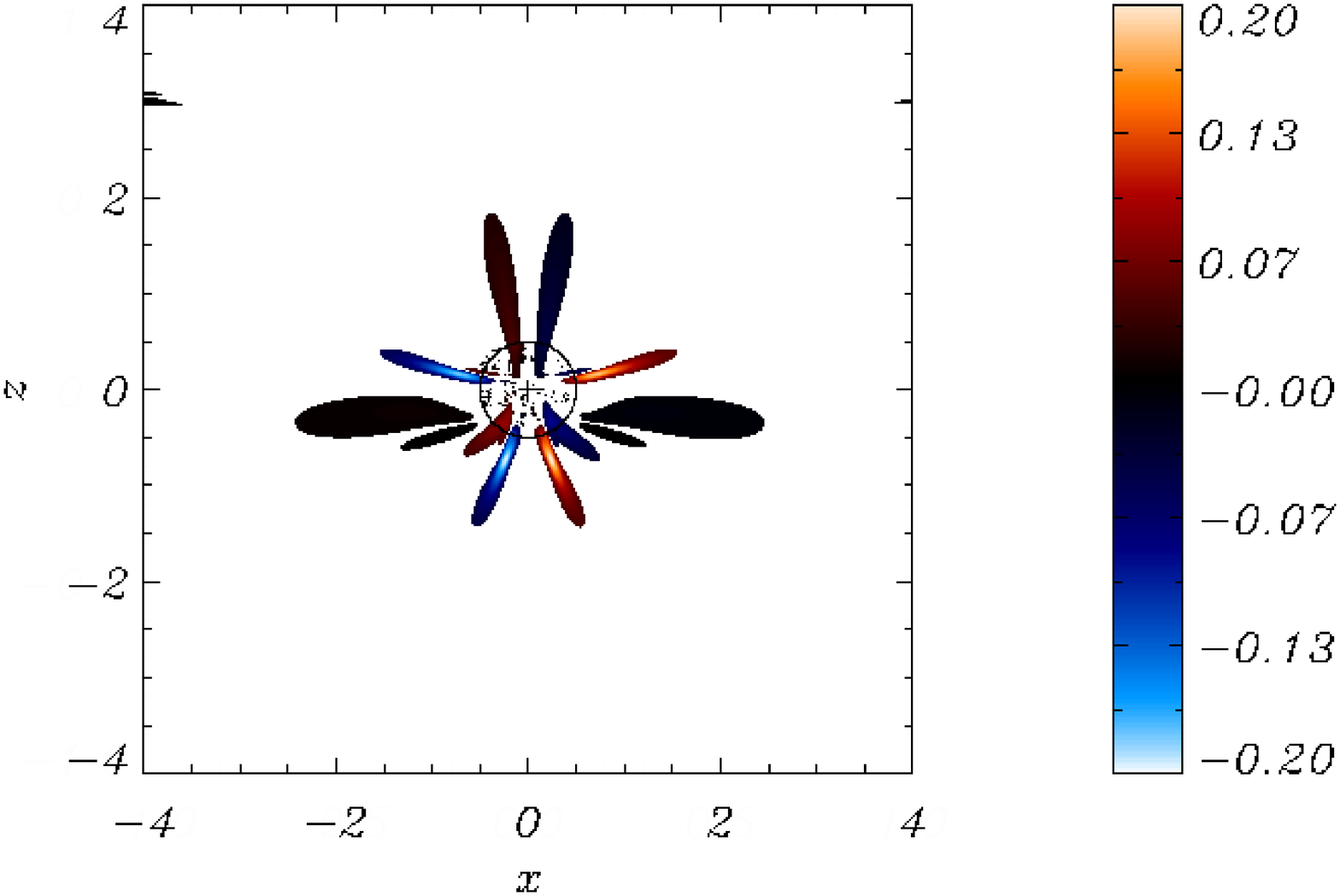}
\caption{Contours of $\rm{v}_\parallel$ for  numerical simulation for 
a fast wave sent in from upper boundary for
$-4 \leq x \leq 4 $ and $\beta_0=0.25$ and its resultant propagation 
at times $(a)$ $t$=0.33, $(b)$ $t$=1.0, $(c)$
$t$=2.0, $(d)$ $t$=2.33, $(e)$ $t$=2.67, $(f)$ $t$=3.0, $(g)$ 
$t=$3.33, $(h)$ $t$=3.67 and $(i)$ $t$=4.0,
labelling from top left to bottom right. The black circle indicates 
the position of the $c_s=v_A\:$ layer. The
cross denotes the null point in the magnetic configuration.}
\label{fig:P3.2}
\end{center}
\end{figure*}
\end{center}

The behaviour of  $\rm{v}_\parallel$ can be seen  in Figure
\ref{fig:P3.2}. We see that it is an antisymmetric wave about the
$z$ axis, and that it is focusing in towards the null. This pulse
travels at the same speed as the perpendicular component wave,
namely the fast wave speed. This is not surprising since, as
mentioned previously, $\rm{v}_\perp$ drives $\rm{v}_\parallel$.
Thus, mathematically this leading wave can be thought of as the
particular solution to an (inhomogeneous) equation. Note that
the range of the colour scale indicates that the maximum amplitude
of $\rm{v}_\parallel$ is $0.2$ and this is approximately the same
size as $\beta_0$. We use this relatively weak coupling to
construct an approximate analytic solution (see Appendix A).
There is also a complementary function part that satisfies
the homogeneous equation
\begin{equation}
{\partial^2 {v_{\parallel}}_{CF}\over \partial t^2} = {\gamma
\beta_0 \over 2}\left (\bf{B}_0 \cdot \nabla\right )^2 \left
({{v_{\parallel}}_{CF}\over B_0^2}\right )\;.
\end{equation}
This has a solution that involves a function of the variable
\begin{displaymath}
t + \sqrt{{2\over \gamma \beta_0}}\log z \;.
\end{displaymath}

\begin{figure*}[t]
\begin{center}
\includegraphics[width=3.1in]{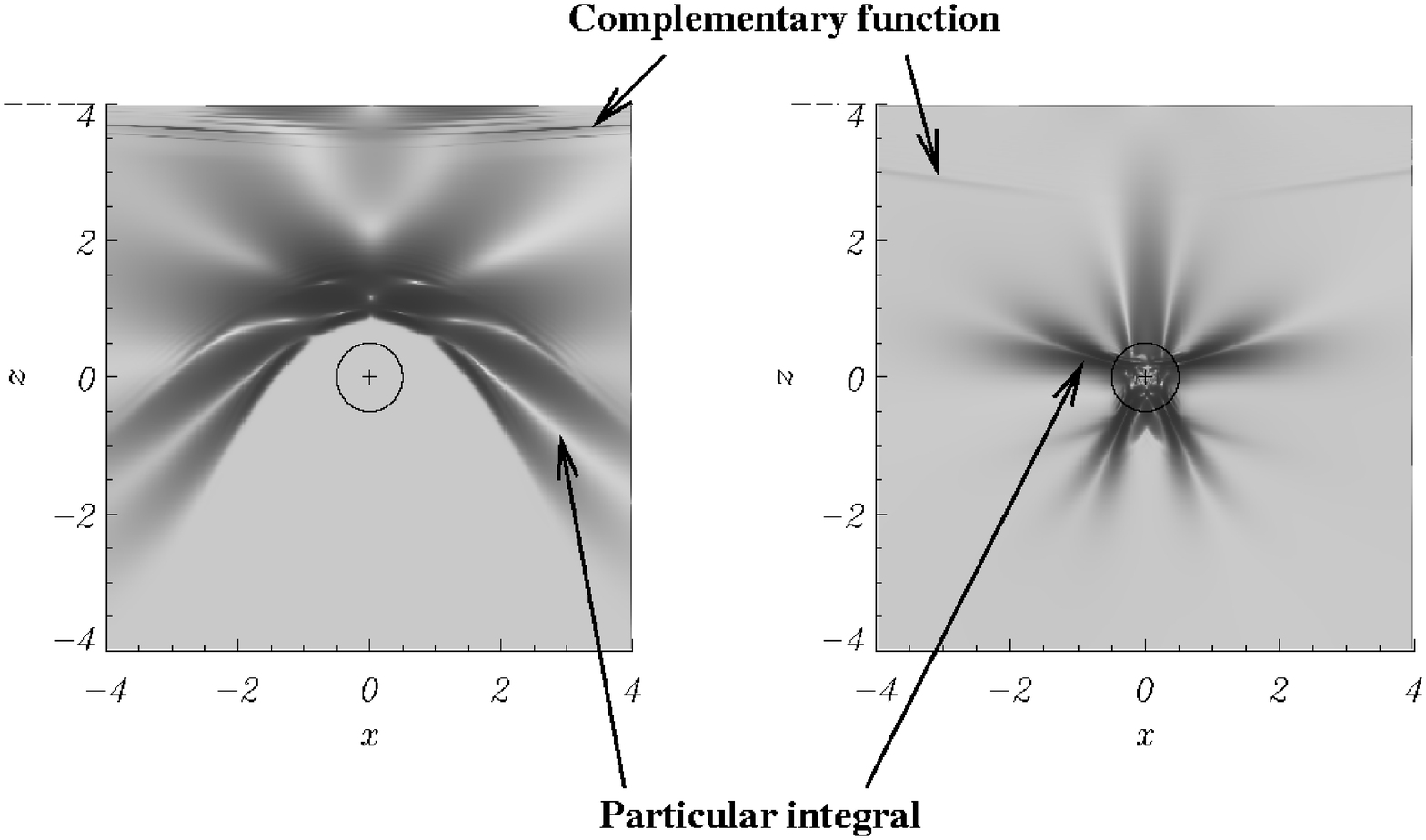}\\
\includegraphics[width=1.7in]{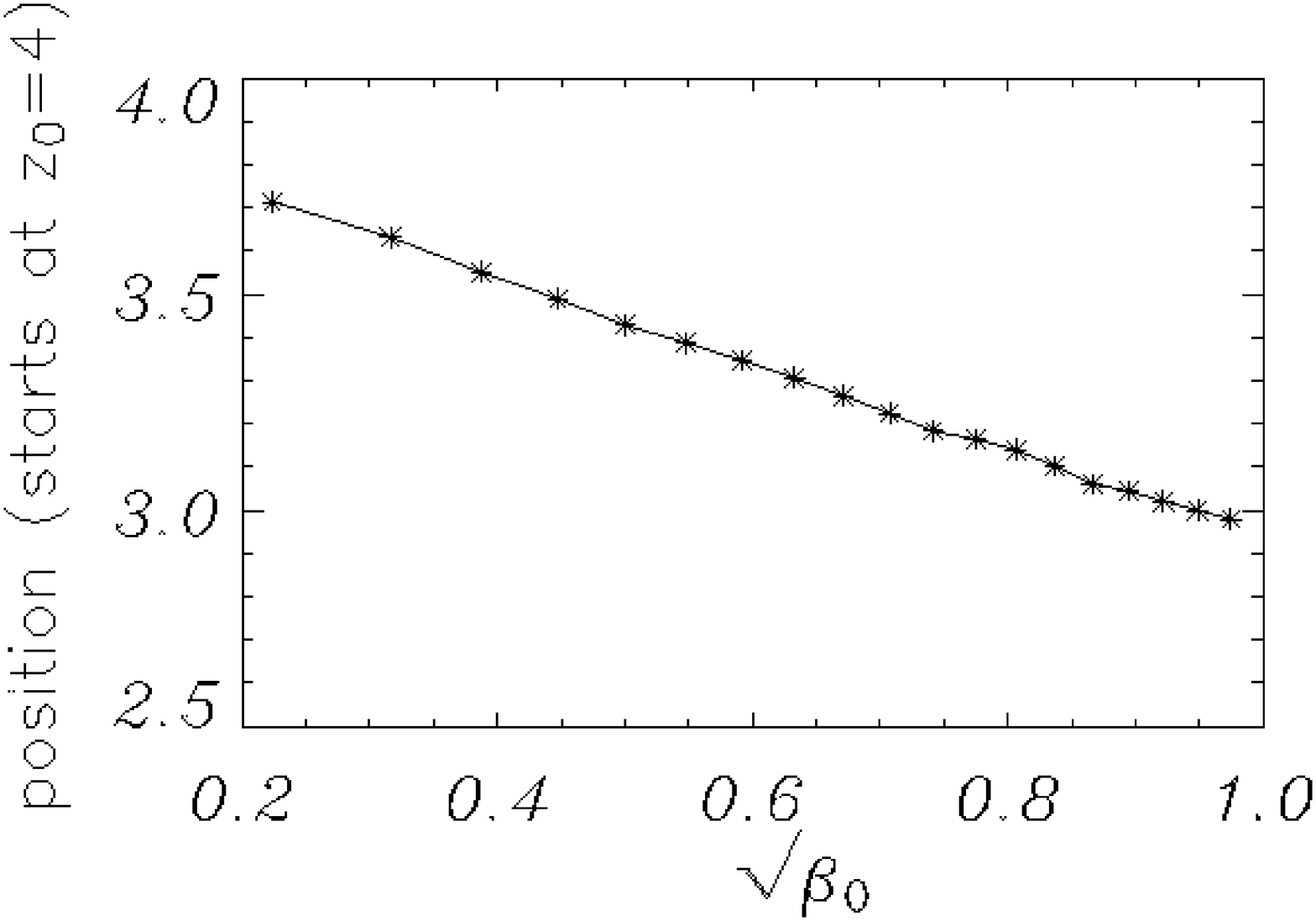}
\hspace{0.2in}
\includegraphics[width=1.7in]{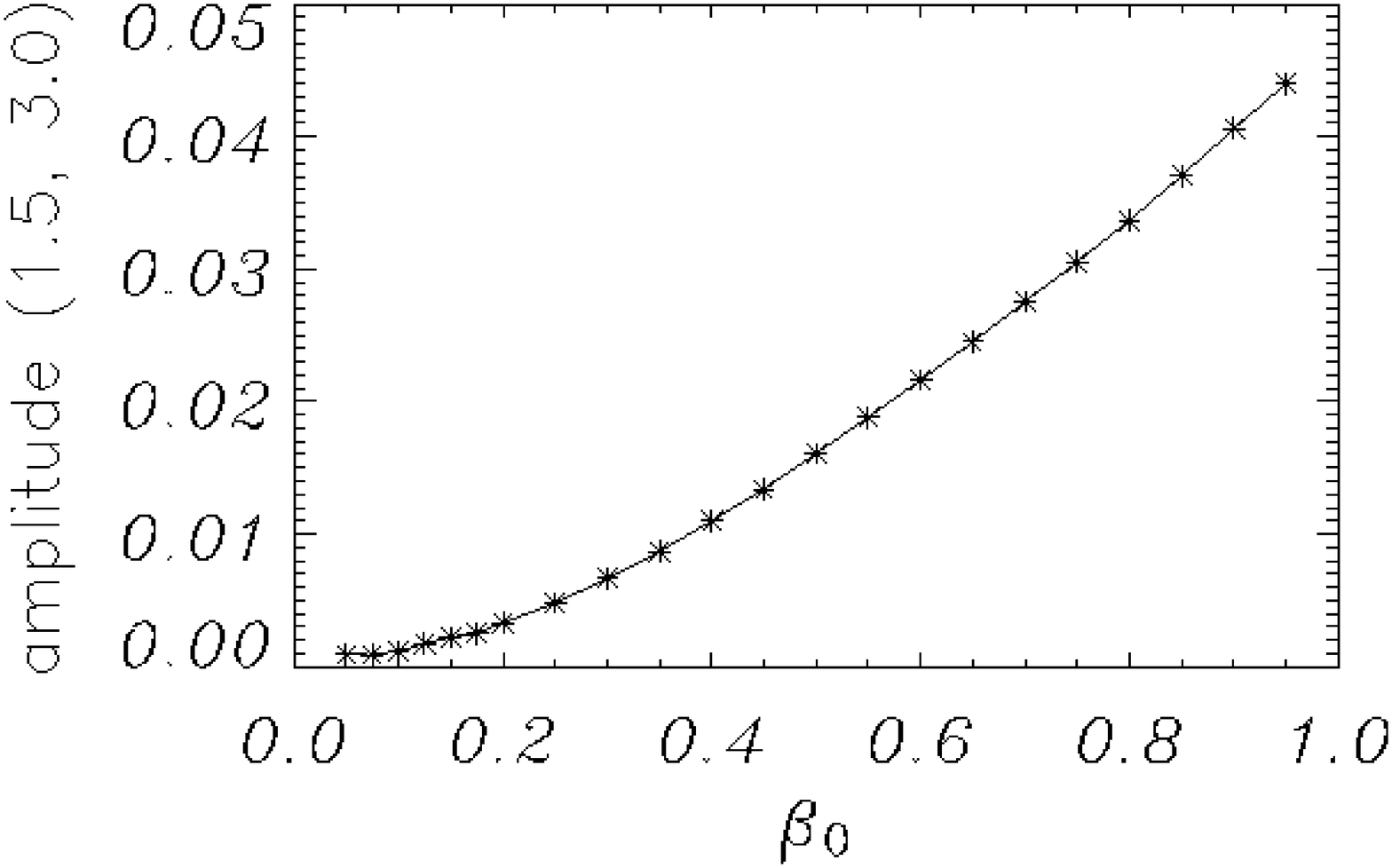}
\caption{Shaded surface  of $\rm{v}_\parallel$ for  numerical
simulation for a fast wave (pulse) sent in from upper boundary for
$-4 \leq x \leq 4 $ and $\beta_0=0.25$ at time $t=1.5$ (top left)
and $t=3.5$ (top right). The bottom left figure shows the position
reached by the complementary function  along $x=1.5$ after $t=1.5$
(the gradient of the straight line based on the first and last
points is $-0.98$. The bottom right figure shows the amplitude of
the complementary function as a function of $\beta_0$. The
amplitude is proportional to $\beta_0^{1.6}$, although $\beta_0^{1.5}$ is
to be expected from the boundary conditions.}
\label{electricseven}
\end{center}
\end{figure*}

The complementary function part can be seen in Figure 
\ref{electricseven}. This second part of the solution lags behind the
particular solution and propagates at the slow wave speed. In a low 
$\beta$ plasma this is approximately the
sound speed and hence is proportional to $\sqrt{\beta_0}$. From the 
lower subfigures of Figure \ref{electricseven},
we see that the complementary function does indeed travel with a 
speed proportional to $\sqrt{\beta_0}$, and has an
amplitude that varies as $\beta_0^{1.6}$. The complementary function 
has a discontinuous shape (as is also shown in
Appendix A) and its amplitude is small, $\mathcal{O}(\beta_0^{1.6})$, 
compared to the particular integral,
$\mathcal{O}(\beta_0)$, and is thus difficult to see in Figure \ref{fig:P3.2}.

There are also several other noteworthy aspects to  Figure 
\ref{fig:P3.2}. Firstly, it is obviously more complicated than the 
$\rm{v}_\perp$
wave. Secondly, this component has eight lobe-like structures by the 
time $t=2.0$, when the leading edge of the wave is
about to cross the region where $c_s = v_A$. The lobes alternate 
between positive and negative values.
This complicated lobe-like structures can be explained by considering 
the same system
but driven by a circular boundary condition. This is  explained in 
Appendix A, where the main
conclusions is that the equilibrium magnetic configuration
{\textit{naturally}} leads to a $\sin{4\theta}$ dependence in 
$\rm{v}_\parallel$.
This explains the complicated, lobe-like structuring, i.e. the lobes
come from the equilibrium magnetic configuration (but the wave is no 
longer circular so the  $\sin{4\theta}$ is less apparent).

\section{Interpretation}\label{Interpretion}

Roberts (1985) showed that ${\bf{v}}_{\rm{Alfv{\acute{e}}n}}$,  ${\bf{v}}_{\rm{slow}}$ and
${\bf{v}}_{\rm{fast}}$  form an orthogonal basis of vectors for
the linearised MHD equations in a uniform plasma. In this paper, we 
do not consider
the Alfv\'en wave and so our 2D system can be described in terms
of the vectors  ${\bf{v}}_{\rm{fast}}$ and ${\bf{v}}_{\rm{slow}}$.
Due to the form of the equilibrium magnetic field near the null
point, we choose to work in the directions parallel and
perpendicular to the magnetic field and thus we may represent
these two vectors in terms of ${\bf{v}}_{\rm{fast}}$ and
${\bf{v}}_{\rm{slow}}$, i.e.
\begin{eqnarray*}
{\bf{v}}_\perp = \mathcal{A} {\bf{v}}_{\rm{fast}} + 
\mathcal{B}{\bf{v}}_{\rm{slow}}\;\; ,\quad {\bf{v}}_\parallel = 
\mathcal{C} {\bf{v}}_{\rm{fast}}
+ \mathcal{D}{\bf{v}}_{\rm{slow}}
\end{eqnarray*}
where $\mathcal{A}$, $\mathcal{B}$, $\mathcal{C}$ and $\mathcal{D}$ 
are unknown functions that depend upon
the magnetic geometry (and possibly  the plasma $\beta$).
Equivalently, we may represent ${\bf{v}}_{\rm{fast}}$ and 
${\bf{v}}_{\rm{slow}}$ in terms of ${\bf{v}}_{\perp}$ and
${\bf{v}}_\parallel$.
This representation is only possible because both 
${\bf{v}}_{\rm{fast}}$ \& ${\bf{v}}_{\rm{slow}}$ and 
${\bf{v}}_\perp$ \&
  ${\bf{v}}_\parallel$  are linearly independent vectors. Note that in 
a low $\beta$ region or
when the plasma is nearly uniform, $\mathcal{B}$ will be very small. 
Thus in these
  regions, we interpret ${\bf{v}}_\perp$ as predominately a fast wave 
(i.e. ${\bf{v}}_\perp \approx \mathcal{A} {\bf{v}}_{\rm{fast}}$).

However, we must be careful. The concepts of fast and slow waves were 
derived for a unidirectional magnetic field (see
\cite{Edwin1983}) and so these ideas may not carry over to more 
complicated geometries quite as simply as claimed here.
Nonetheless, we shall continue to use terms such as fast and slow 
wave in our interpretation of the waves in this paper (i.e.
we shall use the terminology and intuition gained from the 
unidirectional magnetic field model). Thus, {\it{we choose to interpret the waves seen in our perpendicular and parallel velocities using the terminology of fast and slow waves}}.

\section{Mode conversion across the $c_s=v_A\:$ layer}\label{modeconversion}

\begin{center}
\begin{figure}
\begin{center}
\includegraphics[width=1.75in]{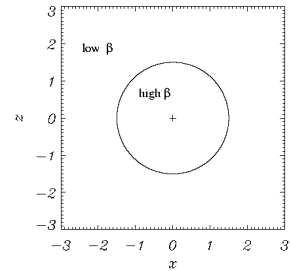}
\caption{Regions of high and low $\beta$ in our equilibrium magnetic 
field, where $\beta= \frac{\beta_0}{x^2+z^2}$. The black
circle indicates the position of the  $c_s=v_A\:$ layer and the cross 
denotes the null point.}
\label{fig:areasofhighandlow}
\end{center}
\end{figure}
\end{center}

The $c_s=v_A\:$ layer is of critical importance to our system. 
{When the Alfv\'en speed
and sound speed are dissimilar, there is negligible coupling
between the fast and slow magnetoacoustic waves. However, near the
$c_s=v_A\:$ layer, the two waves can resonantly interact with each 
other and strong mode
coupling can occur.}
Since $\beta= \frac{\beta_0}{x^2+z^2}$ (where $\beta_0$ is a
constant of our choosing), our system consists of a region of low
$\beta$ plasma outside the $c_s=v_A\:$ layer  and a region of high
$\beta$ plasma within (see Figure \ref{fig:areasofhighandlow}).
Fast and slow waves have differing properties depending on if they
are in a high or low $\beta$ environment (see e.g. Bogdan \emph{et
al.} 2003). See Table 1 for a summary of these properties.

\begin{table*}[h]\label{FSprop}
\begin{center}
\begin{tabular}{|c|c|c|}
\hline
  & Fast Wave & Slow Wave\\
\hline\hline
High $\beta$ & \begin{tabular}{c}
Behaves like isotropic sound wave\\
  (speed $c_s$)
\end{tabular} &
\begin{tabular}{c}
Guided along ${\bf{B}}_0$ \\
Transverse wave propagating at speed $v_A$\\
Dominant velocity component is ${\rm{v}}_\perp$
\end{tabular}\\
\hline
Low $\beta$ & \begin{tabular}{c}
Propagates roughly isotropically \\
  (speed $v_A$)\\
  Dominant velocity component is ${\rm{v}}_\perp$
\end{tabular} & \begin{tabular}{c}
Guided along ${\bf{B}}_0$\\
  Longitudinal wave propagating at speed $c_T$\\
  Dominant velocity component is ${\rm{v}}_\parallel$
\end{tabular}\\
\hline
\end{tabular}
\end{center}
\caption{Properties of fast and slow waves depending on their environment.}
\end{table*}

Thus, in a low $\beta$ region, the velocity vector of the fast wave 
is predominately perpendicular to the equilibrium
magnetic field, although the wave can propagate almost isotropically. 
The slow wave has a velocity component along
the field and propagates along ${\bf{B}}_0$. In the high $\beta$ 
region, the fast wave again propagates isotropically
but this time the velocity vector is parallel to the wavenumber 
vector. Hence, the fast wave now has components of
velocity that are perpendicular and parallel to the field. The slow 
wave is again guided by the field but the velocity
is perpendicular to ${\bf{B}}_0$ \footnote{Note that the high $\beta$ 
slow wave shows up in the perpendicular
component because it is a transverse wave (${\bf{v}}\cdot{\bf{k}}=0$) 
and not because it is propagating across
the field; the slow wave cannot propagate across the field!}.

In our numerical simulations, a straight wave pulse in the 
perpendicular component is sent in from the low $\beta$ region at the 
top boundary.
This disturbance is a low
$\beta$ fast wave. At some point the wave will cross the $c_s=v_A\:$ 
layer and enter the high $\beta$ environment.
Because of the above properties, the incoming fast wave is initially 
perpendicular to ${\bf{B}}_0$ but, on crossing
the $c_s = v_A\;$ layer, part of this perpendicular component remains 
as a fast wave and part is converted into the
slow wave. The first point of contact occurs along $x=0$ and at later 
times for increasing values of $x^2$. Thus, we  have a low
$\beta$ wave approaching the layer, coupling and mixing inside the 
layer and emerging as a mixture of high $\beta$ fast and slow waves.
Later on, the fast wave leaves the high $\beta$ region and re-enters 
the low $\beta$ region. Mode conversion can occur
at this crossing as well.

\begin{center}
\begin{figure*}[t]
\begin{center}
\hspace{0in}
\vspace{0.1in}
\hspace{0.2in}
\includegraphics[width=1.5in]{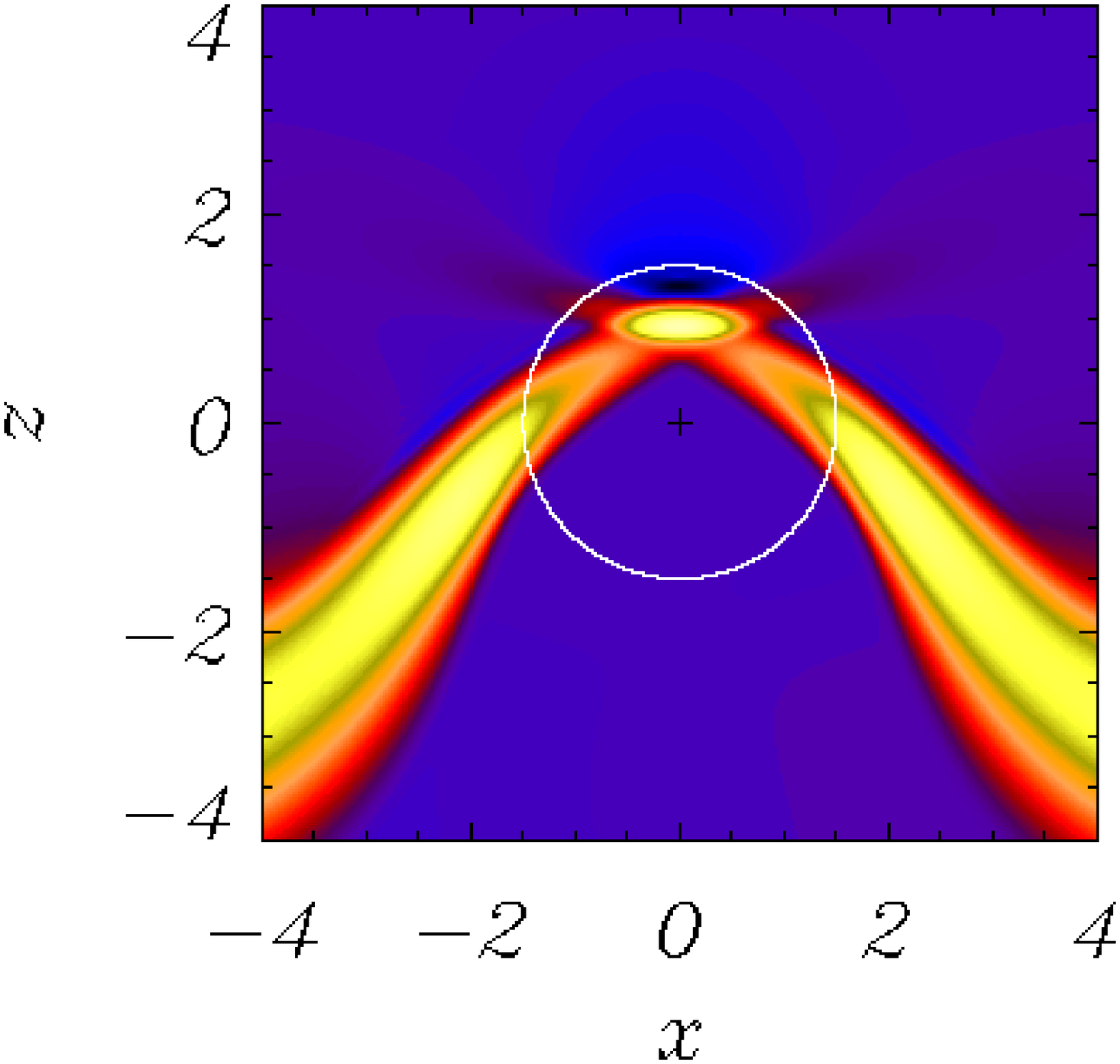}
\hspace{0.2in}
\includegraphics[width=1.5in]{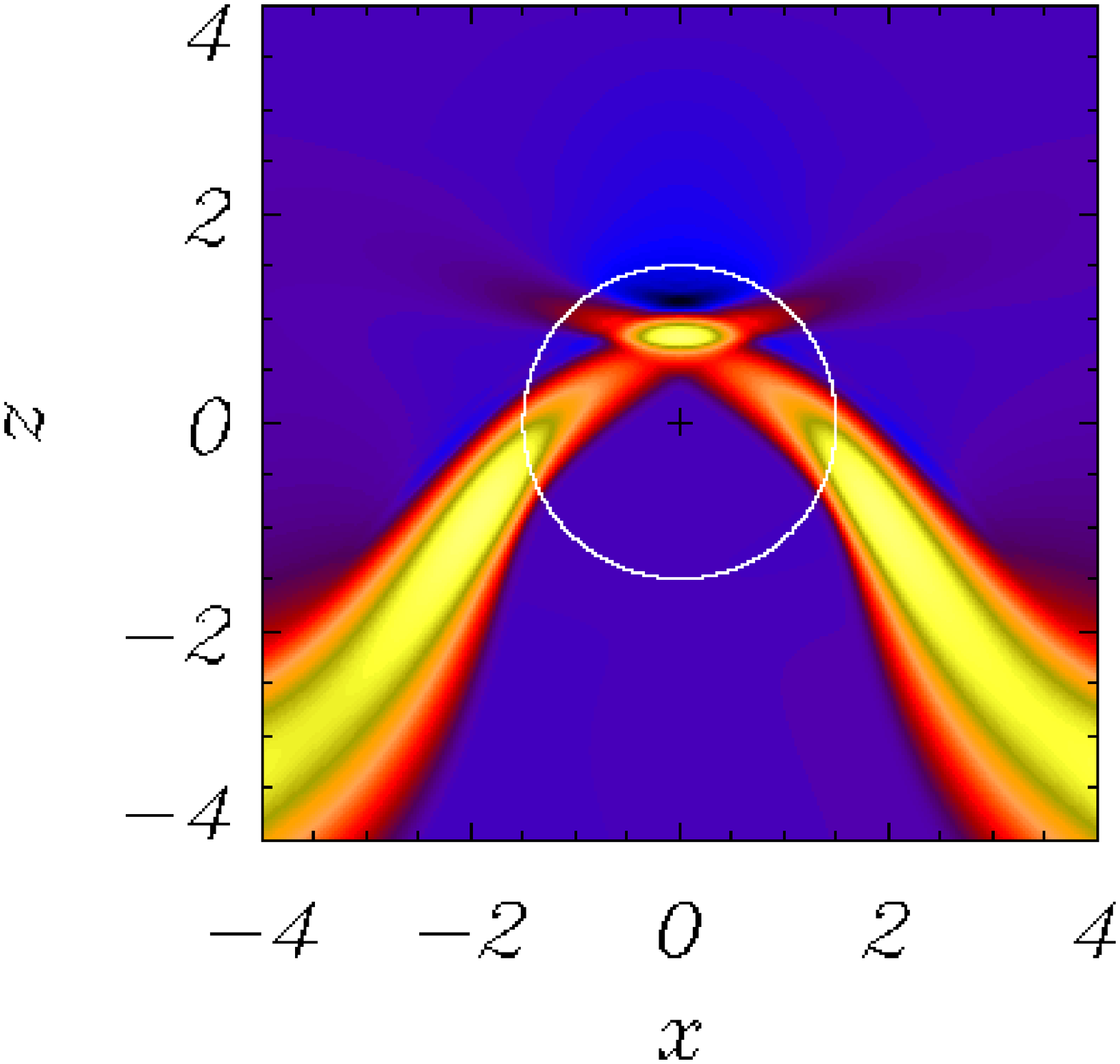}
\hspace{0.2in}
\includegraphics[width=1.5in]{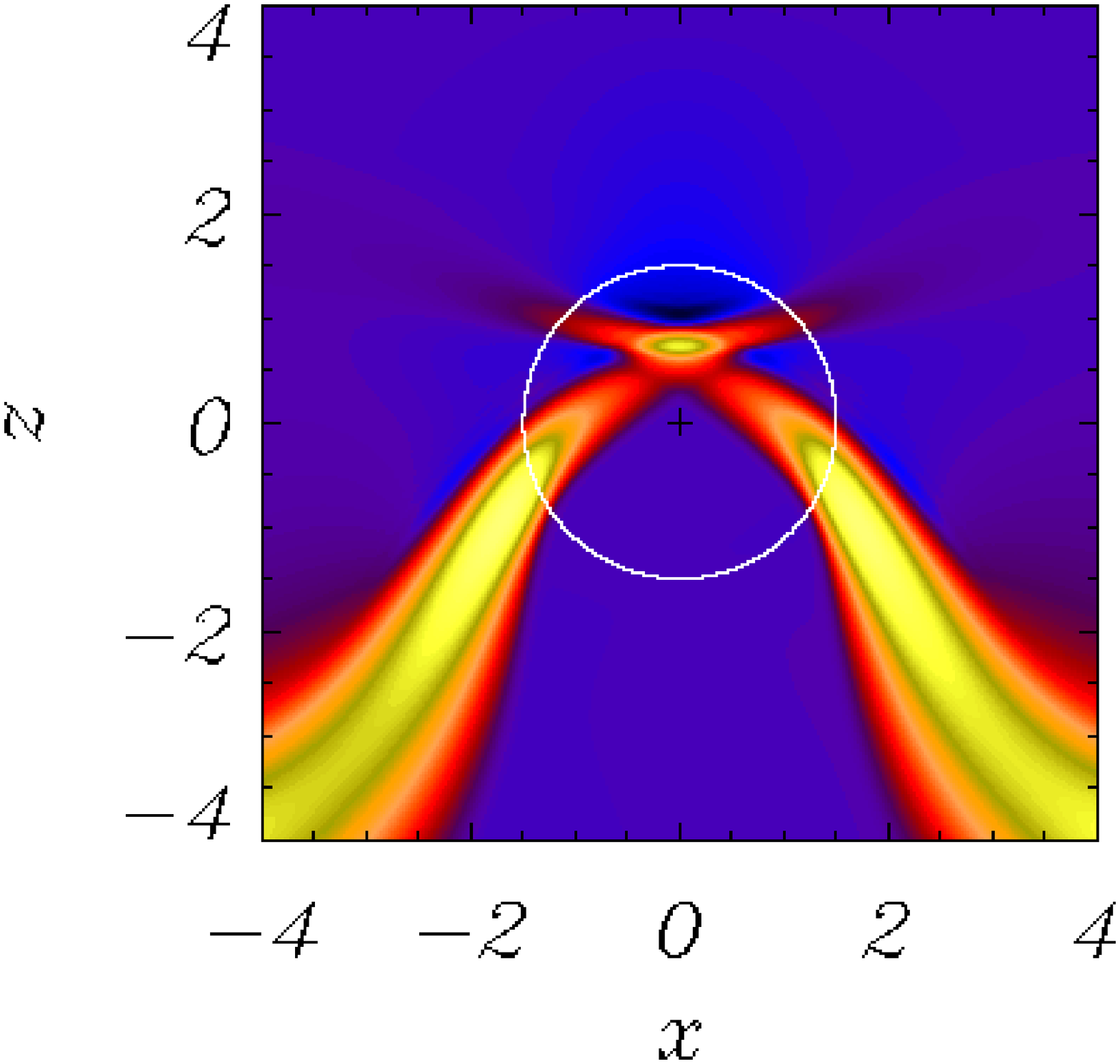}\\
\hspace{0in}
\vspace{0.1in}
\hspace{0.2in}
\includegraphics[width=1.5in]{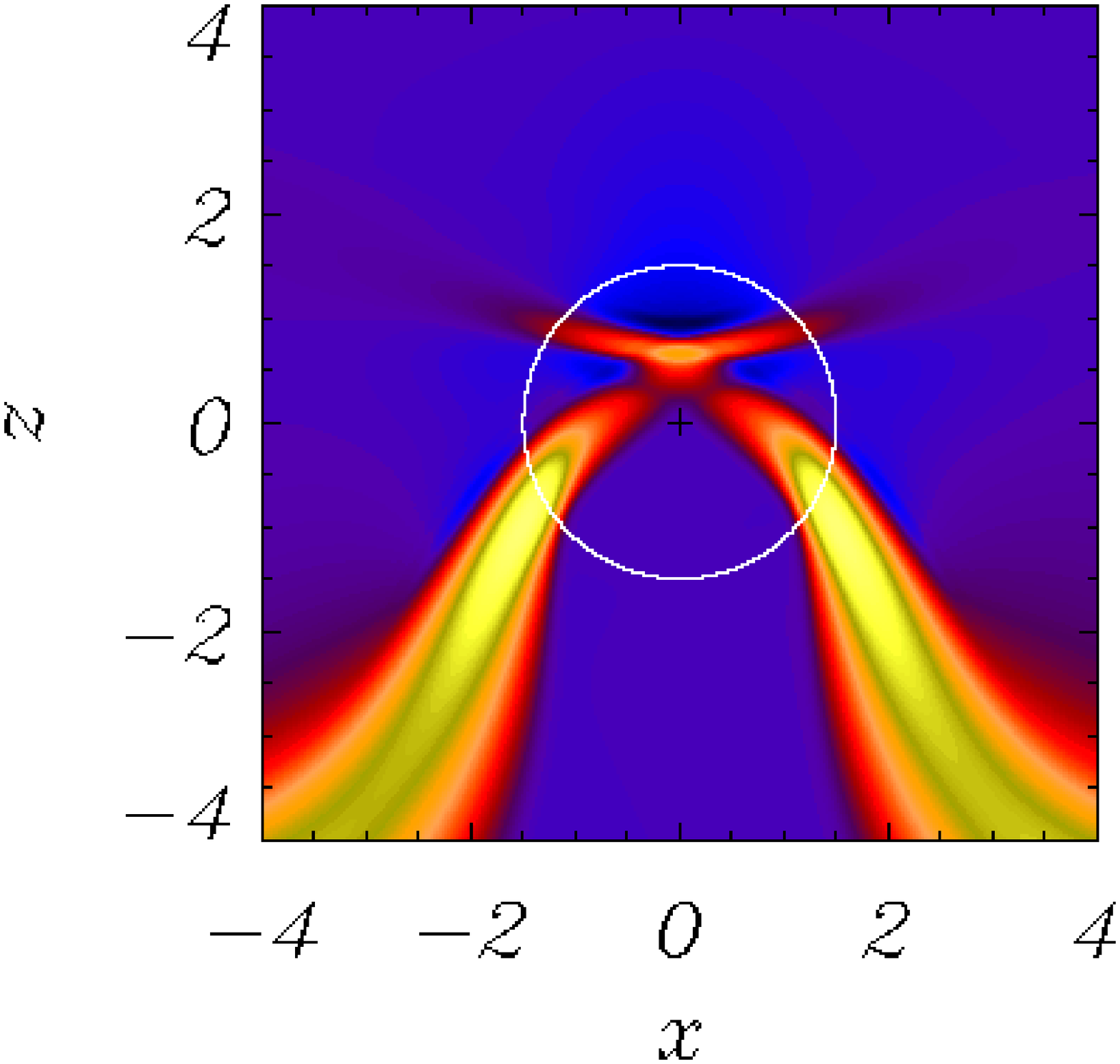}
\hspace{0.2in}
\includegraphics[width=1.5in]{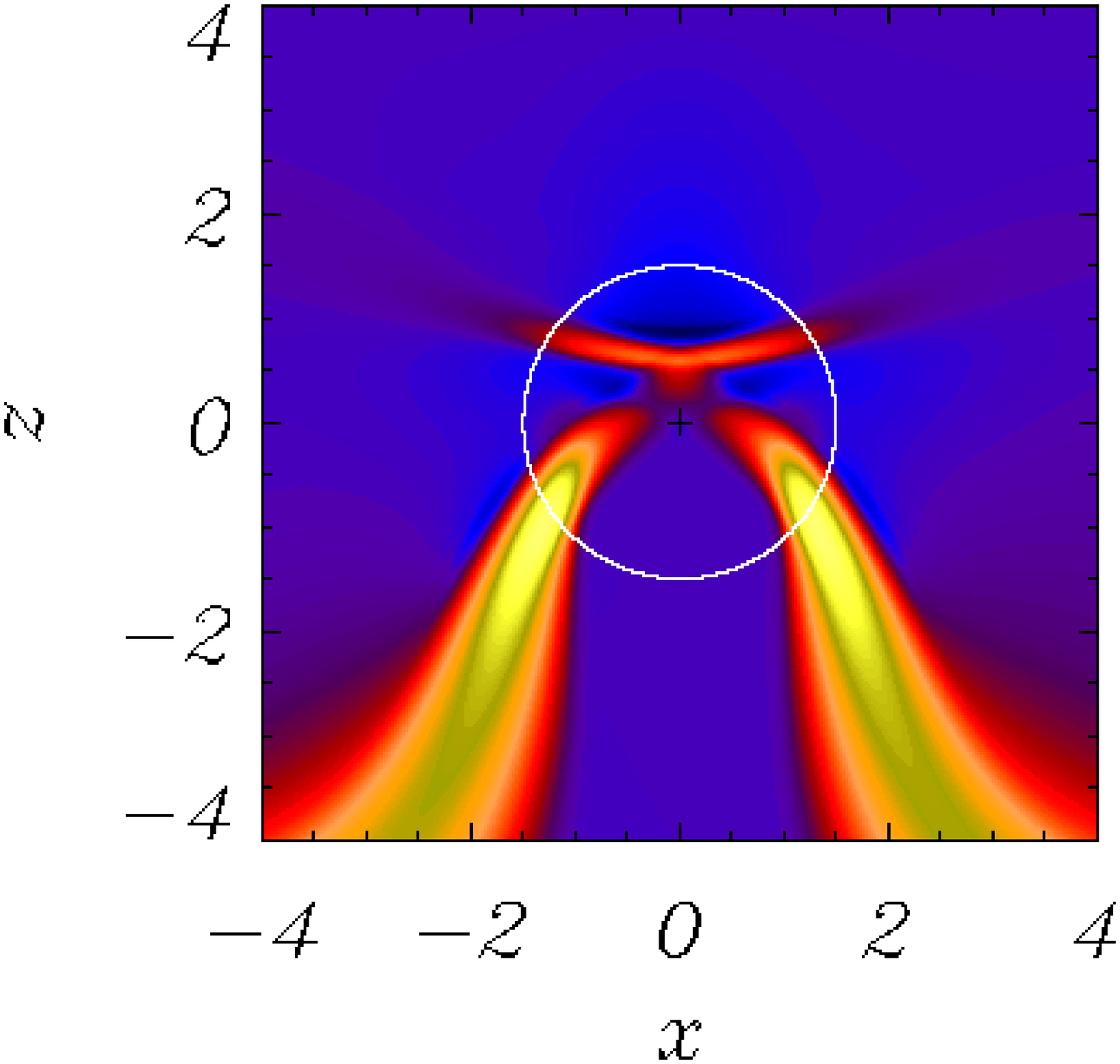}
\hspace{0.2in}
\includegraphics[width=1.5in]{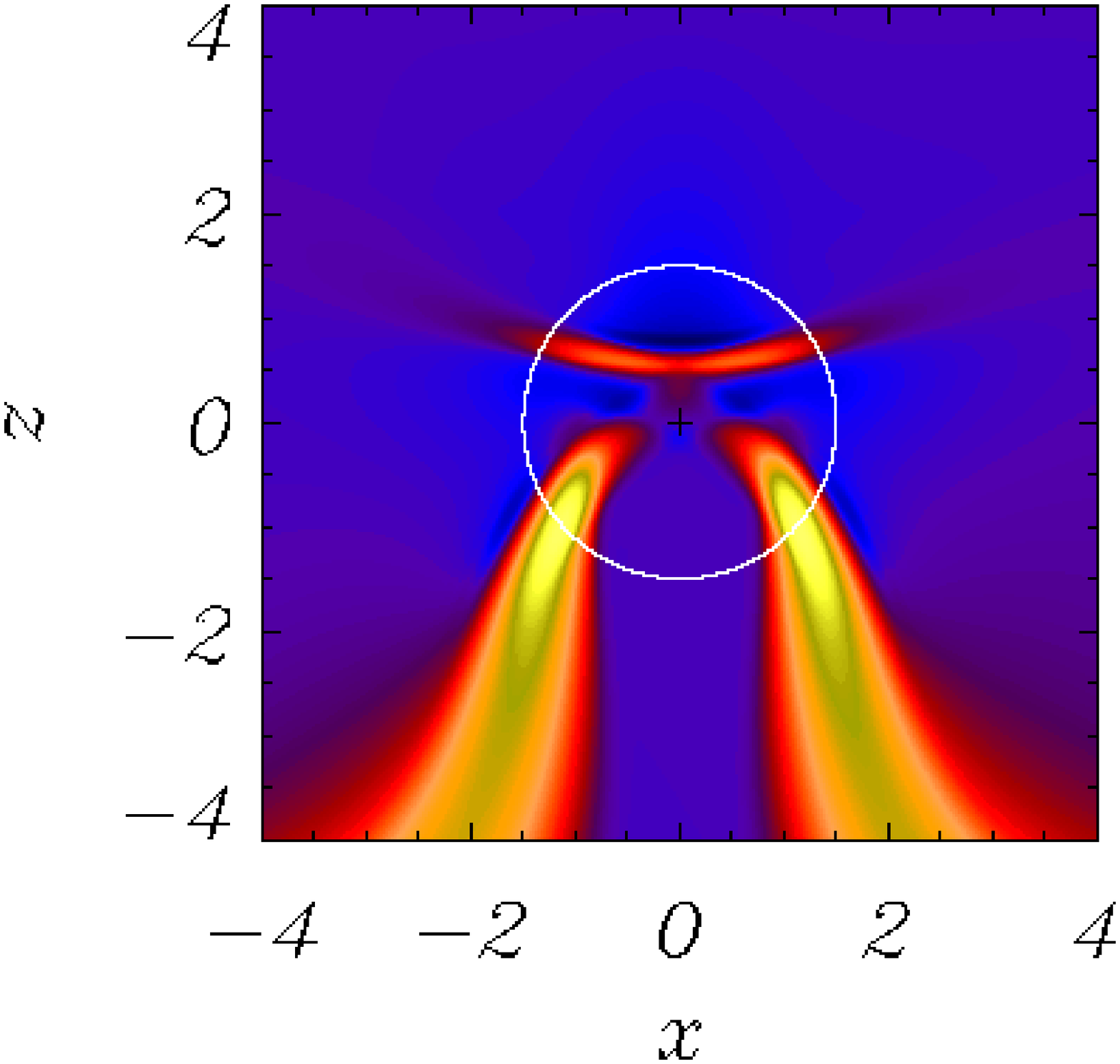}\\
\vspace{0in}
\hspace{0.2in}
\includegraphics[width=1.5in]{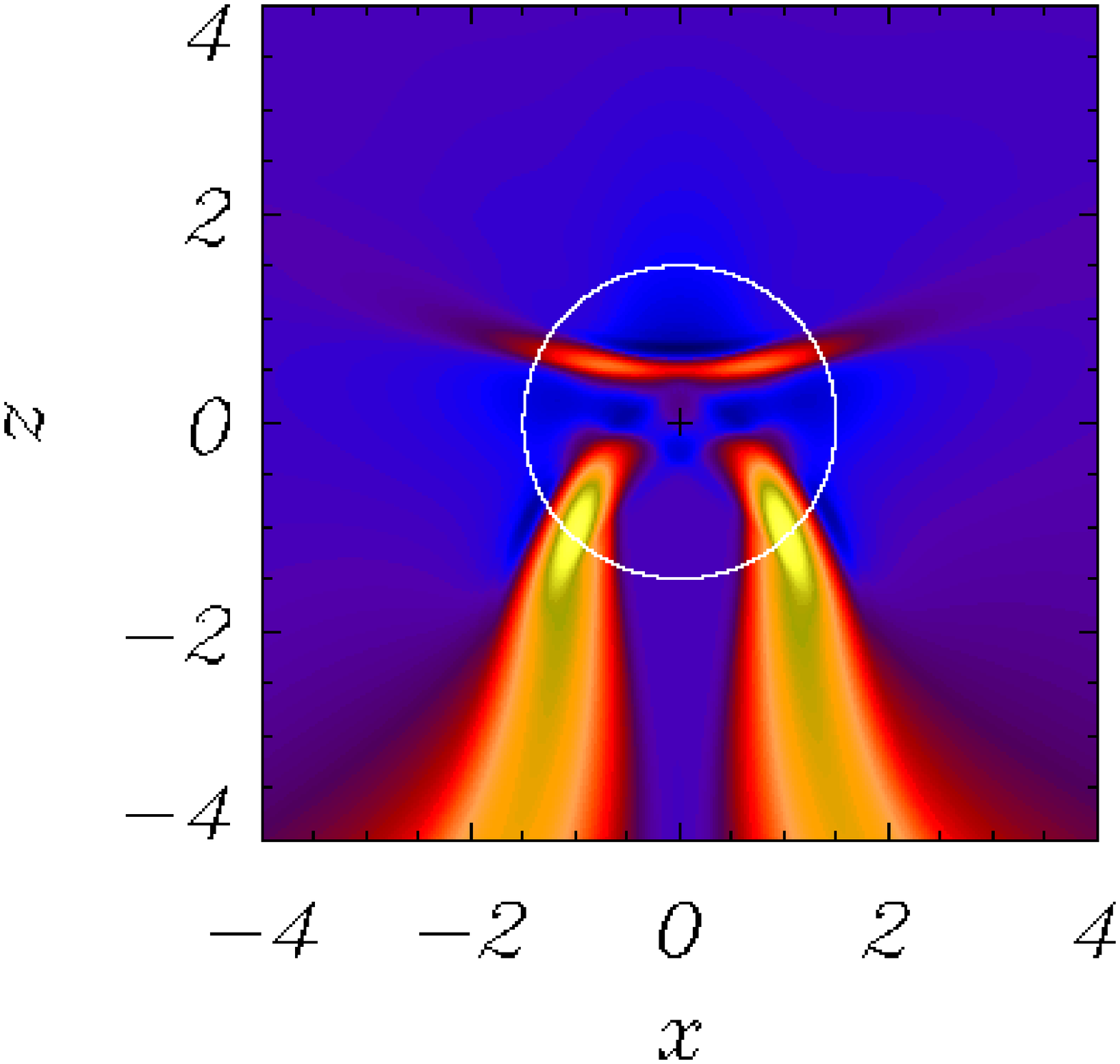}
\hspace{0.2in}
\includegraphics[width=1.5in]{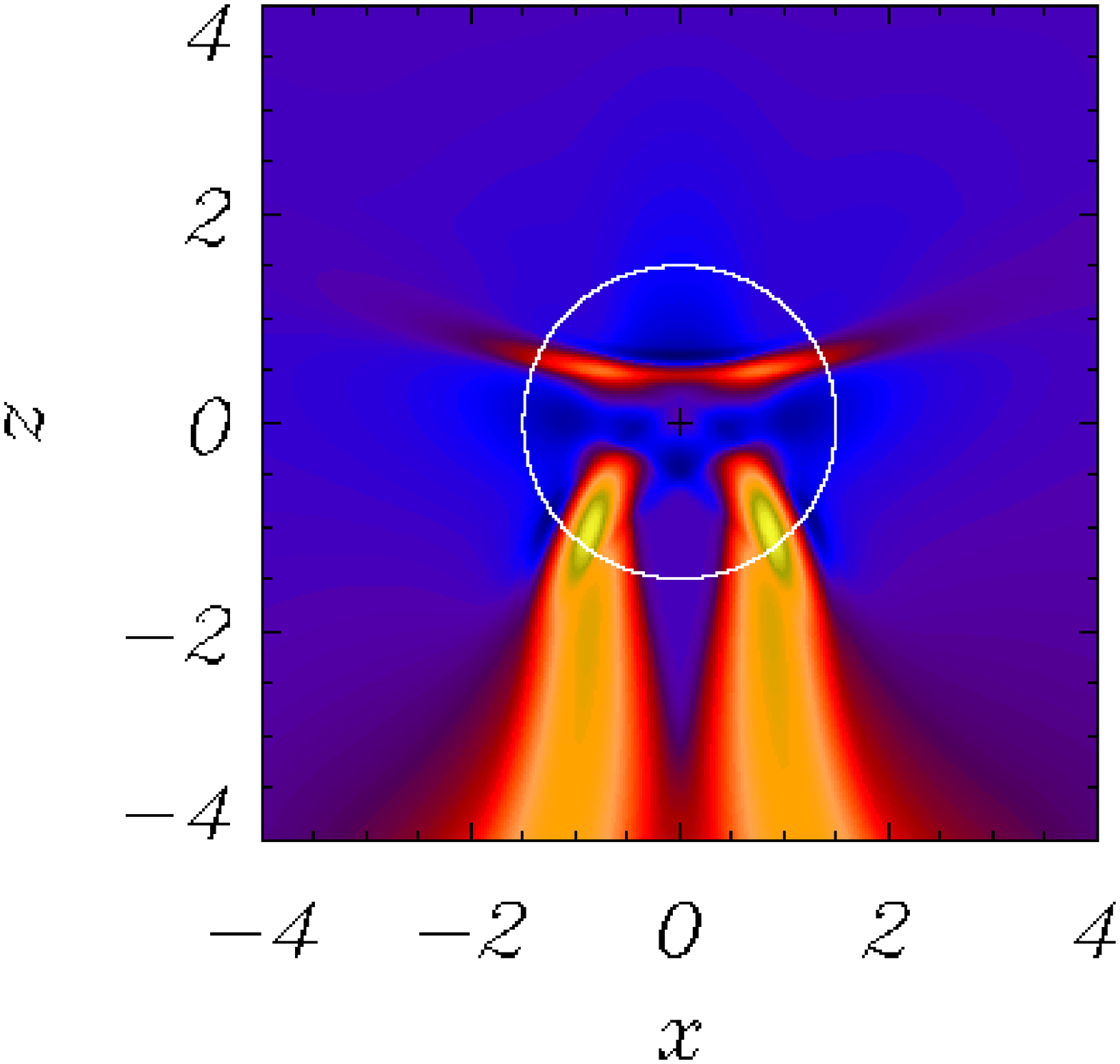}
\hspace{0.2in}
\includegraphics[width=1.5in]{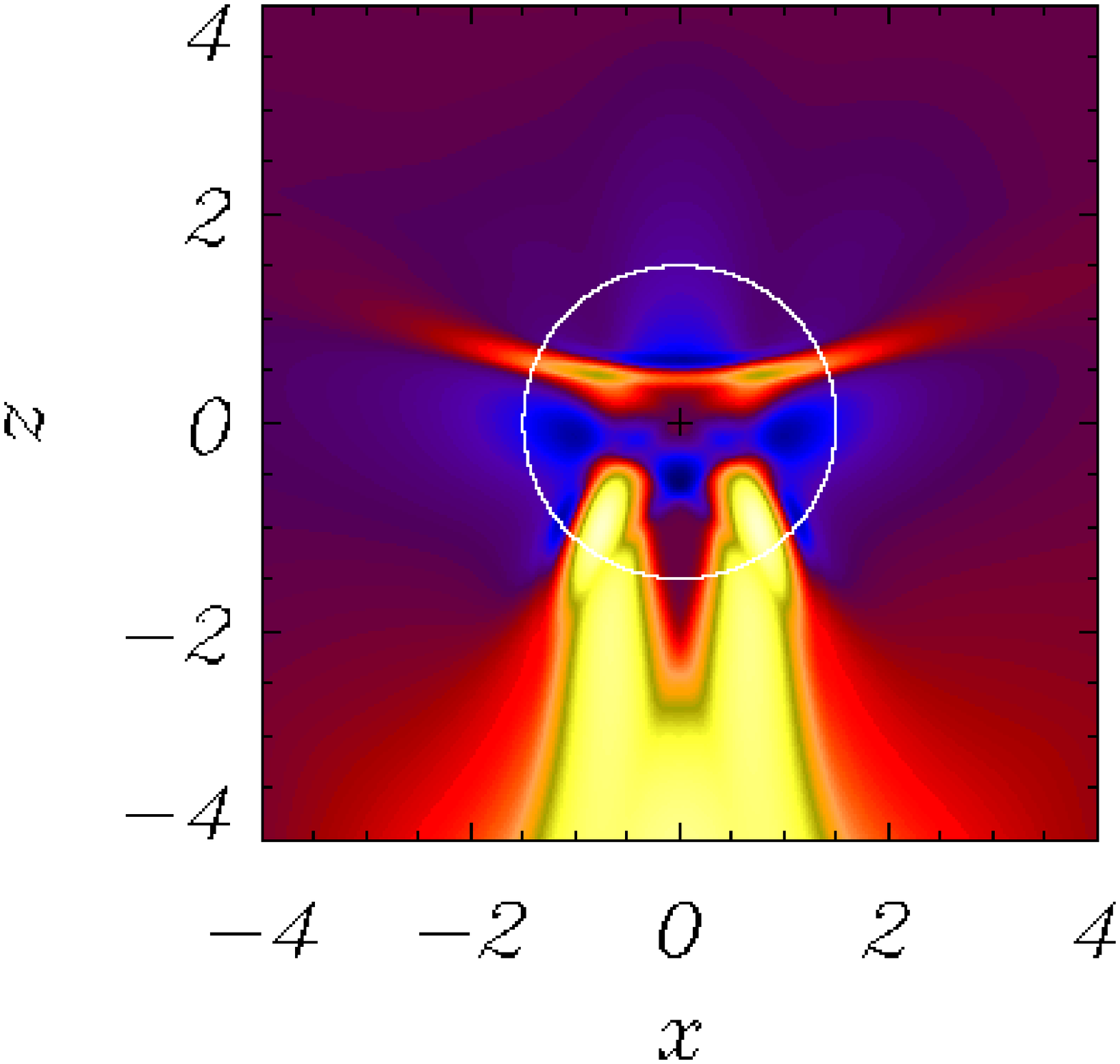}
\caption{Contours of $\rm{v}_\perp$ for  numerical simulation for a 
fast wave sent in from upper boundary for $-4 \leq x \leq 4 $
and $\beta_0=2.25$ and its resultant propagation at times $(a)$ 
$t$=1.7, $(b)$ $t$=1.8, $(c)$ $t$=1.9, $(d)$ $t=2.0$, $(e)$ $t$=2.1,
$(f)$ $t=2.2$, $(g)$ $t$=2.3, $(h)$ $t$=2.4 and $(i)$ $t$=2.5, 
labelling from top left to bottom right. The white circle indicates
the position of the $c_s=v_A\:$ layer and the cross denotes the null 
point in the magnetic configuration.}
\label{MEATLOAF_1_blowup}
\end{center}
\end{figure*}
\end{center}

\section{MHD wave propagation with $\beta_0=2.25$}\label{sec:3.1}
In this section, the linearised equations (\ref{e}) are again solved 
numerically but
for $\beta_0=2.25$ so that the $c_s=v_A\:$ layer is now at a larger
radius than above. All other boundary conditions remain the same.
The behaviour of the perpendicular component of the  magnetoacoustic 
wave when it is inside the  $c_s=v_A\:$ layer can be seen
in Figure \ref{MEATLOAF_1_blowup}.
  Again, we find that the linear, magnetoacoustic wave travels towards 
the neighbourhood of the null point and begins to wrap around it
(as it did for $\beta_0=0.25$). In this case, since the $\beta 
\approx 1$ is now reached earlier, the wave has had less time to 
refract.
Using our terminology and interpretation from Sections 
\ref{Interpretion} and \ref{modeconversion},
we  identify this as the low $\beta$ fast wave. However, when the 
wave passes through the $c_s=v_A\:$ layer, a secondary wave
becomes apparent; we identify this as the high $\beta$ slow wave. 
Thus, the low $\beta$ fast wave has transformed into
two high $\beta$ waves (the split can be followed closely in Figure 
\ref{MEATLOAF_1_blowup}).

The propagation now proceeds in three ways:
\begin{itemize}
\item{The generated high $\beta$ slow wave spreads out along the
field lines and accumulates along the separatrices. {This slow 
wave cannot cross the separatrices.}}

\item{The
high $\beta$ fast wave continues to refract and some of the fast
wave (located very close to the null point) \emph{passes through
the null}. This effect was not seen in the $\beta=0$ model. The
fast wave can now pass through the null because there is now a 
non-zero fast mode speed, $c_{fast}$, at the origin.
Thus, the  high $\beta$ fast
wave has passed (slowly) through the null. This crossing still
creates a large accumulation of current near the null. However,
unlike the $\beta=0$ case, it remains finite. The high $\beta$
fast wave continues to propagate downwards and  leaves the
$c_s=v_A\:$ layer, converting to a  low $\beta$ fast wave and
spreading out (low $\beta$ fast wave propagates roughly
isotropically).}

\item{Finally, the rest of the fast wave located
away from the null (low $\beta$ fast wave) is not greatly affected
by the non-zero sound speed (as $v_A^2 \gg
c_{\textrm{\scriptsize{sound}}}^2$) and so continues to refract
around the null. In fact, as the wings of the low $\beta$ fast
wave wrap around below the null point, they encounter the part of
the fast wave that has travelled through the null and is leaving
the  $c_s=v_A\:$ layer . This results in a complicated
interference pattern. Nonetheless, it appears that the two fast
waves passes through each other without influencing each other
(due to the linear nature of the system). A full non-linear
treatment of the equations may reveal a different behaviour.}
\end{itemize}
Thus, we have a different behaviour than that seen for the
$\beta_0=0.25$ case: With a large $\beta_0$, the perpendicular
component of the velocity has passed through the null (interpreted
as a high $\beta$ fast wave crossing the null), whereas with a
small  $\beta_0$, we did not see any significant wave pass through
the null by the end of the simulation.

So what is the relationship between the $\beta_0=2.25$
and $\beta_0=0.25$ cases; are they not just scaled versions of
each other? This will be discussed in the conclusions.

\subsection{Quantifying mode conversion}

\begin{figure}[ht]
\begin{center}
\includegraphics[width=2.0in]{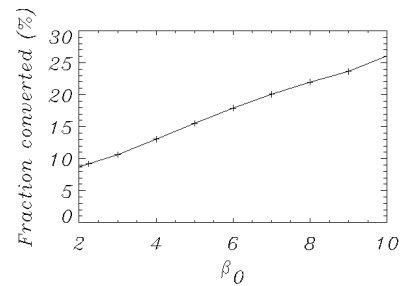}
\caption{{{Fraction of original disturbance converted to high 
$\beta$ slow wave against parameter
$\beta_0$, (gradient of straight line is approximately $2.0$). Note 
that for $\beta_0 < 2.0$, the two
wave types do not separate enough for us to measure them individually.}}}
\label{fig:volume}
\end{center}
\end{figure}

At present, there does not exist a robust set of rules connecting
low and high $\beta$ waves across the $c_s=v_A\:$ layer or across
the $\beta=1$ layer (\cite{CS1999}). Here we specifically
investigate what happens when a low $\beta$ fast wave crosses the
$c_s=v_A\:$ layer  and becomes part high $\beta$ fast wave and
part high $\beta$ slow wave (or alternatively what happens when it
crosses the $\beta \approx 1$ layer a little time earlier).
An obvious question to ask is how much of the incident wave is
converted to high $\beta$ slow wave. This is difficult to quantify
in our $\beta_0=0.25$ simulation, as the high $\beta$ fast and
slow waves never really separate enough for us to measure them
individually. However, the waves do completely separate for the
$\beta_0=2.25$ simulation, for example in subfigure $7(g)$ the
slow wave is clearly separated from the other wave types (i.e. the
high and low $\beta$ fast waves). Thus, we can measure the amount
of this slow wave part by integrating ${\rm{v}}_\perp$ over
the area of the slow wave
(to get the volume or, since the equilibrium density is uniform,
the total momentum of the slow wave) and compare it
to the original incident wave pulse (given by the boundary
conditions; equations (\ref{BC})). {{This shows that $9.2\%$ of the
initial low $\beta$ fast wave disturbance is converted to high
$\beta$ slow wave (for the given boundary conditions with $\beta_0=2.25$).
We can also work out how the proportion of initial disturbance
converted to high $\beta$ slow wave changes with $\beta_0$. This
is shown in Figure \ref{fig:volume} where we can see that the 
fraction of original
disturbance converted to high $\beta$ slow wave is proportional to
$\beta_0$.}}

\section{Conclusions}\label{sec:5}

This paper extends the investigations of McLaughlin \& Hood (2004) into 
the nature of magnetoacoustic waves in the
neighbourhood of a null point. We have conducted two experiments, 
varying the choice of  $\beta_0$ in each, where the choice  dictated 
the
location of the $c_s=v_A\:$ layer.

In the first experiment (Section \ref{sec:2.1}), we set 
$\beta_0=0.25$. We find that the linear
magnetoacoustic wave travels towards the neighbourhood of the null 
point and begins to wrap around it. This occurs
due to the spatially varying Alfv\'en speed, $v_A^2 \left( x,z 
\right)$$=x^2+z^2$, and has been observed before
(Nakariakov \& Roberts  1995; McLaughlin \& Hood  2004). In this 
experiment, we drive a wave pulse in the perpendicular
velocity component and we identify this pulse as  a low $\beta$ fast 
wave.  However, once the wave reaches and
crosses the $c_s=v_A\:$ layer, part of the wave is transformed into a 
high $\beta$ slow wave and
the majority into the high $\beta$ fast wave. The slow wave part
  spreads out along the fieldlines. Meanwhile, the rest of the (low 
$\beta$) fast wave continues to wrap around the
origin. This refraction effect continues again and again, but each 
time part of the fast wave is converted to a slow
wave as it crosses the $c_s=v_A\:$ layer. The majority of the current 
build-up occurs very close to the null.

In our second set of simulations  (Section \ref{sec:3.1}), we set 
$\beta_0=2.25$, thusplacing
the $c_s=v_A\:$ layer at a larger radius than before (since in this 
magnetic geometry, the  $c_s=v_A\:$ layer
occurs at a radius $r= \sqrt{\frac{\gamma \beta_0}{2}}$). In this 
experiment, we find that the magnetoacoustic wave travels
towards the neighbourhood of the null point and begins to wrap around 
it (as before). However, when the wave crosses the
$c_s=v_A\:$ layer, a secondary wave (high $\beta$ slow wave) becomes 
apparent (we find that the fraction of incident
wave converted to slow wave is proportional to $\beta$). The 
propagation now proceeds in three
ways. Firstly, the generated slow wave spreads out along the 
fieldlines and accumulates along the separatrices. Secondly,
the remaining part of the fast wave inside the $c_s=v_A\:$ layer 
continues to refract and some of it (located close to
the null point) \emph{passes through the origin}. We identify this 
part as a high $\beta$ fast wave. The high $\beta$ fast
wave can pass through the origin because, although $v_A (0,0) =0$, 
there is now a non-zero sound speed there
(and $c_{\textrm{\scriptsize{fast}}}^2 \approx v_A^2 + 
c_{\textrm{\scriptsize{sound}}}^2$). This passing causes a large 
current
accumulation near the origin. After it has crossed the null, the high 
$\beta$ fast wave  continues downwards and leaves the
  $c_s=v_A\:$ layer. As it emerges, it becomes a low $\beta$ fast wave 
and  spreads out (since the low $\beta$ fast wave
propagates almost isotropically). Finally, the fast wave located
away from the null and $c_s=v_A\:$ layer  (the 'wings' of the low
$\beta$ wave) are not affected by the non-zero sound speed (as
$v_A^2 \gg c_{\textrm{\scriptsize{sound}}}^2$) and so for them the
refraction effect dominates. In fact, as these wings wrap around
below the null point, they encounter the high $\beta$ fast wave as
it is emerging from the $c_s=v_A\:$ layer. This results in a
complicated interference pattern, but it appears that the two
waves passes through each other without any lasting effect on each
other (due to the linear nature of the system). The $\beta_0=2.25$
numerical simulations also showed good agreement with a WKB
approximation, until the breakdown point when the neighbouring
rays cross each other.

The part of the fast wave that  goes through the null effectively 
escapes the refraction effect of the null point. It seems as
if the smaller the value of $\beta_0$ used, the slower the high 
$\beta$ fast wave can cross the null. This can be understood since
the  high $\beta$ fast wave travels close to the sound speed, which 
is related to $\beta_0$ (recall from section
\ref{sec:1.4} that  $c_s^*= \sqrt{\frac{\gamma}{2} \beta_0}$).
Thus, the behaviour of the  $\beta_0=2.25$ and $\beta_0=0.25$
cases can be understood; in the $\beta_0=0.25$ case, the high
$\beta$ fast wave {\emph{can}} pass through the null, but it
travels at such a slow speed (as $c_s$ varies as $\sqrt{\beta_0}$)
that the refraction effect in the low $\beta$ region dominates.
Hence, as the $c_s=v_A\:$ layer gets closer to the null, less and
less of the fast wave can pass through the null in a given time,
and the parts that can travel through do so at a slower and slower
speed. Also, these parts of the wave that do pass through tend to
be be swamped by the wings of the rest of the wave wrapping round
below the null (repeatedly).

This explains the relationship between our $\beta_0=0.25$ and
$2.25$ investigations. These experiments show scaled  versions of
each other, but with the larger value of  $\beta_0$ it is clearer
to see that there are actually two competing phenomena;  a
\emph{refraction} effect caused by the varying Alfv\'en speed and
a  non-zero sound speed at the null which allows the fast wave to
pass through. It  is the value of $\beta_0$ that dictates which
effect dominates.

Thus, two extremes occur. The first occurs when $\beta_0
\rightarrow 0$; in which case the refraction effect infinitely
dominates over the other effects (and we recover the results of
McLaughlin \& Hood 2004) and the second when $\beta_0
\rightarrow \infty$ and the system becomes hydrodynamic. In this
case, the fast wave reduces to an acoustic wave and so completely
passes through the null (in effect it does not even see the
magnetic field, since  $v_A^2 \ll c_s^2$). Thus, we can understand
the whole spectrum of values of the parameter $\beta_0$.

 From this work, it has been seen that a warm plasma introduces many 
new effects not seen in the cold plasma
limit, most notably the introduction of fast \emph{and} slow waves to 
the system. It also appears that the
choice of $\beta_0$ is of critical importance, since the two 
experiments yield different results. However, the key
choice here is not in picking $\beta_0$, since this just determines 
where the $c_s=v_A\:$ layer will occur. The
choice of $\beta_0$ is equivalent to choosing where to set the 
boundaries of our box, and so the critical parameter
in our system is in choosing  \emph{the distance between the initial 
disturbance and the $c_s=v_A\:$ layer}.
This is because these experiments do not simply show scaled versions 
of each other, they show two competing phenomena;
a \emph{refraction} effect caused by the varying Alfv\'en speed, and 
a non-zero sound speed at the null which allows
the fast wave to pass through.

If the  $c_s=v_A\:$ layer is close to the null (small choice of 
$\beta_0$ and hence there is not much coupling
to the parallel velocity and pressure terms) then the refraction 
effect will dominate, resulting in a scenario
similar to $\beta_0=0.25$, the waves remain trapped near the null 
point and heating will occur close to the null.
The extreme case is $\beta_0
\rightarrow 0$, in which case the refraction effect infinitely 
dominates over the other effects and we recover the results of 
\cite{McLaughlin2004}.

However, if the $c_s=v_A\:$ layer is far from the null, a portion of 
the fast wave will be able to pass through
the origin and escape the system (i.e. will not deposit its energy 
near the null point). In this case, the remaining
wave energy of the generated slow waves is dissipated along the separatrices.
So heating will occur in both
systems, although the nature will be different.

This is all very interesting for its mathematical sake, but how does 
the competition of these two effects manifest
itself in the corona? The plasma $\beta$ parameter is defined as the 
ratio of the thermal
plasma pressure to the magnetic pressure. In most parts of the 
corona, $\beta \ll 1$
and hence the pressure gradients in the plasma can be neglected. 
Values of $\beta=0.01$ are often quoted (e.g. Priest 1982).
However, near null points the plasma $\beta$ can become very large, 
so the (true) plasma $\beta$ varies through the whole
region. However as seen above, it is the distance between the initial 
(planar) pulse and the  $c_s=v_A\:$ layer
that is of critical importance. We believe that coronal disturbances 
will propagate for some distance
before they encounter a (coronal) null point. Hence, the fast wave 
part of the disturbance will feel the refraction
effect of the null and begin to refract around it. By the time the 
$c_s=v_A\:$ layer is reached or the  sound
speed becomes important, the fast wave disturbance will be (almost) 
circular in nature and will have thinned and
dissipation will be extracting the energy from the wave. It is true 
that some of the wave may pass through the null or
be converted into high $\beta$ slow waves, but it is likely that the 
majority of the wave energy will accumulate
close to the null, causing large current accumulation and heating there.

\section*{Appendix A: Lobe generation}\label{Appendix}
\begin{figure*}[t]
\begin{center}
\includegraphics[width=2.1in]{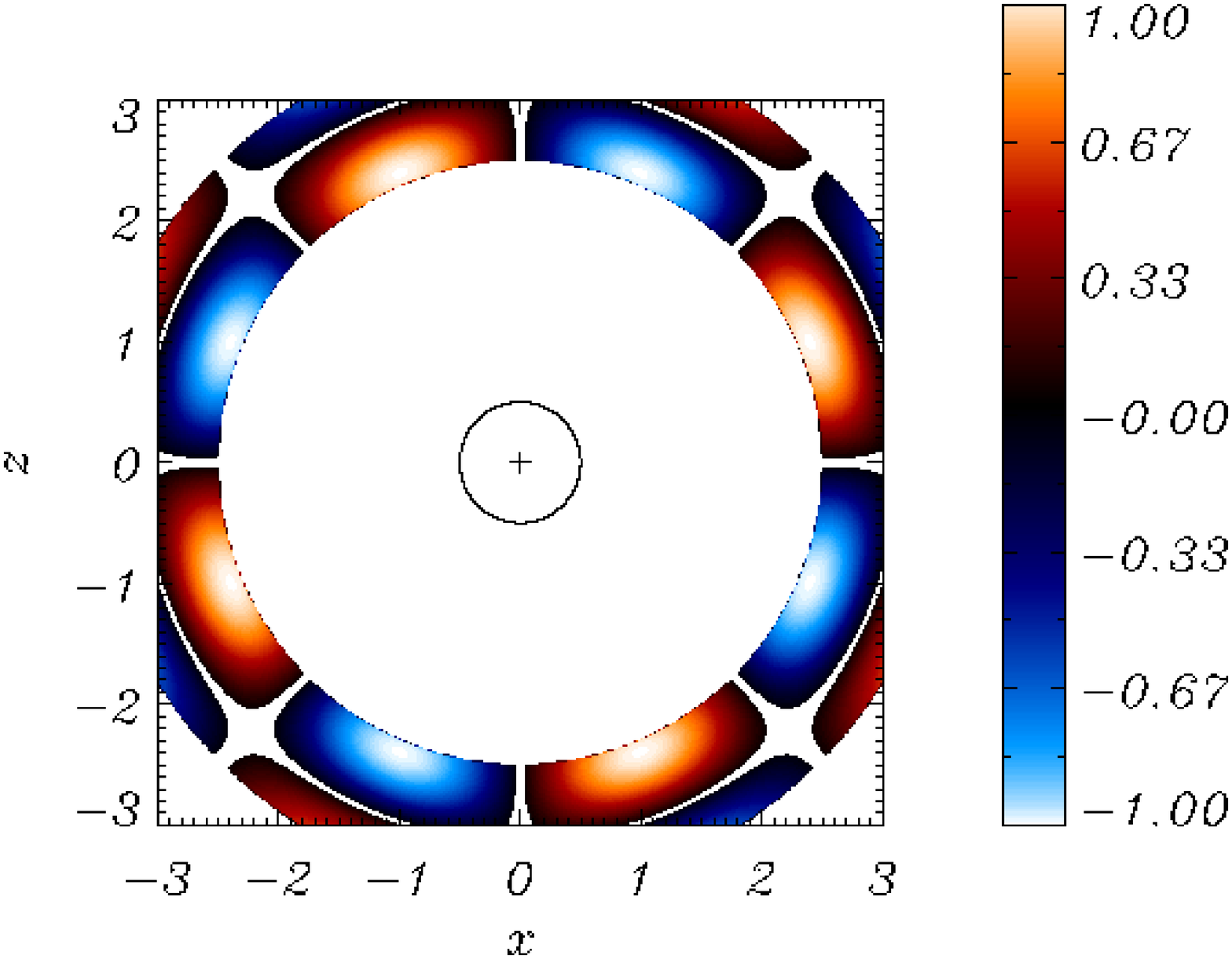}
\hspace{0.1in}
\includegraphics[width=1.7in]{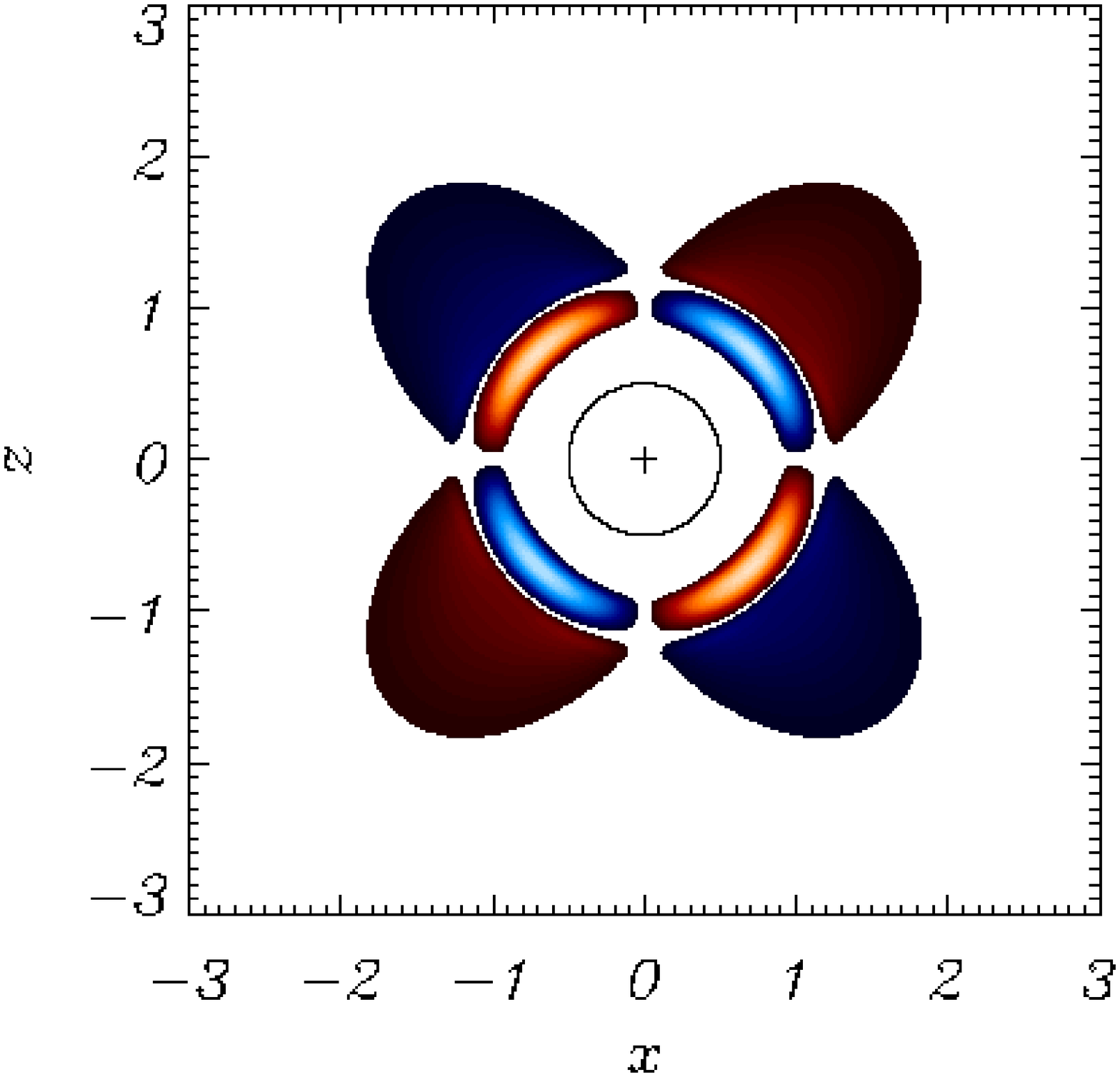}
\caption{({\emph{Left}}) Contour of    $\frac{\gamma}{4}\sin{4 \theta}
\frac{\sqrt{3}}{r^2}\left[ \left( 8-\pi^2 r^2\right) \sin{\pi(r-2.5)} 
\right.$$\left.- 5\pi r \cos{\pi(r-2.5)}  \right]$
for $2.5 \leq r \leq 3.5$. ({\emph{Right}}) Contours of the numerical 
simulation of $p_1$ for a fast wave pulse initially
located  about a radius $\sqrt{x^2+z^2}=3$ and its resultant 
propagation after time $t$=1.0.
The black circle indicates the position of the  $c_s=v_A\:$   layer 
and the cross denotes the null point in the
magnetic configuration. $p_1$ has an alternating form, where orange 
represents $p_1>0$ and blue $p_1<0$.}
\label{sinfourtheta}
\end{center}
\end{figure*}

We can gain insight into the $\beta \neq 0$ system using a small 
$\beta_0$ expansion of the linearised ideal MHD equations.
In polar coordinates, our magnetic field is
\begin{eqnarray*}
{\bf{B}}_0 &=& {-r \cos{2\theta} \:{\bf{\hat{r}}}} + {r 
\sin{2\theta}\:} {  { \bf{{\hat\theta}} }  }
\end{eqnarray*}
We now assume ${{\rm{v}_\perp}}$ and ${{\rm{v}_\parallel}}$ can be 
expanded in powers of $\beta_0$ such that:
\begin{eqnarray*}
{\rm{v}_\perp} = {{\rm{v}_\perp}_1}(r,t) + \beta_0^2 \: 
{{\rm{v}_\perp}_2}(r,\theta,t) \;,\quad {\rm{v}_\parallel} =
\beta_0\: {{\rm{v}_\parallel}_1}(r,\theta,t) \;\;.
\end{eqnarray*}
Substituting these forms into equations (\ref{e}) gives:
\begin{eqnarray}
  \frac{\partial^2 }{\partial t^2 }  {{\rm{v}_\perp}}_1  &=& r 
\frac{\partial }{\partial r }\left(r  \frac{\partial }{\partial r }
{{\rm{v}_\perp}}_1 \right) + \mathcal{O}(\beta_0)\;,\nonumber\\
   \frac{\partial }{\partial t }  {{\rm{v}_\parallel}}_1   &=& - 
{\frac {1}{2}} \left( \mathbf{B}_0  \cdot \nabla  \right) p_1
+ \mathcal{O}(\beta_0)   \;,\nonumber\\
\frac{\partial   }{\partial t } p_1 &=& -\gamma \nabla \cdot \left[ 
\left( \frac{ \sin {2\theta}}{r} {{\rm{v}_\perp}}_1\right)
  \:{\bf{\hat{r}}} +  \left( \frac{ \cos {2\theta}}{r} {{\rm{v}_\perp}}_1\right)
\:{\bf{\hat{\theta}}}\right] + \mathcal{O}(\beta_0)\;\nonumber \\
&=& \gamma \:{\sin {2 \theta}} \: r                   \frac{\partial 
}{\partial r}   \left(
\frac{{{\rm{v}_\perp}}_1}  {r^2}           \right)+ 
\mathcal{O}(\beta_0)\label{gavin}
\end{eqnarray}
Substituting $\frac{\partial   }{\partial t } p_1$ into the equation for
$\frac{\partial }{\partial t }  {{\rm{v}_\parallel}}_1$ and ignoring 
terms of order $\beta_0$  gives:
\begin{eqnarray}
  \frac{\partial^2 }{\partial t^2 }  {{\rm{v}_\parallel}}_1 &=& 
\frac{\gamma}{2} \left\{ -r \cos{2\theta}\sin{2\theta}
\frac{\partial   }{\partial r}r \left[ \frac{\partial   }{\partial r} 
\left(    \frac{{{\rm{v}_\perp}}_1}  {r^2}
\right)\right] \right. \nonumber\\
&+& \left. r \sin{2\theta}   \cos{2\theta}     \left( 
\frac{{{\rm{v}_\perp}}_1}  {r^2}           \right)
\right\}\nonumber\\
&=& \frac{\gamma}{4}\sin{4 \theta} \:r^3  \frac{\partial   }{\partial 
r}\frac{1}{r} \left[ \frac{\partial   }{\partial r}
\left(    \frac{{{\rm{v}_\perp}}_1}  {r^2}           \right)\right] = 
\frac{\gamma}{4}\sin{4 \theta} \:\mathcal{F}(r,t)\label{viva}
\end{eqnarray}
where $\mathcal{F}(r,t)$ is a known function that depends only on
$r$ and $t$. This function is determined solely by the behaviour
of ${{\rm{v}}_\perp}_1$. Although the shape of the ${v_\perp}_1$
pulse may be continuous, the derivatives in (\ref{viva}) mean that
$v_\parallel$ may have discontinuous leading and trailing edges.
Thus, we can see that, with our choice of magnetic null point and
by driving ${\rm{v}}_\perp$, the form of ${\rm{v}}_\parallel$
naturally develops a $\sin{4 \theta}$ dependence. Assume an
initial circular pulse
\begin{displaymath}
{{\rm{v}}_\perp}_1 = \left \{\begin{array}{cc}
   \sin \pi \left (r-2.5\right ) & 2.5 < r < 3.5 \\
   0 & \hbox{elsewhere} \\
\end{array}
\right .
\end{displaymath}
The right hand side of (\ref{viva}) is $\frac{\gamma}{4}\sin{4
\theta} \frac{\sqrt{3}}{r^2}\left[ \left( 8-\pi^2 r^2\right)
\sin{\pi(r-2.5)} \right.$$\left.- 5\pi r \cos{\pi(r-2.5)} \right]$
and a contour plot of this is shown in Figure \ref{sinfourtheta}.
However, the contour shown is not a true integral of such an
initial condition (i.e. is not a solution of (\ref{viva})) but it
does clearly demonstrate the $\sin{4 \theta}$ dependence. From
equation (\ref{gavin}), we can also see that $p_1$ develops a
$\sin{2 \theta}$ behaviour. This behaviour can be seen on the
right-hand side of  Figure \ref{sinfourtheta}, which shows the
contours of $p_1$ from a numerical simulation.

\section*{Appendix B: Analytical work}\label{sec:4a}

In order to gain more insight into the numerical simulations, we use 
a geometrical optics WKB solution to obtain an approximate analytical 
solution.
By combining equations
(\ref{eq:2.3}), (\ref{eq:2.4}) and (\ref{eq:2.5}), we can form:
\begin{eqnarray}
\frac{\partial^2 \mathbf{v} } {\partial t^2} = \frac {\gamma 
p_0}{\rho_0} \nabla \left( \nabla \cdot \mathbf{v} \right) +
\left\{ \nabla \times \left[  \nabla \times \left( \mathbf{v} \times 
{\mathbf{B}}_0 \right) \right] \right\} \times \frac
{{\mathbf{B}}_0 } {\mu \rho_0}\label{10}
\end{eqnarray}
We substitute $\mathbf{v} = {\bf{a}} e^{i \phi (x,z) } \cdot e^{-i 
\omega t}$  into (\ref{10}) where ${\bf{a}}$ is a constant
amplitude vector in the geometrical optics approximation. The 
physical optics approximations would allow the amplitude vectors
to vary with position as well as the wavenumber $\phi$. Thus, we 
follow the ray paths. We now take the dot product with ${\bf{B}}_0$ and $\nabla \phi$ to 
get the two velocity components:
\begin{eqnarray*}
\left[\begin{array}{cc}
\omega^2 & - \frac{\gamma p_0}{\rho_0} ({\bf{B}}_0 \cdot {\nabla \phi})\\
       \frac{1}{\mu \rho_0} (     {\bf{B}}_0 \cdot {\nabla 
\phi})\left|{\nabla \phi}\right|^2 &  \omega^2  -
\left( \frac{\gamma p_0}{\rho_0} +    \frac{ 
\left|{\bf{B}}_0\right|^2  }{\mu \rho_0}  \right) \left|{\nabla \phi} 
\right|^2
\end{array} \right]  \left(\begin{array}{c}
{\bf{v}}\cdot {\bf{B}}_0\\
{\bf{v}}\cdot {\nabla \phi}
\end{array} \right) = \left(\begin{array}{c}
0\\
0
\end{array} \right)
\end{eqnarray*}
These two coupled equations must have zero determinant to prevent a 
trivial solution, and so by evaluating the determinant,
substituting $c_s^2= \frac{\gamma p_0}{\rho_0}$ and 
${\rm{v}}_0^2=\frac{B^2}{\mu \rho_0}$, and making the   WKB 
approximation such that
$\omega \sim \phi \gg 1$ leads to a first order equation of the form:
\begin{eqnarray*}
\omega^4 - \omega^2\left[ c_s^2 + {\rm{v}}_0^2\left(x^2  \right. \right. &+& \left. \left.   z^2\right) 
\right]\left(p^2+q^2\right)  \\
   &+& c_s^2 {\rm{v}}_0^2 
\left(p^2+q^2\right)
\left(xp-zq\right)^2=0\qquad \qquad \qquad \\
\Rightarrow \mathcal{F} \left( x,z,\phi,p,q \right) &=& 0 \nonumber \\ 
&=& {\frac{1}{2}}\left[ \omega^4 - \omega^2\left[  c_s^2 + {\rm{v}}_0^2
\left(x^2+z^2\right) \right]\left(p^2+q^2\right)   \right. \\
&+& \left. c_s^2 
{\rm{v}}_0^2 \left(p^2+q^2\right) \left(xp-zq\right)^2\right]
\end{eqnarray*}
where  $p=\frac {\partial \phi} {\partial x}$ and  $q=\frac {\partial 
\phi} {\partial z}$ and $\mathcal{F}$ is a
non-linear PDE. Also note that the Alfv\'en speed $v_A^2= 
{\rm{v}}_0^2\left(x^2+z^2\right)$.
This PDE can also be written as
\begin{eqnarray*}
2\omega^2 = \left[   c_s^2+{\rm{v}}_0^2\left(x^2+z^2\right)  \right] 
\left(   p^2+q^2 \right) \quad \quad\quad \quad \quad \quad\quad \quad\\
\pm \sqrt{    \left(  p^2+q^2 \right)^2
\left[ {c_s^2+{\rm{v}}_0^2}\left(x^2+z^2\right) \right]^2 -4\:  c^2_s 
{\rm{v}}_0^2  \left(  p^2+q^2 \right) \left(   xp-zq   \right)^2 }
\label{ppo}
\end{eqnarray*}
This equation is reminiscent of the dispersion relation for 
magnetoacoustic waves (e.g. \cite{roberts1985}).
This equation contains information about the two wave types. The 
method to solve this equation is to assume it to be of the form
$(\omega^2 - \omega_{\textrm{\scriptsize{slow}}}^2) \: (\omega^2 - 
\omega_{\textrm{\scriptsize{fast}}}^2)=0$. Considering the
fast wave (so   $\omega^2 \neq 
\omega_{\textrm{\scriptsize{slow}}}^2$), we can apply the method of 
characteristics to generate
the equations:
\begin{eqnarray*}
\frac {d \phi }{ds} &=& 2 \omega ^2  \nonumber\\
  \frac {dp}{ds} &=&  -Ax - \left( CA^2x - 2pAB c_s^2 
{\rm{v}}_0^2\right) / D \nonumber \\
  \frac {dq}{ds} &=&  -Az - \left( CA^2z + 2qAB c_s^2 
{\rm{v}}_0^2\right) / D\nonumber \\
  \frac {dx}{ds} &=&\; \; Cp + \left( AC^2p - 2pB^2c_s^2{\rm{v}}_0^2 - 
2c_s^2 {\rm{v}}_0^2 x AB \right)  / D \nonumber \\
\frac {dz}{ds} &=& \;\; Cq + \left( AC^2q - 2qB^2c_s^2{\rm{v}}_0^2 + 
2c_s^2{\rm{v}}_0^2 z AB \right)  / D \nonumber 
\label{fastcharacteristics}
\end{eqnarray*}
where $A=p^2+q^2$, $B=xp-zq$, $C=c_s^2 + 
{\rm{v}}_0^2\left(x^2+z^2\right)$,  $D = \sqrt{A^2C^2-4AB^2c_s^2 
{\rm{v}}_0^2 }$ and $\omega$ is the
frequency of our wave and $s$ is some parameter along the characteristic.
These five ODEs were solved numerically using a fourth-order 
Runge-Kutta method. Contours of constant $\phi$ can be thought of as
defining the positions of the edges of the wave pulse, i.e. with 
correct choices of $s$, the WKB solution represents the front,
middle and back edges of the wave.

\begin{center}
\begin{figure*}[t]
\begin{center}
\vspace{0.05in}
\includegraphics[width=1.5in]{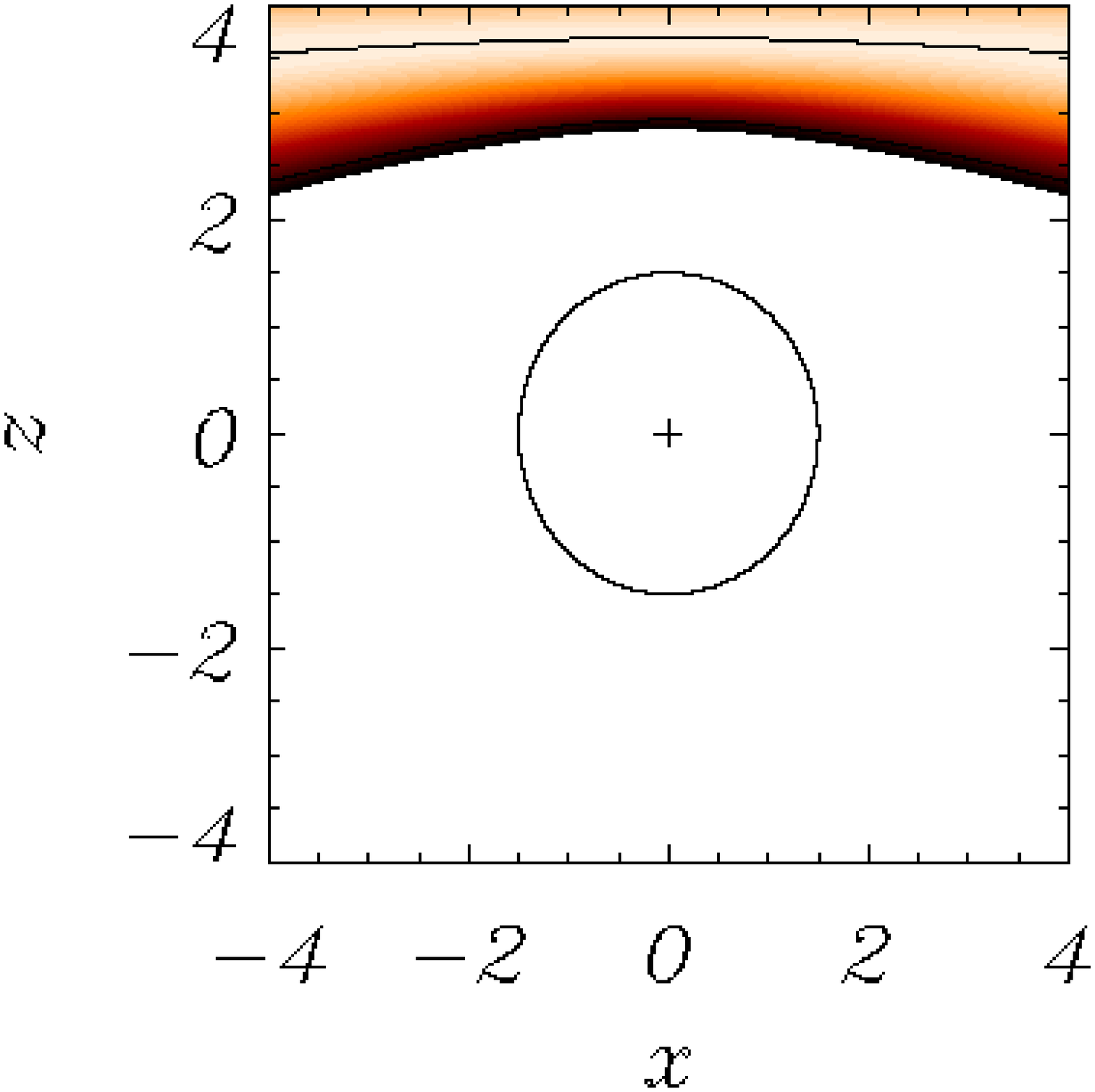}
\hspace{0.1in}
\includegraphics[width=1.5in]{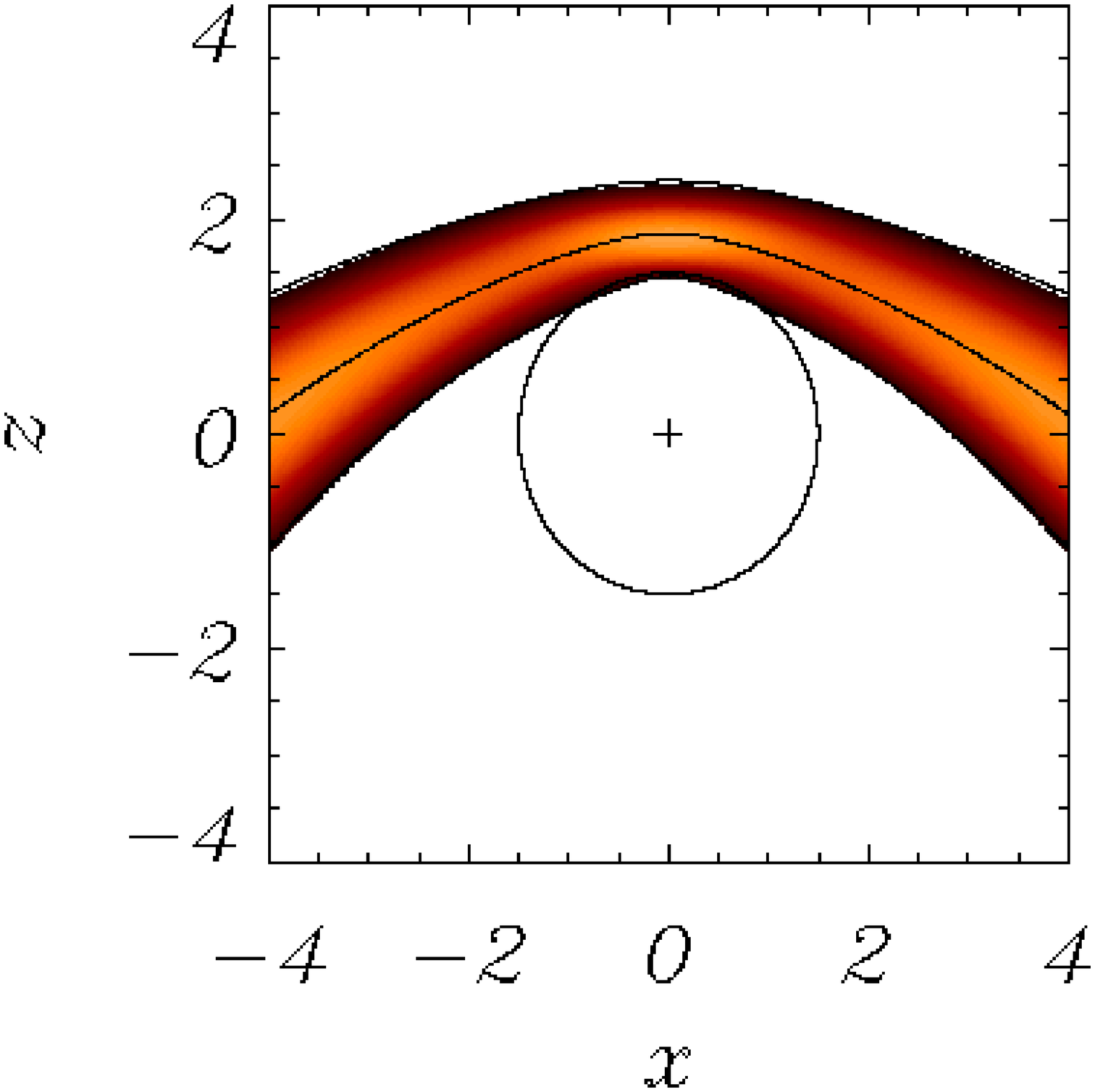}
\hspace{0.1in}
\includegraphics[width=1.5in]{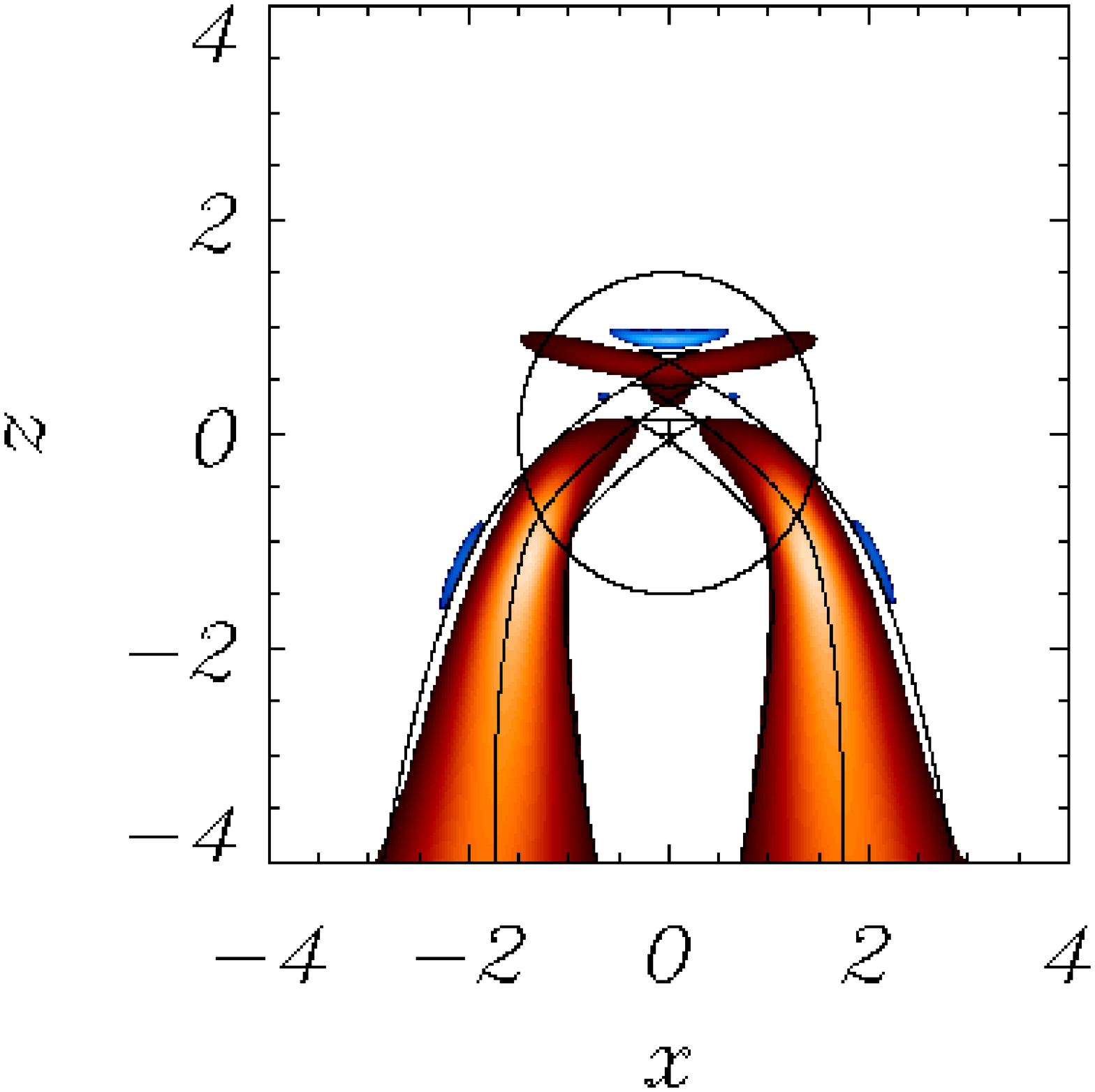}\vspace{0.05in}\\
\includegraphics[width=1.5in]{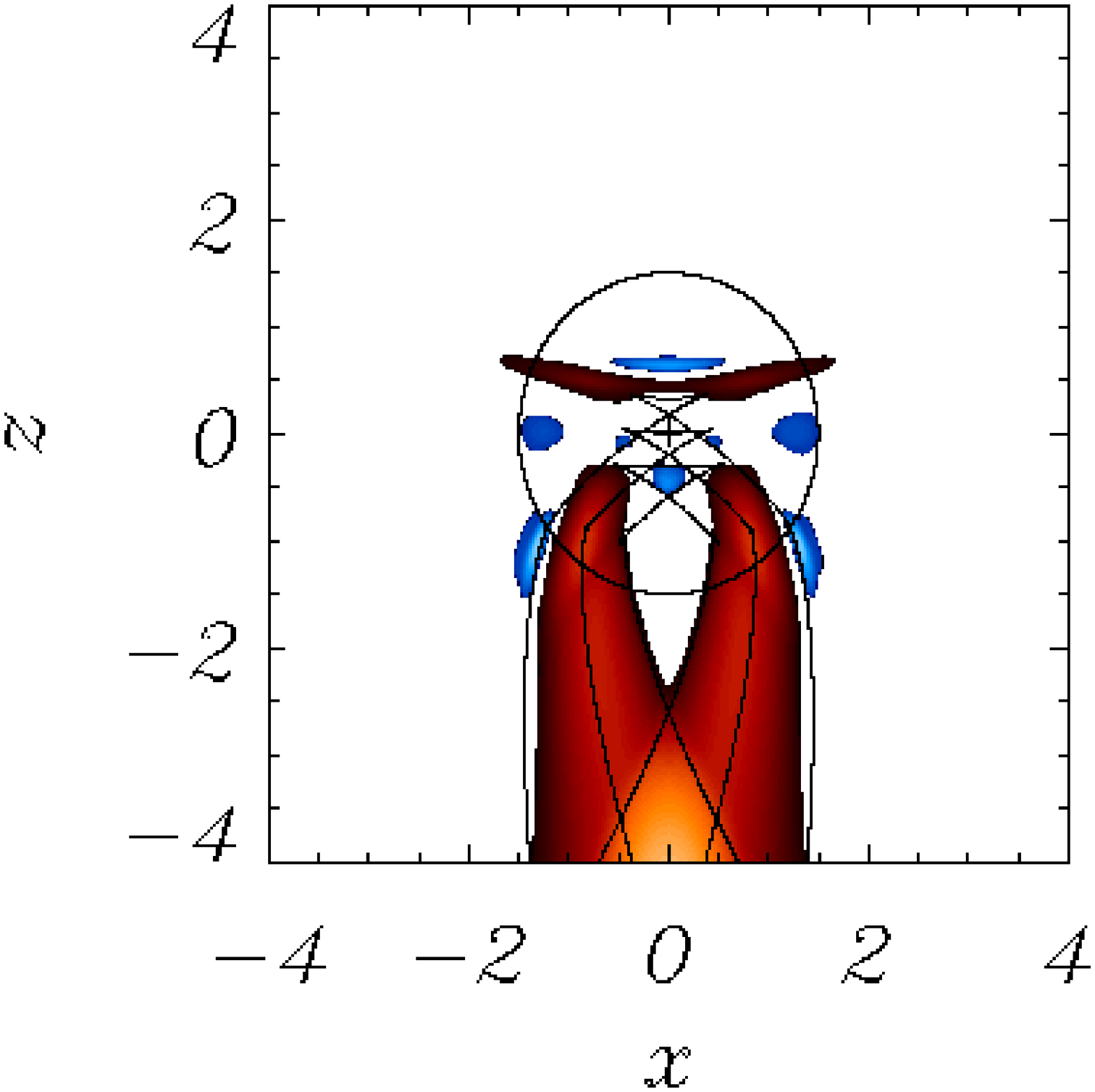}
\hspace{0.1in}
\includegraphics[width=1.5in]{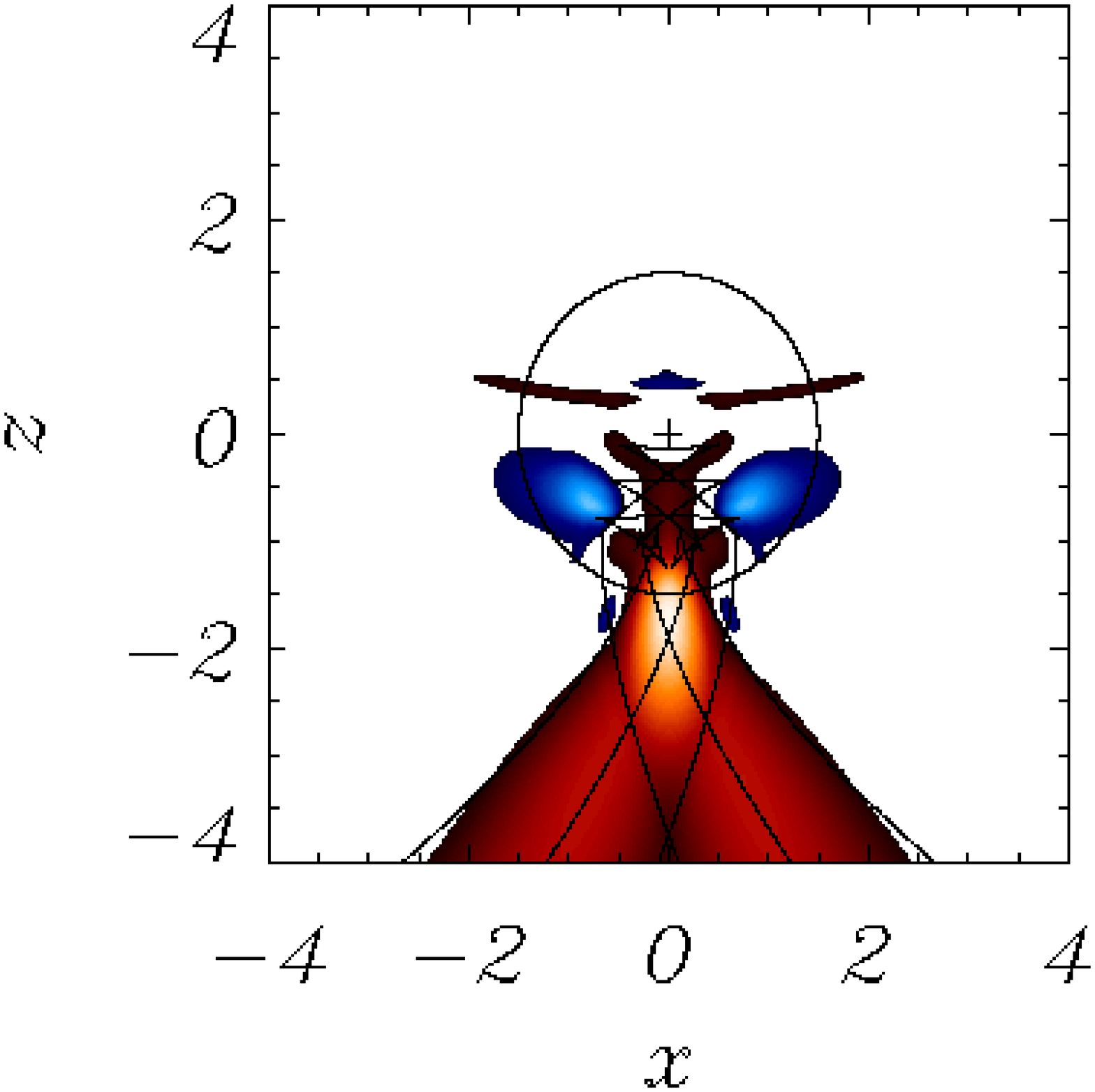}
\hspace{0.1in}
\includegraphics[width=1.5in]{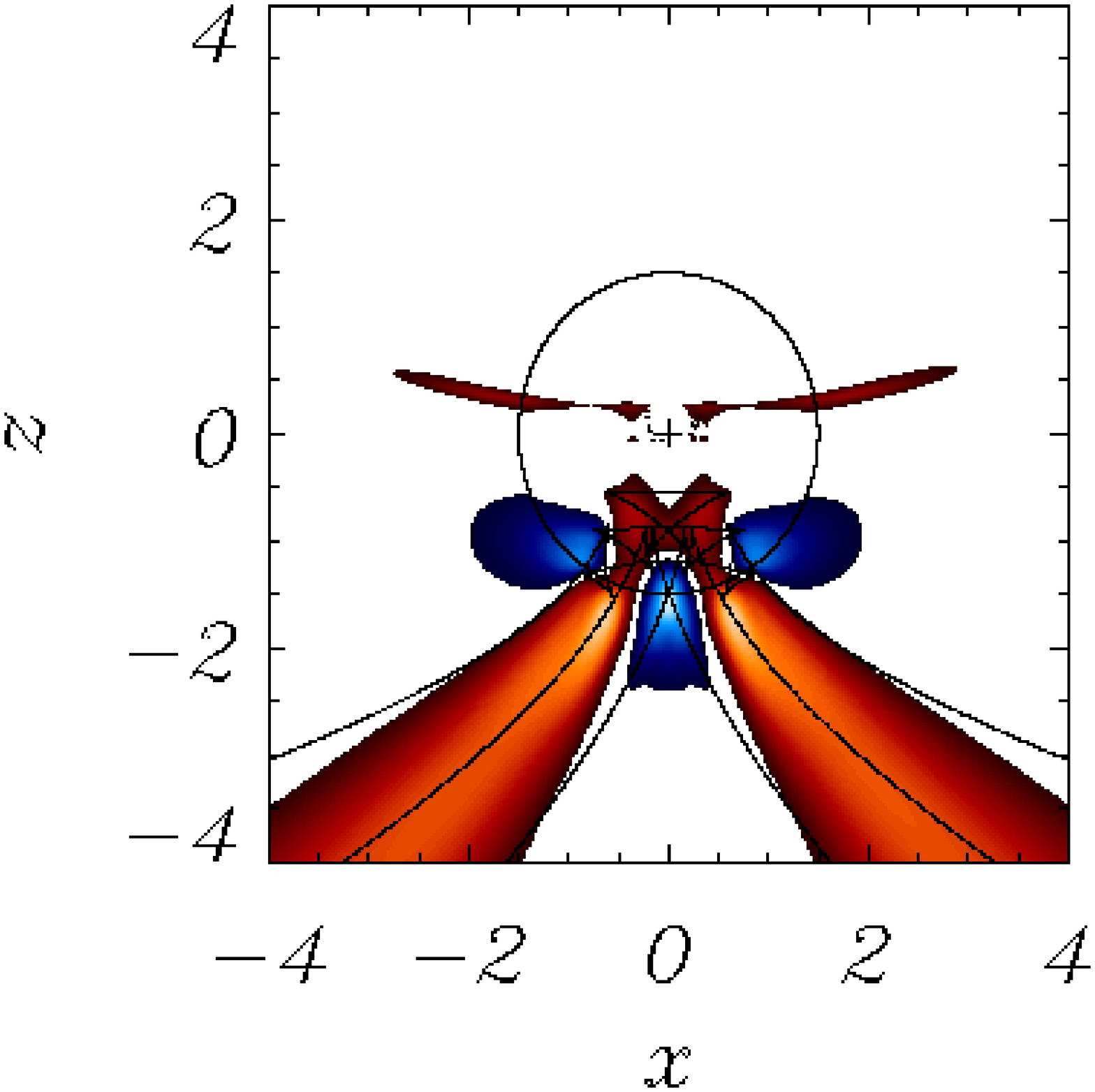}\vspace{0.05in}\\
\includegraphics[width=1.5in]{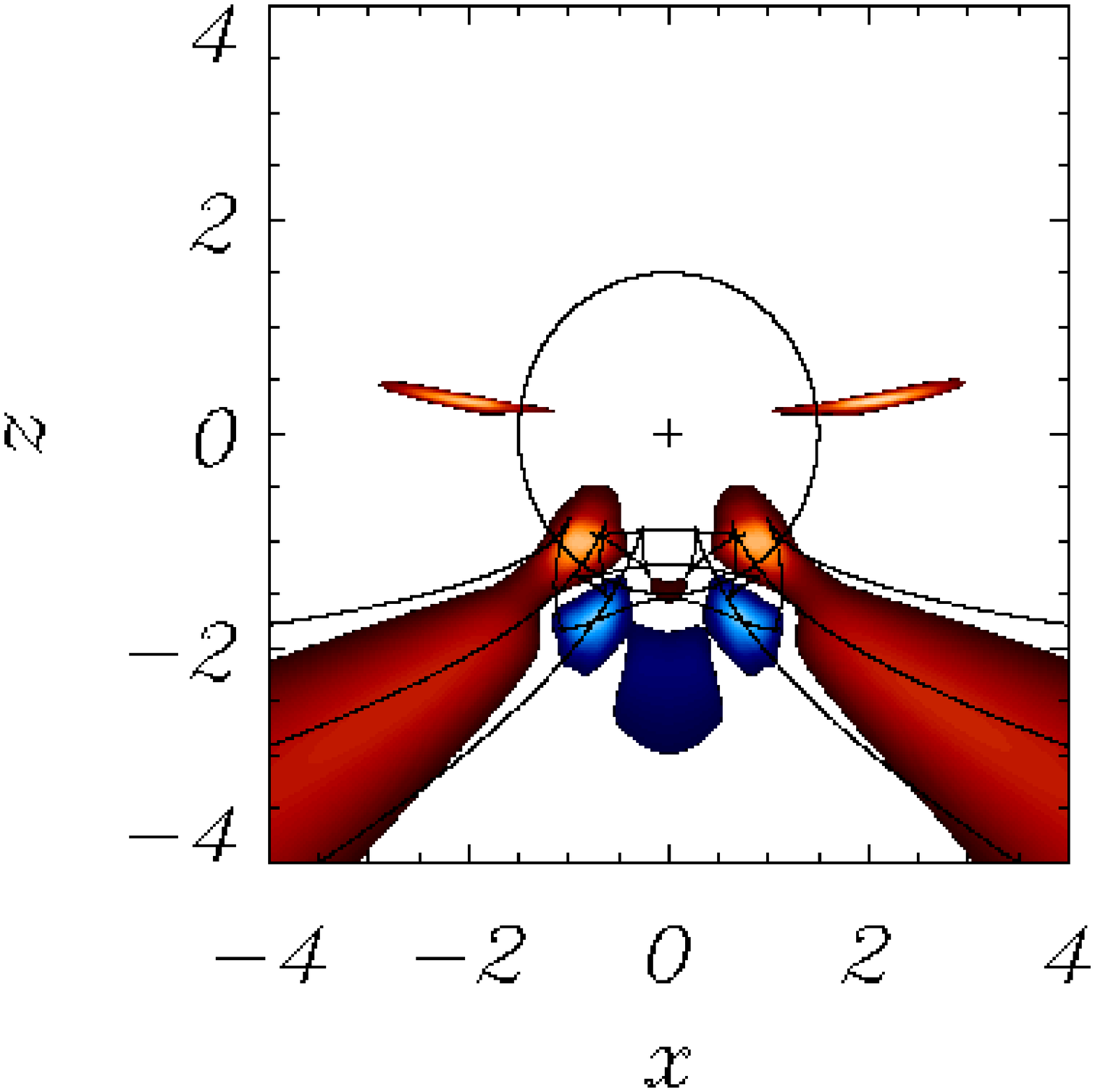}
\hspace{0.1in}
\includegraphics[width=1.5in]{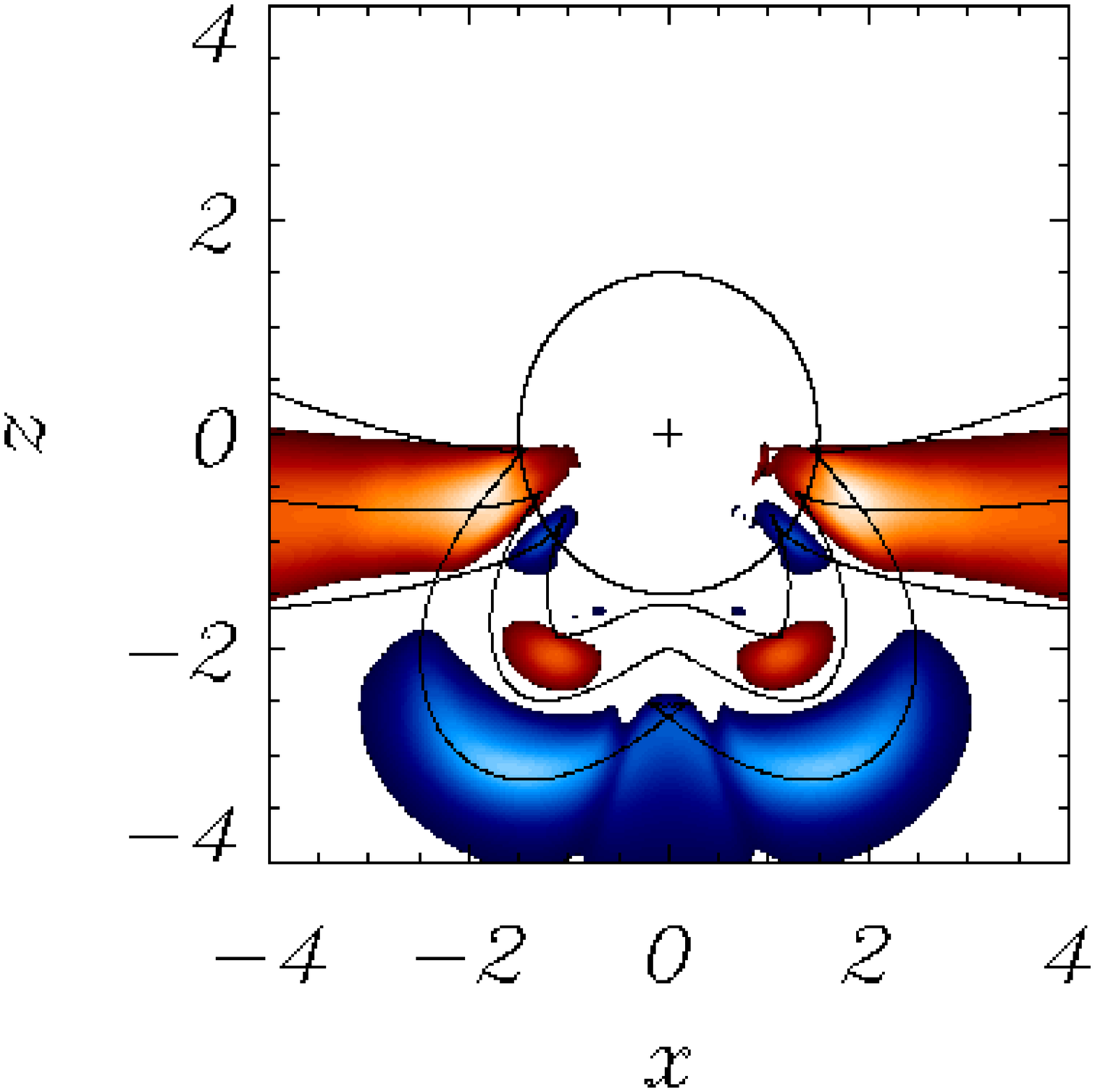}
\hspace{0.1in}
\includegraphics[width=1.5in]{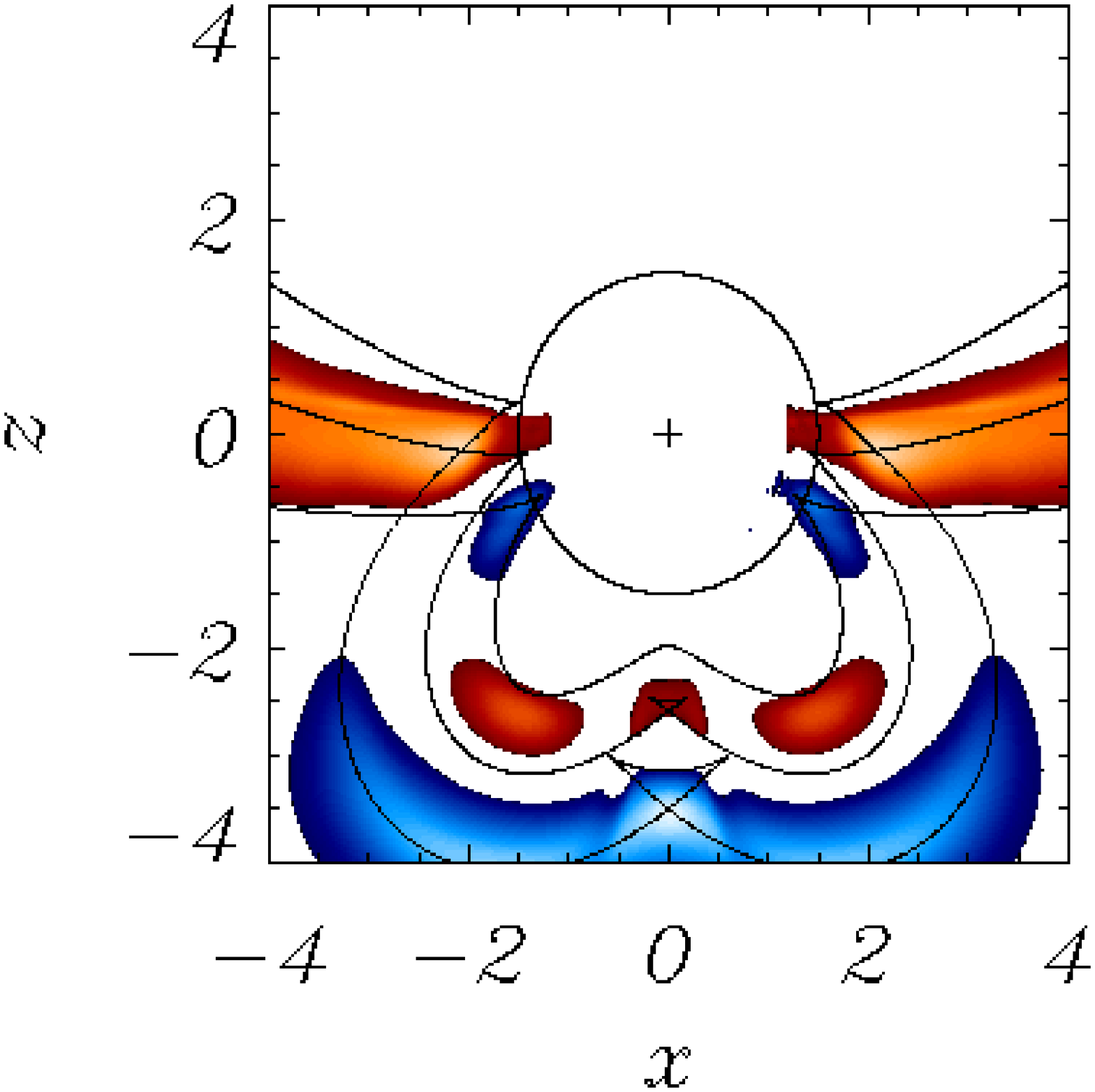}
\caption{{{Comparison  of $\rm{v}_\perp$ from  numerical 
simulation and analytical solution for a fast wave sent in from upper
boundary for $-4 \leq x \leq 4 $
and $\beta_0=2.25$ and its resultant propagation at times  $(a)$ 
$t$=0.33, $(b)$ $t$=1.0, $(c)$ $t$=2.0, $(d)$ $t$=2.33, $(e)$
$t$=2.67, $(f)$ $t$=3.0, $(g)$ $t=$3.33, $(h)$ $t$=3.67 and $(i)$ 
$t$=4.0, labelling from top left to bottom right. The lines
represent the front,  middle and back edges of the wave. The circle 
indicates the position of the $c_s=v_A\:$ layer and the cross
denotes the null point in the magnetic configuration.}}}
\label{newnew}
\end{center}
\end{figure*}
\end{center}

\begin{center}
\begin{figure*}[t]
\begin{center}
\vspace{0.05in}
\includegraphics[width=1.5in]{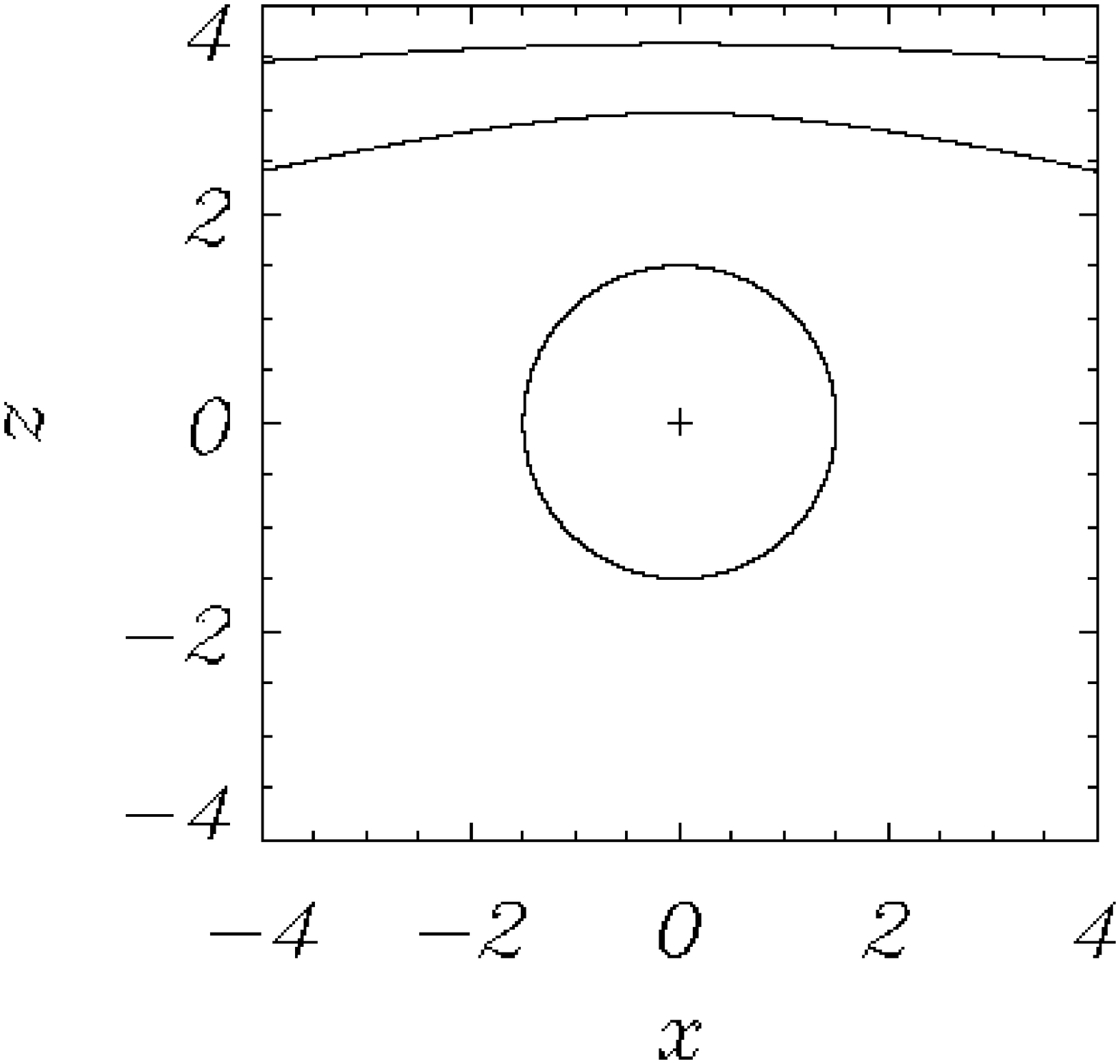}
\hspace{0.1in}
\includegraphics[width=1.5in]{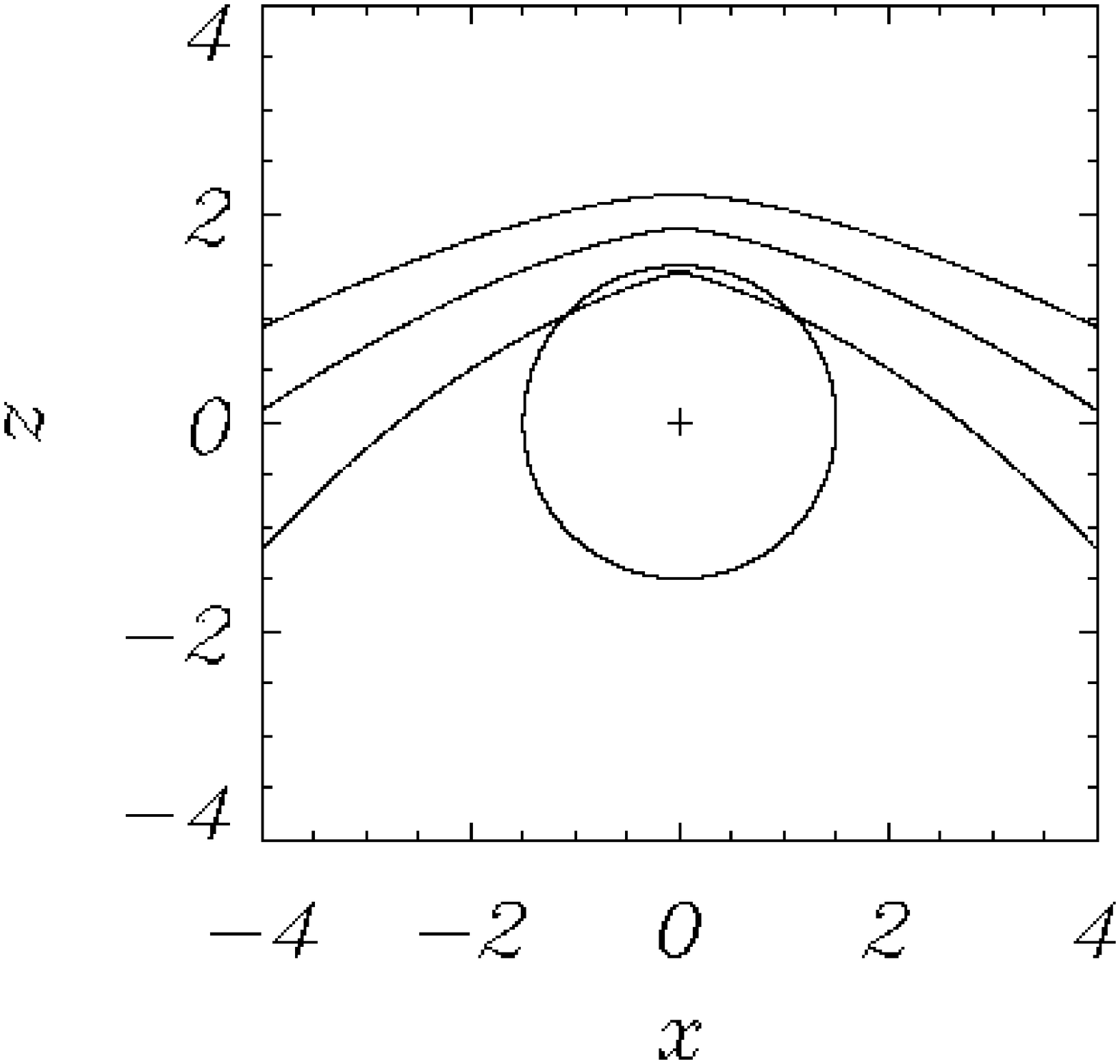}
\hspace{0.1in}
\includegraphics[width=1.5in]{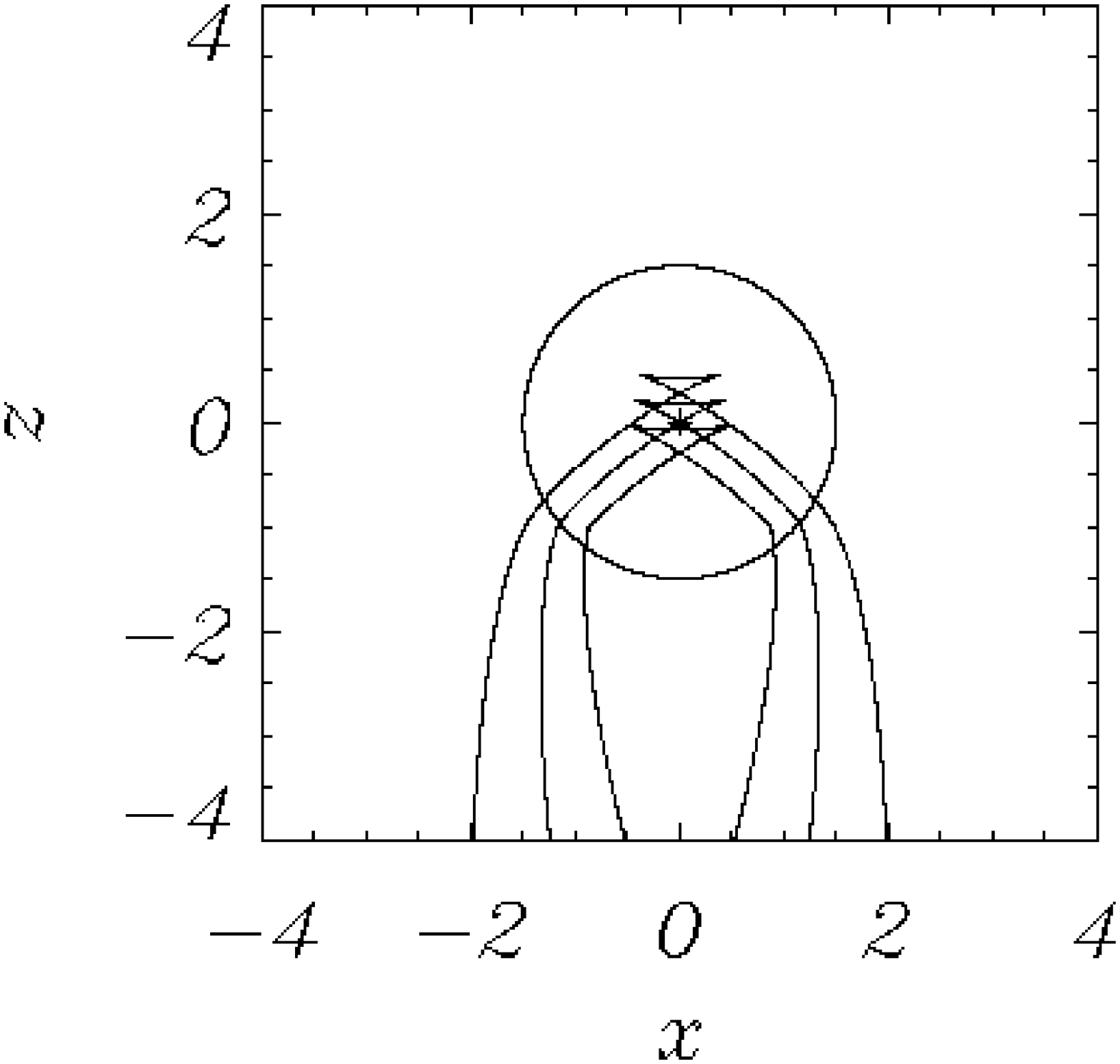}\vspace{0.05in}\\
\includegraphics[width=1.5in]{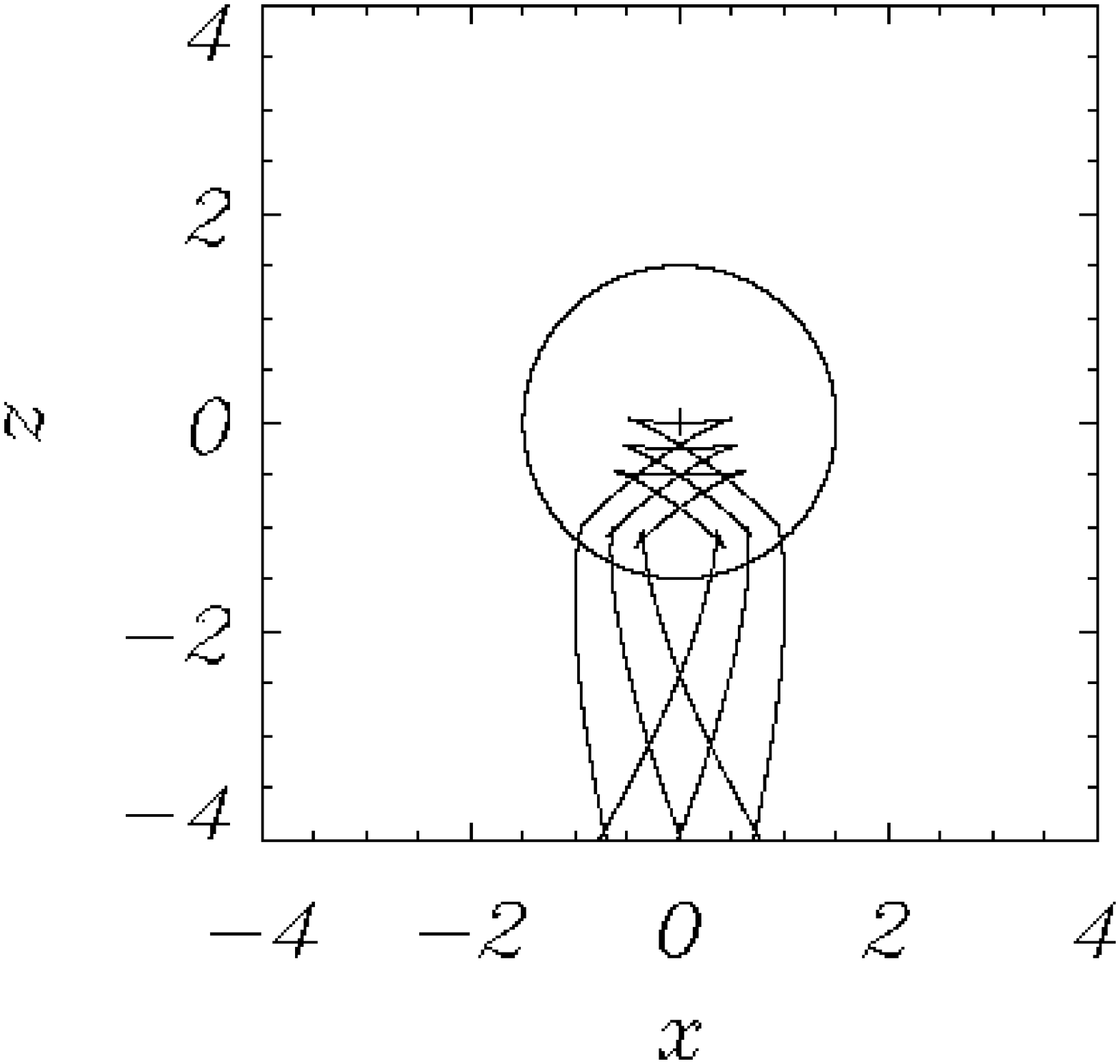}
\hspace{0.1in}
\includegraphics[width=1.5in]{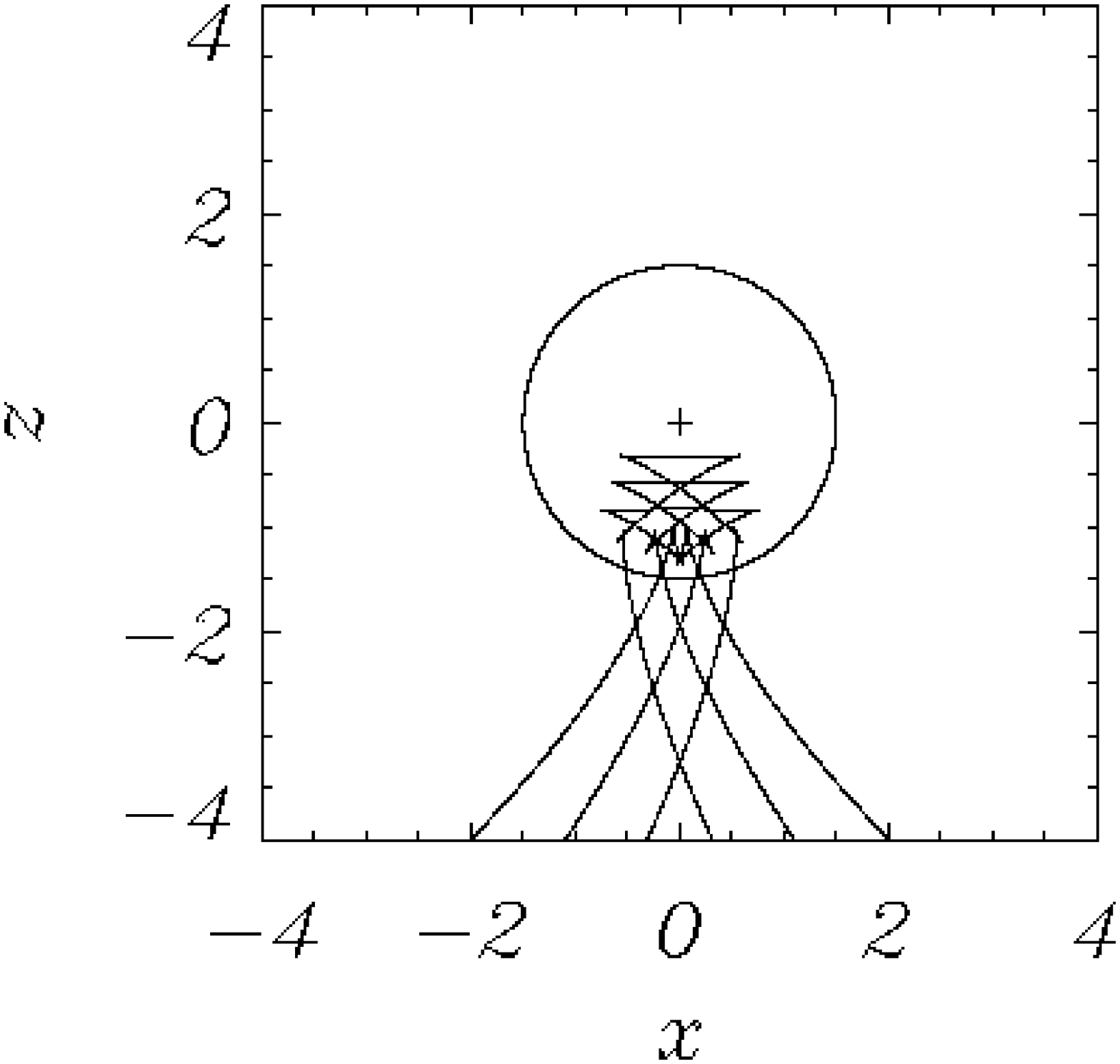}
\hspace{0.1in}
\includegraphics[width=1.5in]{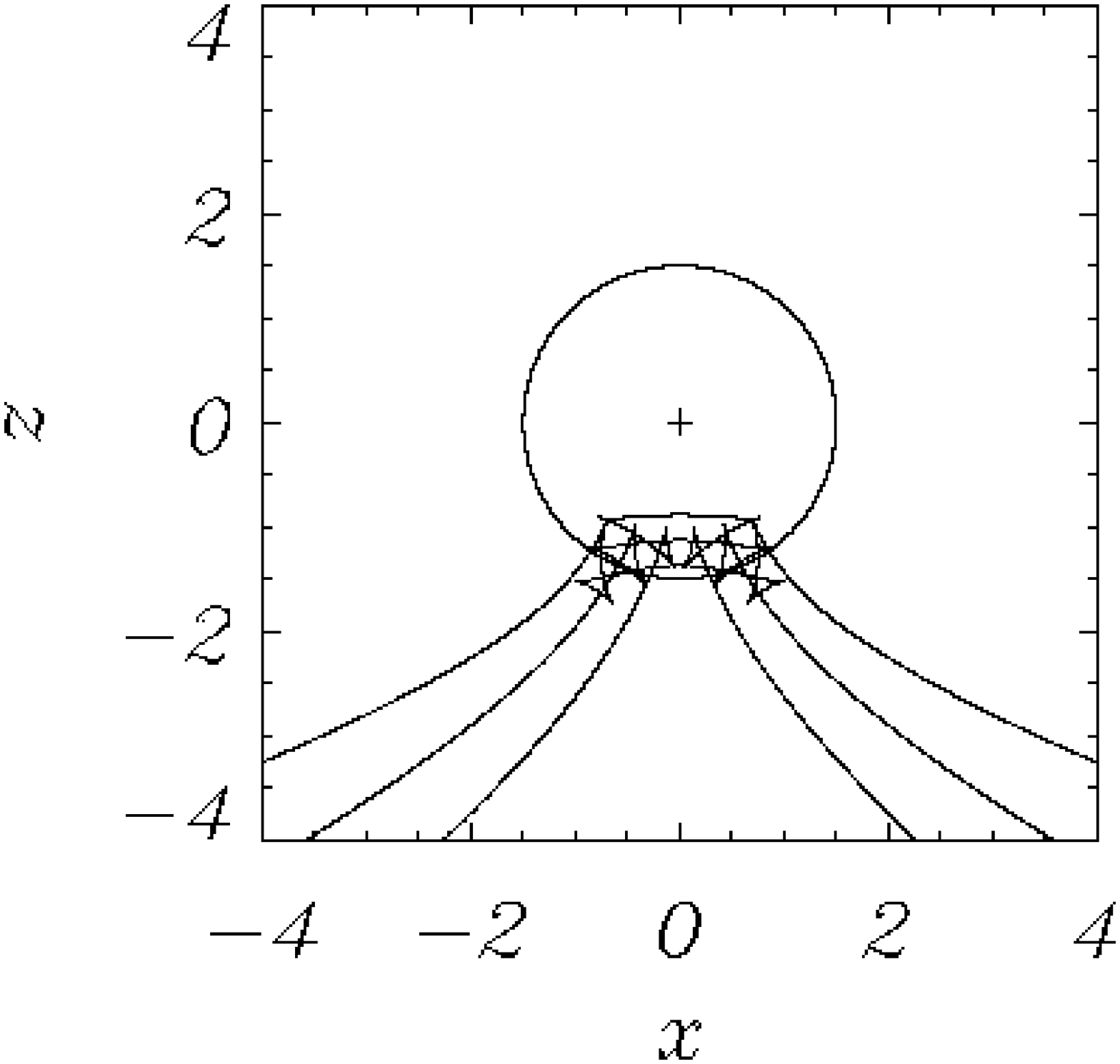}\vspace{0.05in}\\
\includegraphics[width=1.5in]{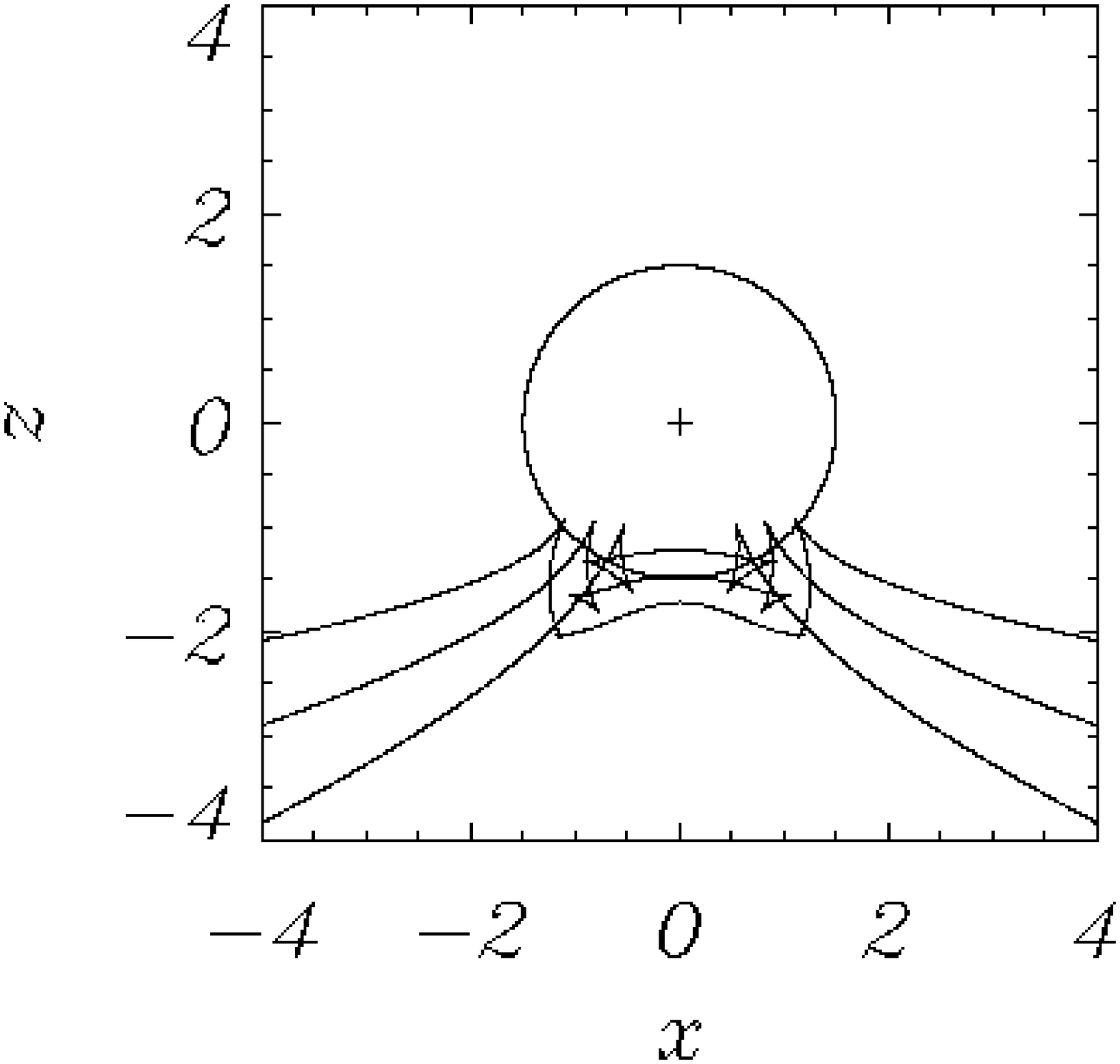}
\hspace{0.1in}
\includegraphics[width=1.5in]{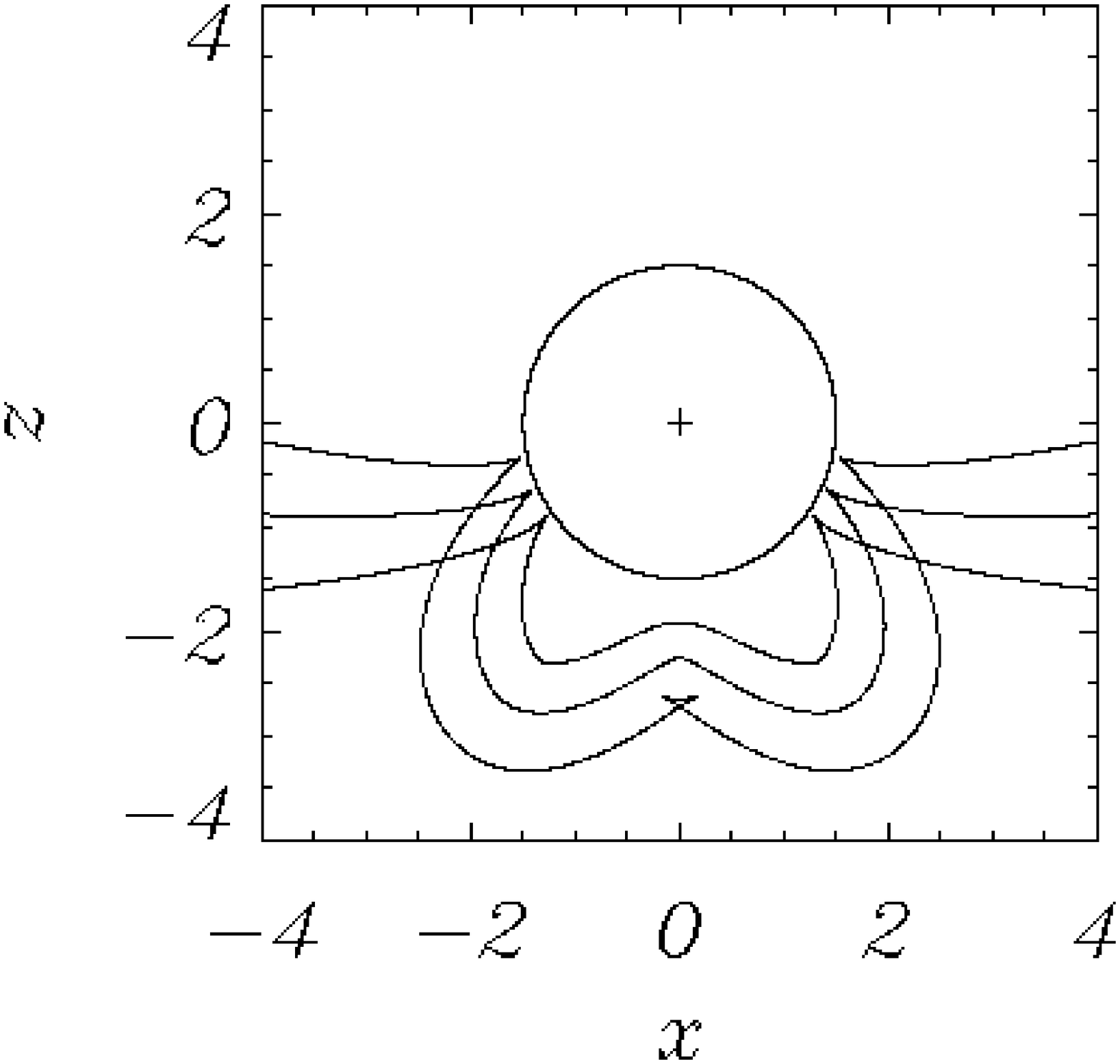}
\hspace{0.1in}
\includegraphics[width=1.5in]{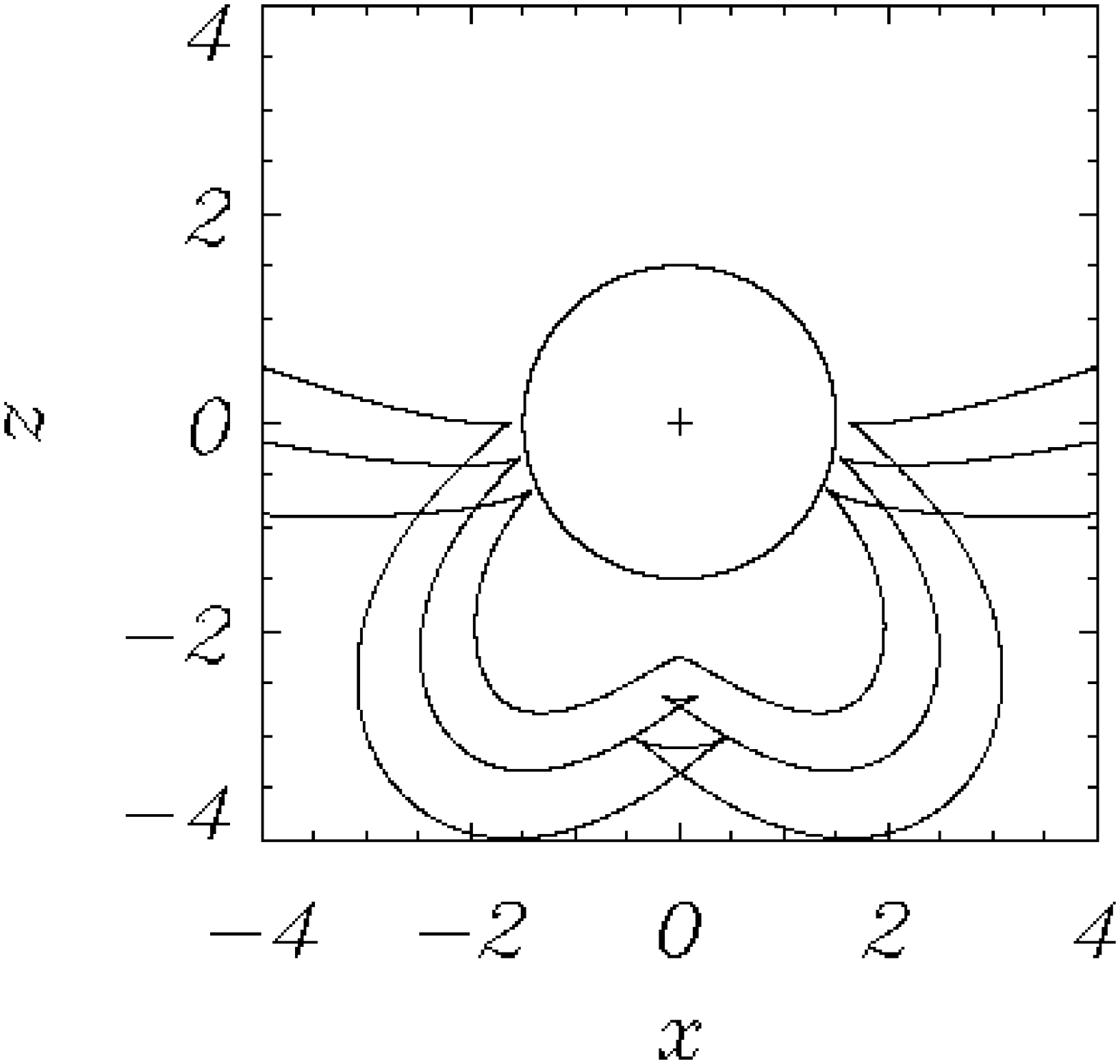}
\caption{Plots of WKB solution for a wave sent in from the upper 
boundary and its resultant positions at times
$(a)$ $t$=0.33, $(b)$ $t$=1.0, $(c)$ $t$=2.0, $(d)$ $t$=2.33, $(e)$ 
$t$=2.67, $(f)$ $t$=3.0, $(g)$ $t=$3.33, $(h)$
$t$=3.67 and $(i)$ $t$=4.0, labelling from top left to bottom right. 
The lines represent the front,
  middle and back edges of the wave. The black circle indicates the 
position of the $c_s=v_A\:$ layer and the cross denotes
the null point in the magnetic configuration.}
\label{WKB_system_beta_nonzero}
\end{center}
\end{figure*}
\end{center}

\begin{figure*}
\begin{center}
\includegraphics[width=1.5in]{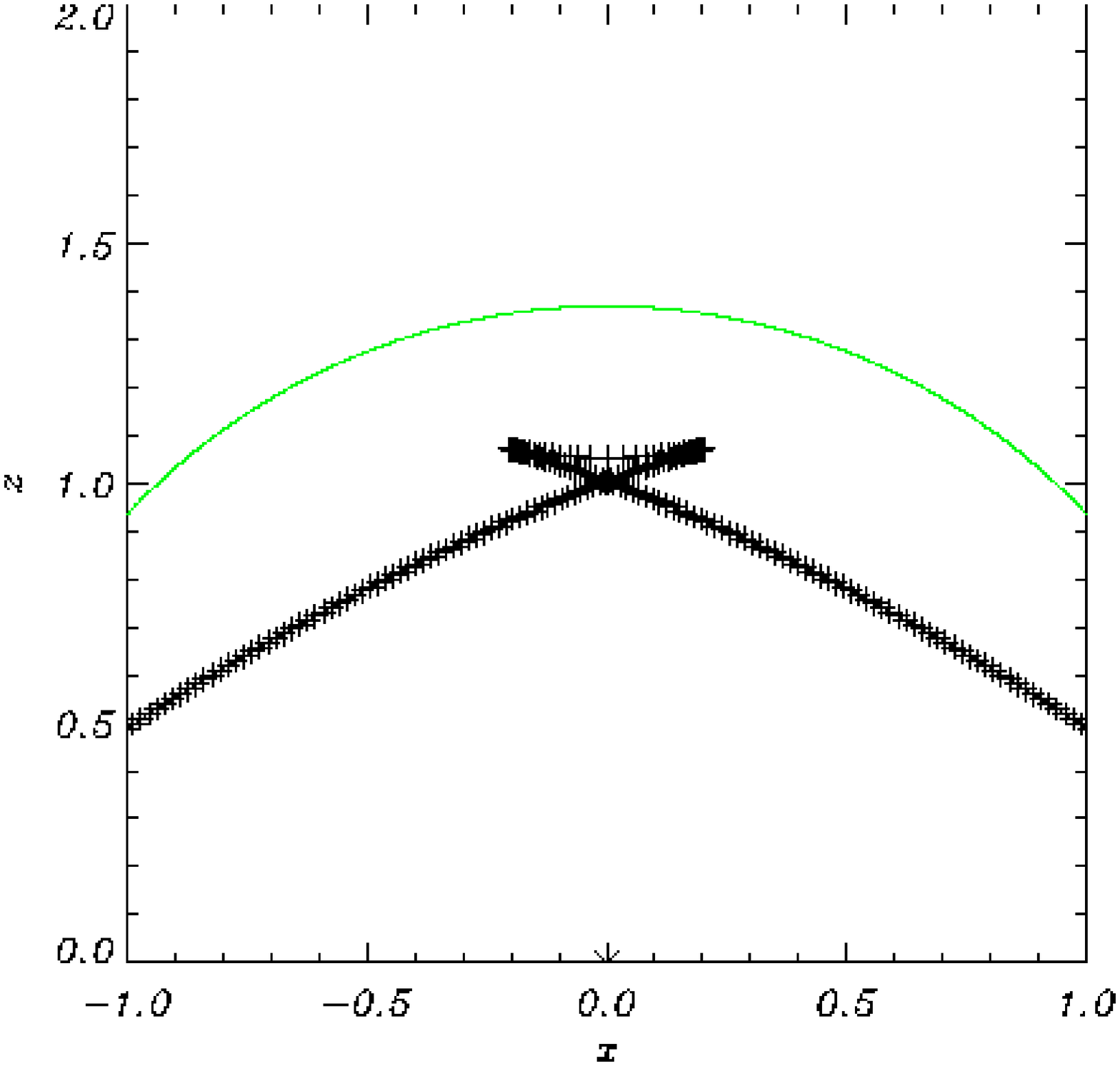}
\hspace{0.05in}
\includegraphics[width=1.5in]{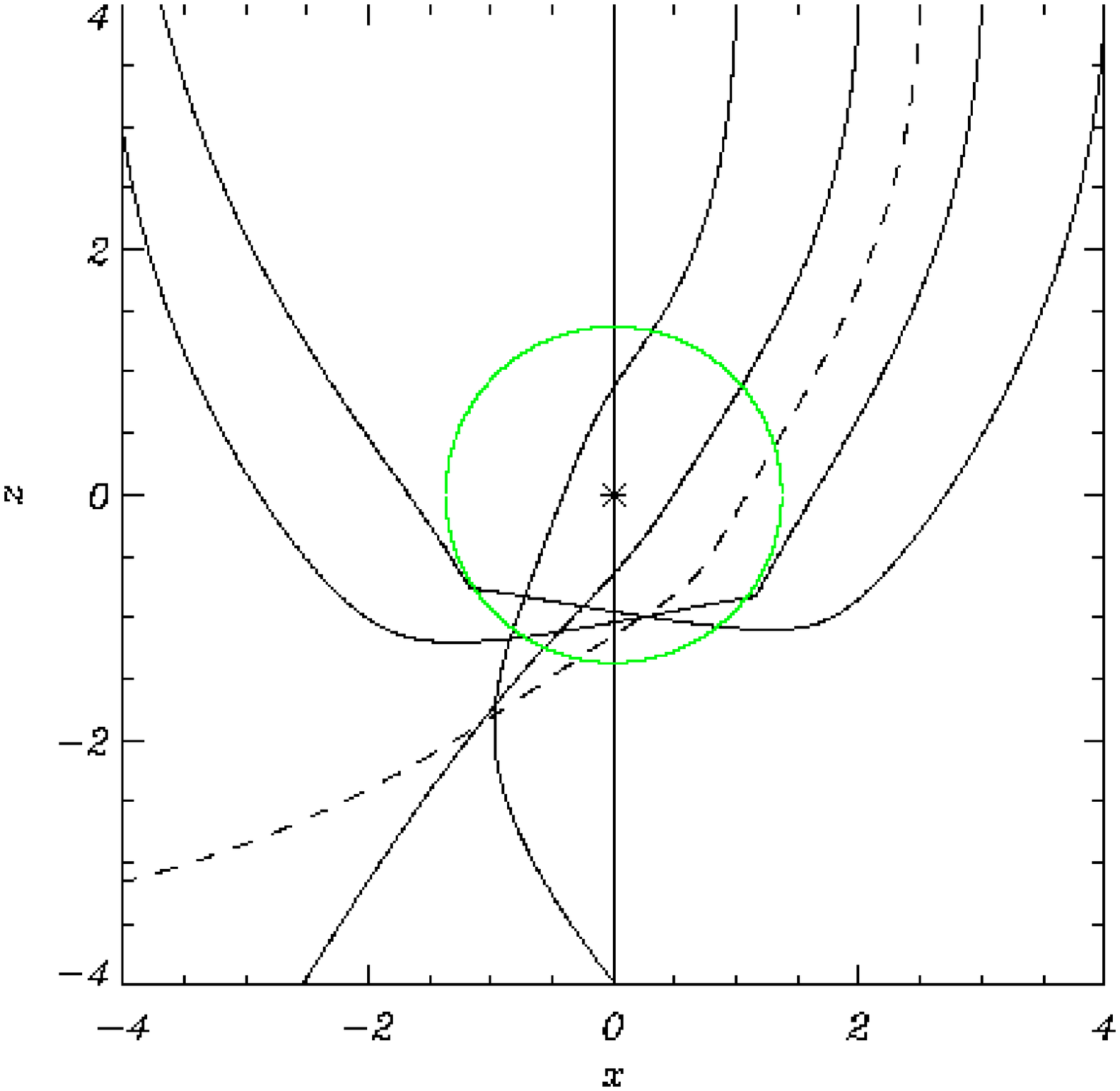}
\hspace{0.05in}
\includegraphics[width=1.5in]{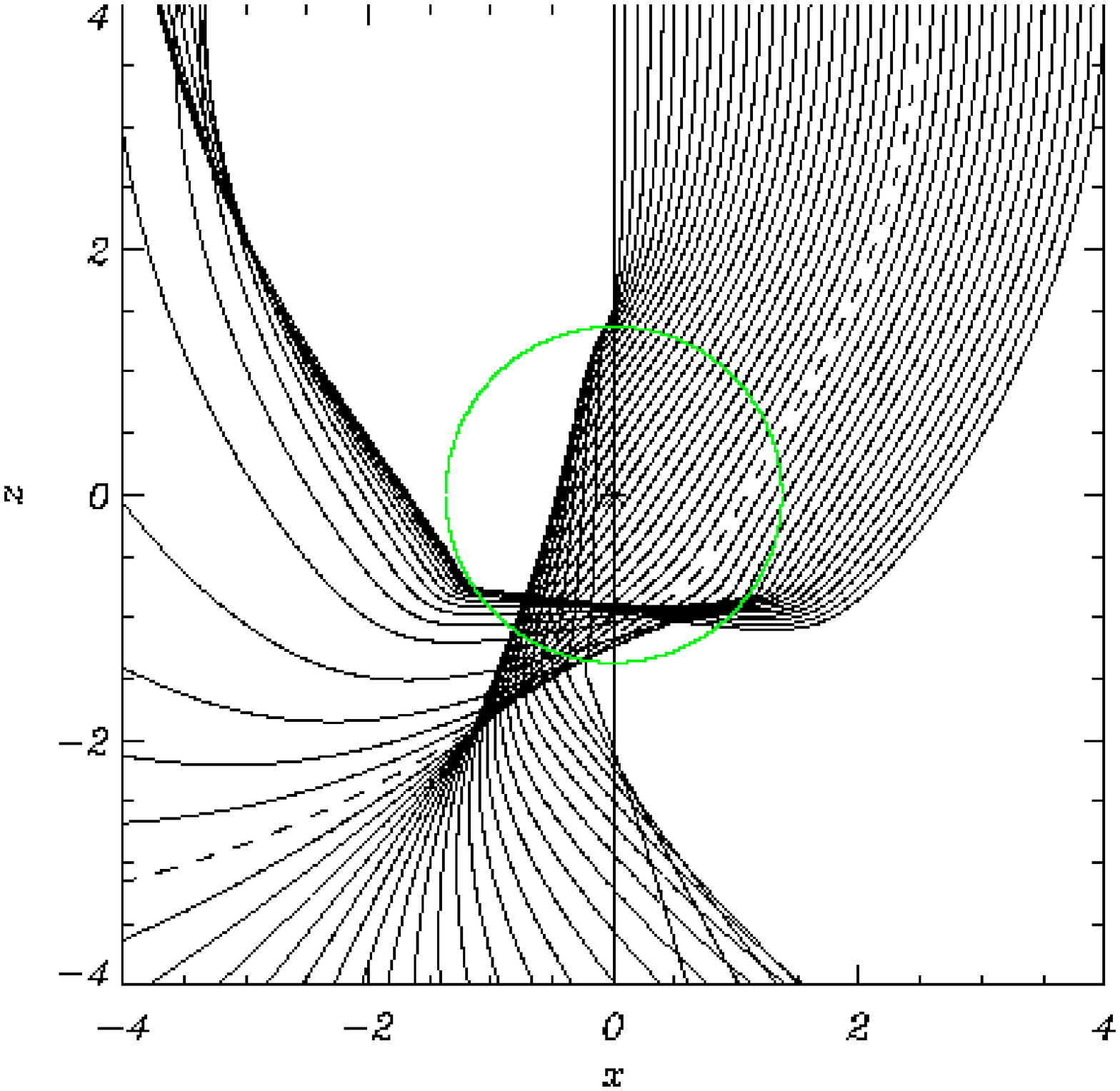}
\caption{Graphs showing various aspects of the analytical WKB 
approximation. {\emph{Left}} shows blow-up of wavefront at time
$t=0.65$. {\emph{Center}} shows rays for WKB solution for a wave sent 
in from the upper boundary for starting points of
$x=0$, $1$, $2$, $2.5$, $3$ and $4$ along $z=4$. {\emph{Right}} shows 
same rays but for $x \in[0,4]$ at intervals of 0.1.
  The green circle indicates the position of the $c_s=v_A\:$ layer and 
the star denotes the null point.}
\label{figuretwelve_JAMES}
\end{center}
\end{figure*}

We see that the agreement between the numerical simulation and the 
analytical approximation is quite good ({{see Figure 
\ref{newnew}}}). In Figure
\ref{WKB_system_beta_nonzero}, we see the main features of the 
numerical simulation;  the wave refracts as it approaches
the null and that part of the wave passes  through the null (at a 
slower speed than the 'wings' outside the
$c_s=v_A\:$ layer). This central part (that passed through the null) 
then  emerges from the  $c_s=v_A\:$ layer and
spreads out isotropically. Meanwhile, the wings continue to refract 
around the null. Also, as the wave crosses the
$c_s=v_A\:$ layer, the wavefront overlaps with itself (forming small 
triangular shapes, see Figure \ref{figuretwelve_JAMES});
this may explain why we see the perpendicular wave form a sharp edge 
as it crosses the   $c_s=v_A\:$ layer. However,
the crossing of neighbouring rays frequently indicates the breakdown 
of the WKB solution, as happens at caustics. This is
almost certainly where the conversion of part of the fast wave into a 
slow wave occurs and this process requires a more
detailed study of the WKB equations. Instead of the triangular shape 
propagating at the fast speed it should propagate at
the slow speed.

The rays of the WKB solution can be seen in Figure 
\ref{figuretwelve_JAMES}. The left graph shows the smalltriangular 
shape
formed as the wavefront overlaps with itself when it crosses the 
$c_s=v_A\:$ layer, indicating that this is
not just a discontinuity in the wavefront (i.e. since it can be 
resolved). The central graph  shows the
rays for starting points of $x=1$, $2$, $2.5$, $3$ and $4$ along 
$z=4$. We see that for a starting point of $x=3$ and $x=4$
($z=4$), the ray is deflected substantially by the null point, 
whereas the deflection is less severe for  $2$ and $1$.
A starting point of $x=2.5$  seems to be the critical starting point 
that determines if  $\frac {dz}{ds}$ changes sign (at
least in the range $-4 \leq x \leq 4$). We also see that $x=0$ is not 
deflected at all.  The right hand  graph  shows the
rays for for $x \in[0,4]$ in intervals of 0.1. The dashed line 
represents $x=2.5$; for $x > 2.5$, the ray path is
deflected so much that $\frac {dz}{ds}$ changes sign.  For  $x < 
2.5$, the rays are deflected but escape from the
opposite corner of the system. This nature reflects the numerical 
solution; part of the wave passes through the null and part
refracts around it.

However, the analytical WKB approximation does not give a full
description of the numerical simulation. Firstly, the WKB
approximation breaks down when neighbouring rays cross and this
requires a more detailed study to demonstrate exactly how the fast 
and slow waves interact.
Secondly, the WKB approximation can also be applied
to the $\omega^2 - \omega_{\textrm{\scriptsize{slow}}}^2=0$ equation 
but this does not greatly add to our understanding of the system.

\section*{Acknowledgements}
James McLaughlin acknowledges financial assistance from the Particle 
Physics and Astronomy Research Council (PPARC).
He also wishes to thank Tom Bogdan, Toni D\'iaz and Erwin Verwichte 
for helpful and insightful discussions.

\end{document}